\documentclass{aa}  
\usepackage{txfonts}
\usepackage{graphicx}
\usepackage{mathrsfs}
\usepackage[layout=inline,draft]{fixme} 
\usepackage{siunitx}
\usepackage{amssymb,amsfonts,hyperref,subfig}
\usepackage{soul}
\usepackage[dvipsnames]{xcolor}\usepackage{comment}
\usepackage{siunitx}

\title{Planet-disk-wind interaction: the magnetized fate of protoplanets}
\titlerunning{Planet-disk-wind interaction: the magnetized fate of protoplanets}
\author{
    Gaylor Wafflard-Fernandez\inst{\ref{i:ipag}}%
    \thanks{\email{gaylor.wafflard@univ-grenoble-alpes@fr}}
\&
    Geoffroy Lesur\inst{\ref{i:ipag}}
}
\institute{
    Univ. Grenoble Alpes, CNRS, IPAG, 38000 Grenoble, France\label{i:ipag}
}
\authorrunning{GWF \& GL}

\date{Received xxx; accepted xxx}

\begin{document}

\abstract
{
Models of planet-disk interaction are mainly based on two and three dimensional viscous hydrodynamical simulations. In such models, accretion is classically prescribed by an $\alpha_\nu$ parameter which characterizes the turbulent radial transport of angular momentum in the disk. This accretion scenario has been questioned for a few years and an alternative paradigm has been proposed that involves the vertical transport of angular momentum by magneto-hydrodynamical (MHD) winds.
}
{
We revisit planet-disk interaction in the context of MHD wind-launching protoplanetary disks. In particular, we focus on the planet's ability to open a gap and produce meridional flows. Accretion, magnetic field and wind torque in the gap are also explored, as well as the evaluation of the gravitational torque exerted by the disk onto the planet.
}
{
We carry out high-resolution 3D global non-ideal MHD simulations of a gaseous disk threaded by a large-scale vertical magnetic field harboring a planet in a fixed circular orbit using the GPU-accelerated code Idefix. We consider various planet masses ($10$ Earth masses, $1$ Saturn mass, $1$ Jupiter mass and $3$ Jupiter masses for a Solar-mass star) and disk magnetizations ($10^4$ and $10^3$ for the $\beta$-plasma parameter, defined as the ratio of the thermal pressure over the magnetic pressure).
}
{
We find that gap-opening always occurs for sufficiently massive planets, typically of the order of a few Saturn masses for $\beta_0=10^3$, with deeper gaps when the planet mass increases and when the initial magnetization decreases. We propose an expression for the gap opening criterion when accretion is dominated by MHD winds. We show that accretion is unsteady and comes from surface layers in the outer disk, bringing material directly towards the planet poles. A planet gap is a privileged region for the accumulation of large-scale magnetic field, preferentially at the gap center or at the gap edges for some cases. This results in a fast accretion stream through the gap, which can become sonic at high magnetizations. The torque due to the MHD wind responds to the planet presence in a way that leads to a more intense wind in the outer gap compared to the inner gap. More precisely, for massive planets, the wind torque is enhanced as it is fed by the planet torque above the gap's outer edge whereas the wind torque is seemingly diminished above the gap's inner edge due to the planet-induced deflection of magnetic field lines at the disk surface. This induces an asymmetric gap, both in depth and in width, that progressively erodes the outer gap edge, reducing the outer Lindblad torque and potentially reversing the migration direction of Jovian planets in magnetized disks after a few hundreds of orbits. For low-mass planets, we find strongly fluctuating gravitational torques that are mostly positive on average, indicating a stochastic outward migration.
%
}
{The presence of MHD winds strongly affects planet-disk interaction, both in terms of flow kinematics and protoplanet migration. This work illustrates the tight dependence between the planet torque, the wind torque and magnetic field transport that is required to get the correct dynamics of such systems. In particular, many of the predictions from "effective" models that use parameterized wind torques are not recovered (such as gap formation criteria, migration direction and speed) in our simulations.}


\maketitle

\section{Introduction}

The question of accretion and its origin in the context of protoplanetary disks is an utmost step if we want to develop reliable and realistic models, whether it be of the evolution of the disk itself \citep[][]{Pascucci2022}, of its constituents such as the gas, the dust and the planets, but also of their interactions \citep[see, e.g.,][]{Rabago&Zhu2021}.

Observational evidence of an ultraviolet excess in the spectral energy distribution gives through continuum fitting methods a quantitative estimate of the stellar accretion rate, with a typical value of $10^{-8}~\rm{M_\sun.yr^{-1}}$ \citep[][]{Venuti2014}. In order to explain such high values of the stellar accretion rate, one needs to rely on mass and angular momentum transport models. The conventional viscous model relies on a turbulent transport of angular momentum. It aims at parametrizing the disk's turbulent viscosity $\nu$ with an $\alpha_\nu$-parameter \citep[][]{Shakura1973}. Turbulence in disks is in general supposed to be the consequence of the non-linear saturation of the Magneto-Rotational Instability \citep[MRI,][]{Balbus&Hawley1991, Lesur2022}. Although the MRI is the most promising instability to explain turbulence-driven accretion, recent global simulations of protoplanetary disks with non-ideal MHD effects have shown that the MRI may not occur in most regions of protoplanetary disks \citep[see, e.g.,][]{Perez-Becker&Chiang2011a, Perez-Becker&Chiang2011b, Bai2013b, Lesur2014}. Beside the MRI, a wide range of hydrodynamical instabilities can play a role in the production of turbulence and therefore accretion, like the Gravitational Instability \citep[GI,][]{Kratter&Lodato2016, Bethune&Latter2022} or the Vertical Shear Instability \citep[VSI,][]{Nelson2013, Stoll&Kley2014} but they rely on either very massive disks (GI) or on very fast cooling timescales (VSI) which are probably not always applicable to Class II objects \citep[][]{Lesur2022}.

The diversity of increasingly resolved observations provides valuable constraints on disk accretion models. In particular, the turbulence-driven accretion model agrees with the inferred disk lifetimes and measured stellar accretion rates if $\alpha_\nu$ lies in the range $10^{-4}-10^{-2}$. Moreover, such hydrodynamical models are able to reproduce flows falling from the disk surface towards the disk midplane observed via $^{12}$CO emission at the radial locations of dark rings, as in \citet{Teague2019}. Such fast meridional flows ($\simeq0.1~c_s$, with $c_s$ the local sound speed) are expected to occur in planet-induced gaps, and would result from the refill of material at the gap edges \citep[see, e.g.,][]{Szulagyi2014, Morbidelli2014, Fung&Chiang2016}. However, other observational constraints tend to challenge this turbulence-driven accretion picture, when trying to estimate the level of turbulence in protoplanetary disks. Indeed, both the turbulent broadening of CO ro-vibrationnal line emissions \citep[][]{Flaherty2015,Flaherty2017} and dust settling measures from rings \citep{Pinte2016} or edge-on disks \citep{Villenave2020} all tend to suggest very low values of turbulence with $\alpha_\nu<10^{-4}-10^{-3}$ close to the midplane, a value much smaller than expected from full-blown MRI-driven turbulence.


Knowing that the gas mass should be efficiently transported from the outer parts of the disk towards the star and yet that the level of turbulence is (much) lower than expected has triggered the development of another paradigm of disk evolution, different from the viscous accretion model. In the MHD wind-driven accretion model, angular momentum is evacuated from the surface of the disk by a magnetized wind \citep[][]{Blandford&Payne1982}, which induces a radial laminar transport of the gas mass following the idea of \cite{Wardle1993}. This model reconciles at once the high stellar accretion rate, the \textit{a priori} low turbulence, but also part of the disk dispersal in the latest stages of protoplanetary disks evolution \citep[][]{Bai&Stone2013,Lesur2014}. Both accretion models (viscous and MHD wind) are not mutually exclusive, and come both from angular momentum conservation. Wind-driven accretion is an even more promising mechanism as it is supported by the detection of outflows coming from the inner parts of protoplanetary disk that are probably too massive to be explained by thermal winds only \citep[e.g.][]{Louvet2018,deValon2020}. See also \citet{Pascucci2022} for a comprehensive review of recent observations probing gaseous outflows.

An important challenge is to have a thorough understanding of the MHD wind-driven accretion mechanisms in order to be able to prescribe the impact of these winds on the gas in phenomenological models in one or two dimensions. It corresponds to the same process that made it possible to carry out 2D hydrodynamical simulations with an $\alpha_\nu$ prescription in viscous disks. For instance, \citet{Tabone2022} derived 1D equations modeling mass and momentum transport by a wind, and parameterized by an $\alpha_\nu$-like dimensionless parameter, noted here $\alpha_{\rm dw}$. Such prescription is necessary in order to study long-term disk evolution or to perform systematic parameter space exploration, as highly resolved 3D MHD global simulations are time and energy consuming.

Recently, more and more studies have investigated disk/planet interactions and planetary migration in wind-driven accretion disks, using approaches similar to \citet{Tabone2022}. For example, \citet{Kimmig2020} have explored how the wind-driven accretion process affects the migration of massive planets (Saturn-mass and Jupiter-mass planets) via 2D hydrodynamical simulations, with prescribed wind torques and ejection rates. They found in particular that planets can undergo episodes of runaway type-III-like outward migration \citep[see, e.g.,][]{Masset2003,Peplinski2008c}, if the wind ejection rate is sufficiently large. Concerning type-I planet migration with a wind-driven mass-loss \citep[][]{Ogihara2015} and with an inward gas flows induced by a wind torque at the midplane \citep[][]{Ogihara2017}, studies have shown via simplified 1D models that type-I migration can also be slowed down and even reversed depending on the wind efficiency, with a positive unsaturated corotation torque that prevents the infall of super-Earths onto the central star. \citet{McNally2020} have focused on the migration of low-mass ($6.7$ Earth-mass) planets in a wind-driven accretion disk, via inviscid 3D hydrodynamical simulations, with a prescribed wind torque localized on the disk surface. They found a decoupling between the accretion layers and the passive planet-bearing midplane, leading to a negligible impact of the wind torque on the migration of the embedded planet. \citet{Lega2022} have conducted a similar study, but focusing on massive planets ($M_j$ and $M_s$). They were able to identify two phases in the migration of giant planets, related to the formation and evolution of a vortex at the gap's outer edge. \citet{Elbakyan2022} have examined via 1D and 2D hydrodynamical simulations the ability of fixed massive planets to open gaps in wind-driven accretion disks parametrized with a prescribed wind torque. They reported that opening deep gaps is easier when angular momentum transport is dominated by magnetized winds, and therefore concluded that gap opening planets may be less massive than currently believed. Finally, \cite{Aoyama2023} have recently shown via 3D global non-ideal MHD simulations that magnetic flux tends to accumulate in the corotation region of massive planets, associated with nearly sonic accretion streamers and an enhanced angular momentum extraction from the gap region. They have shown that the gaps obtained in such MHD simulations are in essence deeper and wider than those obtained in viscous and inviscid simulations. Last, they have found that the Lindblad torques are reduced, and the total gravitational torque exerted by the gas onto the planet is negative and fluctuates stochastically. In this manuscript, we revisit in a more systematic manner and on longer timescales these processes, varying the field strength and including low mass planets which do not open gaps.

Most of these approaches rely on prescribed MHD wind torque (and in some cases also ejection rates) with very simple functional dependencies. However, it is well known that the wind torque and ejection rate depend strongly on the disk magnetization, which is set by the disk surface density and the poloidal magnetic field strength \citep{Lesur2021}. In turn, the surface density and the distribution of poloidal field is expected to evolve as a result of the disk dynamics \citep{Guilet2012,Guilet2013,Guilet2014,Leung2019}. Hence, the planet wake and gap necessarily affect  the disk magnetization which in turn is expected to feedback on the wind properties in a way that is not captured by phenomenological models. Our aim is to circumvent these difficulties by carrying out self-consistent 3D non-ideal MHD simulations of a disk-planet-wind system  where the magnetic field is free to evolve. More specifically, we want to focus on the opening of a gap by a fixed planet in a wind-driven disk, and reciprocally consider the accretion behavior in the vicinity of such planets. We also want to provide some leads to improve the 1D/2D hydrodynamical models that use a wind torque prescription, as well as verify if planet gaps as formed with a wind can reproduce meridional flows.

The plan of this paper is the following. In Section~\ref{sec:method}, we describe the physical model and numerical methods of the non-ideal MHD simulations. The results are then presented in Section~\ref{sec:3d_results}, where we focus on the opening (Section~\ref{sec:GO}) and evolution (Sections~\ref{sec:HM2_magnetic} and \ref{sec:HM2_accretion}) of the planet gap, as well as meridional flows (Section~\ref{sec:HM2_MF}) and gravitational torques (Section~\ref{sec:torques}), important for planet migration. A summary and a discussion follow in Section~\ref{sec:conclusion}.

\section{Numerical methods and setups}
\label{sec:method}

We perform MHD simulations using the \href{https://gricad-gitlab.univ-grenoble-alpes.fr/lesurg/idefix-public}{\fontfamily{cmtt}\selectfont IDEFIX} code \citep{Lesur2023} that integrates the compressible MHD equations via a finite-volume method with a Godunov scheme. For this particular problem, {\fontfamily{cmtt}\selectfont IDEFIX} is run on the Jean Zay cluster at IDRIS (France)  on Nvidia V100 GPUs using a second order Runge-Kutta scheme, second order spatial reconstruction and the HLLD Riemann solver. The solenoidal condition is enforced using the constrained transport method and using an electromotive force reconstruction based on the contact wave upwinding strategy of \cite{Gardiner2005}. Diffusion terms (ambipolar diffusion and Ohmic resistivity) are integrated using the second order Runge-Kutta Legendre (RKL) scheme \citep{Meyer2014}. Note that resistivity is used here only as an inner boundary damping, as indicated in Sections~\ref{sec:non_ideal_MHD} and \ref{sec:boundary}. The RKL scheme helps reduce the time to solution by at least a factor 2.5. We do not use the FARGO orbital advection scheme because the timestep is limited by the Alfvén velocity, not the azimuthal velocity of the gas at the inner radius.

\subsection{Grid and resolution}
\label{sec:method_grid}

We adopt a cylindrical coordinate system ($R$, $\phi$, $z$), with $R$ the radial cylindrical coordinate measured from the central star, $\phi$ the azimuthal angle, and $z$ the vertical direction. We will also make use of the spherical coordinate system ($r$, $\theta$, $\phi$), $r$ being the radial spherical coordinate and $\theta$ the colatitude.

In the azimuthal direction, the grid extends uniformly from 0 to 2$\pi$ with 2048 cells. In the radial direction, we use two blocks with 512 evenly-spaced cells from 0.3$R_p$ to 1.9$R_p$, and an outer stretched grid with 128 cells between 1.9$R_p$ and 6.0$R_p$ where $R_p$ is the (fixed) planet location. Concerning the vertical direction, the grid is refined around the midplane with 256 evenly-spaced cells between -0.4$R_p$ and 0.4$R_p$, and two stretched grids symmetric around the midplane with 64 cells on both sides such that the total vertical extension spans between -6.0$R_p$ and 6.0$R_p$. Around the planet location at $R=R_p=1$, the radial, azimuthal and vertical resolutions correspond to 16 points per disk pressure scale height $H(R)=h R$. The total resolution is therefore $N_R \times N_\phi \times N_z=640\times2048\times384$. The aspect ratio $h$ is considered constant, such that $h=h_0=0.05$.

\subsection{Average conventions}

For a quantity Q, we define the azimuthal average, the temporal average between $T_1$ and $T_2$, and the vertical integral between altitudes $z_-$ and $z_+$ respectively in Eq.~(\ref{eq:average_phi}), (\ref{eq:average_t}) and (\ref{eq:average_z}):
\begin{equation}
    \langle Q \rangle_\phi = \displaystyle \frac{1}{2\pi} \int_{0}^{2\pi}Q d\phi,
    \label{eq:average_phi}
\end{equation}
\begin{equation}
    \langle Q \rangle_{T_1-T_2} = \displaystyle \frac{1}{T_2-T_1}, \int_{T_1}^{T_2}Q dt,
    \label{eq:average_t}
\end{equation}
\begin{equation}
    \overline{Q} = \displaystyle \int_{z_-}^{z_+}Q dz.
    \label{eq:average_z}
\end{equation}

\noindent We will make use of the notation $\langle Q \rangle_{t}$ if we temporally average a quantity over the last $100$ orbits of the simulation. We note $\tilde{Q}$ the value of Q at the disk midplane. Finally, we define as
\begin{equation}
    \displaystyle \left[Q\right]_{-}^{+} = Q|_{z_{\rm w}}-Q|_{-z_{\rm w}}
\end{equation}

\noindent the difference of Q between the disk's upper layer at $z_{\rm w}$ and its lower layer at $-z_{\rm w}$, with $z_{\rm w}\simeq4H(R)$.

\subsection{Planet}
\label{sec:method_planet}

The planet is held on a fixed circular orbit at $r_p=1$ in the disk midplane.  Several planet-to-primary mass ratios $q_p$ are chosen for the planet to explore the planet-disk-wind interaction: $3\times10^{-3}$, $10^{-3}$, $3\times10^{-4}$ and $3\times10^{-5}$, which correspond respectively to $M_p=3M_{j}$, $M_{j}$, $M_{s}$ and $10M_{\oplus}$ for a Solar-mass star. We will make use of the planet's Hill radius, defined as 
\begin{equation}
    R_{\rm hill}=(q_p/3)^{1/3}R_p. 
    \label{eq:rhill}
\end{equation}

\noindent Accretion on the planet is neglected, as well as planet migration. The mass of the planet is gradually increased over its first $N_0=10$ orbits until it reaches $q_p$, according to $q_p(t)=\displaystyle q_p\sin^2{\left(\frac{t}{4 N_0}\right)}$, with $t$ in units of $\Omega^{-1}$. The total gravitational potential $\Phi$ considered in the simulation is:
\begin{equation}
    \Phi = \Phi_{\star} + \Phi_p + \Phi_{\rm ind}.
    \label{eq:total_potential}
\end{equation}

\noindent $\Phi_{\star}$ is the term due to the central star:
\begin{equation}
    \Phi_{\star}(\vec{r}) = -\displaystyle\frac{1}{|\vec{r}|},
\end{equation}

\noindent $\Phi_p$ is the gravitational potential of the planet:
\begin{equation}
    \Phi_p(\vec{r}) = - \displaystyle\frac{q_p}{\sqrt{|\vec{r}-\vec{r_p}|^2+\epsilon^2}},
    \label{eq:planet_potential}
\end{equation}

\noindent where $\epsilon$ is a softening length that avoids divergence at the planet location. $\epsilon$ is set to $0.1H(r_p)$, which is slightly larger than the characteristic size of a grid cell at $r_p$. Finally, we take into account the indirect term $\Phi_{\rm ind}$ arising from the acceleration of the star by the planet:
\begin{equation}
    \Phi_{\rm ind}(\vec{r}) = \displaystyle\frac{q_p(\vec{r}\cdot\vec{r_p})}{|\vec{r_p}|^3}.
    \label{eq:ind_potential}
\end{equation}

\noindent All simulations have been run for at least $200$ orbits, which is a good compromise between the computational cost and the characteristic timescales of disk/planet interactions. 

\subsection{Gas}
\label{sec:method_gas}

The setup that we use for the gas is similar to the one presented in \citet{Riols2020,Martel&Lesur2022}. We summarize here some of the main features of this setup.

\subsubsection{Non-ideal MHD}
\label{sec:non_ideal_MHD}

We solve the non-ideal MHD equations, for the density $\rho$, the velocity field $\vec{v}$ and the magnetic field $\vec{B}$:
\begin{equation}
    \displaystyle \frac{\partial \rho}{\partial t} + \vec{\nabla} \cdot \left(\rho \vec{v}\right) = 0,
\end{equation}
\begin{equation}
    \displaystyle \frac{\partial \rho \vec{v}}{\partial t} + \vec{\nabla} \cdot \left(\rho \vec{v} \otimes \vec{v} \right) = -\rho\vec{\nabla}\Phi -\vec{\nabla} \rm{P} +\vec{J}\times\vec{B},
\end{equation}
\begin{equation}
    \displaystyle \frac{\partial \vec{B}}{\partial t} = \vec{\nabla}\times\left(\vec{v}\times\vec{B}\right) + \vec{\nabla}\times\left[\eta_{\rm A}\left(\vec{J}\times\vec{e_{\rm b}}\right)\times\vec{e_{\rm b}}-\eta_{\rm O}\vec{J}\right],
\end{equation}

\noindent where $\Phi$ is the total gravitational potential defined in Eq.~(\ref{eq:total_potential}--\ref{eq:ind_potential}), $\rm P$ the gas pressure, $\vec{J}=\nabla\times\vec{B}$ is the current density, $\vec{e_{\rm b}}$ the unit vector parallel to the magnetic field line, $\eta_{\rm A}$ the ambipolar diffusivity and $\eta_{\rm O}$ the Ohmic diffusivity. We neglect the contribution of the Hall effect, and the Ohmic resistivity is included only as a damping process close to the inner radial boundary (see Section~\ref{sec:boundary}).

Regarding ambipolar diffusion, we define the dimensionless ambipolar Elsässer number
\begin{equation}
    \displaystyle A_{\rm m} = \frac{v_A^2}{\Omega_K\eta_{\rm A}},
\end{equation}

\noindent with $v_A = B/\sqrt\rho$ the Alfvén speed and $\Omega_K$ the keplerian angular frequency. To keep the model as simple as possible, we follow \cite{Lesur2021} and prescribe an ambipolar diffusion profile instead of computing a ionisation model which would make the computation prohibitive.  In the simple prescription that we use here, the midplane Elsässer number $\tilde{A}_{\rm m}$ is constant and equal to $1$, similarly to full radiative models \citep[e.g.][]{Thi2019}. At higher altitudes, $A_{\rm m}$ increases abruptly to mimic the ionisation due to X-rays and far UVs at the disk surface.  The diffusivity profile we use is
\begin{equation}
    \displaystyle A_{\rm m} = \max\left\{\tilde{A}_{\rm m} \exp{\left[\left(\frac{z}{\epsilon_{\rm id} H(R)}\right)^4\right]},\frac{1}{10}\left(\frac{v_A}{c_s}\right)^2\right\},
\end{equation}

\noindent with $\epsilon_{\rm id}=6.0$ in all our simulations, which quantifies the thickness of the non-ideal layer. We cap $A_{\rm m}$ in order to save some computational time. Note that we would actually expect the ionization by X-rays from the central star to increase when the gas is depleted in a planet gap, leading to a stronger coupling with magnetic fields, as reported in \cite{Kim&Turner2020}. As we prescribe $A_{\rm m}$, its dependence on $\Sigma$ is not taken into account here. For such $\tilde{A}_{\rm m}$, we expect the MRI to emerge at low amplitude and generate some weak turbulence \citep[see][and Appendix~\ref{sec:appendix_b2}]{Cui&Bai2022}.

The disk is initially threaded by a large-scale vertical magnetic field given by the profile
\begin{equation}
    \displaystyle B_z(R) = h_0\sqrt{\frac{2}{\beta_0}}R^{-5/4},
\end{equation}

\noindent with $\beta_0$ the initial value of the plasma parameter $\beta$, defined as the ratio of the thermal pressure $P$ over the magnetic pressure:
\begin{equation}
    \beta=\displaystyle\frac{2 P}{B_z^2}.
\end{equation}

\noindent Two $\beta_0$ are chosen in order to explore the influence of the magnetization on the interaction between the planet and the disk: $10^4$ and $10^3$.

Our numerical parameters are chosen to be close to more complete thermo-chemical models. For instance, a constant $\tilde{A}_{\rm m}\sim 1$ matches the thermo-chemical model of \cite{Thi2019} with $\tilde{\beta}=10^4$ in the midplane (see, e.g., the green region in the middle left panel of their figure~8). Moreover, Ohmic diffusion is negligible for $R>10~\rm{au}$ as these models have Elsässer number $\gg 1$ (see the top left panel of \citealt{Thi2019} figure~8). Hence, our numerical models with ambipolar diffusion only closely ressemble thermo-chemical models in the $10$-$100~\rm{au}$ range that corresponds also to the regions that are typically unveiled by radio interferometry with ALMA. In particular, sequences of dark and bright rings along with other substructures are commonly detected in such regions at millimeter wavelengths, in the continuum emission and CO emission lines (see, e.g., the DSHARP \cite{Andrews2018} and MAPS \cite{Oberg2021} programs).

\subsubsection{Initial disk properties}

The initial gas volume density arises from the hydrostatic disk equilibrium:
\begin{equation}
    \rho(R,z) = \displaystyle\frac{\Sigma(R)}{\sqrt{2\pi} H(R)} \exp{\left[\frac{1}{\tilde{c_s}^2(R)}\left(\frac{1}{\sqrt{R^2+z^2}}-\frac{1}{R}\right)\right]},
\end{equation}

\noindent with $\tilde{c_s}=h_0/\sqrt{R}$ the sound speed at disk midplane and $\Sigma$ the surface density following
\begin{equation}
    \Sigma(R)=\Sigma_0 \displaystyle R^{-0.5},
\end{equation}


\noindent which is coherent with some spectral lines observations \citep[see ][for the disk HD 163296]{Williams&McPartland2016}. We neglect self-gravity throughout this study, which is \textit{a priori} valid for Class II objects.

The simulation domain can be divided into a cold dense disk at temperature $\tilde{T}(R)=h_0^2/R$ and a low-density hot corona with temperature $T_{\rm cor}(R)=36~\tilde{T}(R)$ to account for the inefficient gas cooling via thermal accommodation on dust particles in the low-density hot corona \citep[see][e.g. their appendix F]{Thi2019}. We model the transition between these two regions with the following smooth vertical profile:
\begin{equation}
\begin{aligned}
    \displaystyle T_{\rm eff}(R,z) & = \frac{1}{2}\left(T_{\rm cor}(R)+\tilde{T}(R)\right) \\
    & + \frac{1}{2}\left(T_{\rm cor}(R)-\tilde{T}(R)\right)\tanh{\left(6\ln{\left|\frac{z}{\epsilon_{\rm id}H(R)}\right|}\right)}.
\end{aligned}
\end{equation}

\noindent We use a fast $\mathcal{B}$-cooling presciption, with $\mathcal{B}=0.1\Omega^{-1}$, in order to quickly relax the temperature $T(R,z)$ towards the prescribed profile $T_{\rm eff}(R,z)$. With such finite but short cooling timescale, we expect to be on the verge of the VSI-unstable regime \citep{Manger2021}. We discuss whether the VSI is present in our simulation in appendix~\ref{sec:appendix_b2}. The pressure and density are related by $P(R,z)=\rho(R,z)T(R,z)$. The cooling prescription is used to solve the energy equation, assuming an ideal equation of state for the gas, with an internal energy density $e(R,z)=\displaystyle\frac{P(R,z)}{\gamma-1}$, and an adiabatic index $\gamma=5/3$. Regarding the initial gas azimuthal velocity, we use the following sub-Keplerian velocity profile 
\begin{equation}
    v_\phi(R,z) = v_K \displaystyle \left(\frac{R}{\sqrt{R^2+z^2}}-1.5h_0^2\right),
\end{equation}

\noindent with $v_K=R\Omega_K=1/\sqrt{R}$ the keplerian speed. Note that the above expression is not an equilibrium in the hot corona because of the larger temperature in this region. This is not a problem since the MHD wind, which is neither a dynamical equilibrium, is launched very rapidly once the simulation has started and clear the initial condition in the corona.

\subsubsection{Boundary and internal conditions}
\label{sec:boundary}

At the inner radius, we copy the gas density and pressure, as well as $B_\phi$ and $B_z$ from the innermost active zone to the ghost zone. We impose $v_z=0$ and a keplerian velocity for $v_\phi$. The boundary condition is reflecting (symmetric) for $v_R$, such that the fluxes are null at the interface with the ghost zone. At the outer radius, we apply power-law expressions for all quantities to ghost zones, with radial power-law exponents of -1.5, -2.5, -0.5, -0.5, -0.5, -5/4 and -5/4 for $\rho$, $P$, $v_R$, $v_\phi$, $v_z$, $B_\phi$ and $B_z$ respectively to mimic self-similar models \citep{Lesur2021}. The boundary conditions are periodic in the azimuthal direction, and we apply outflow conditions for all the quantities in the vertical direction.

To avoid too small time steps, we limit the Alfvén speed in the simulation domain to the value of the keplerian speed at the inner radial boundary of the disk $R_{\rm in}=0.3$. This assumption leads to an overestimation of the density when the criterion $v_A>v_K(R_{\rm in})$ is encoutered, typically in the disk corona. We also add a threshold in density at $\rho_{\rm MIN}=10^{-12}$. Because the planet is kept fixed at $R_p=1$ and far from the outer radial boundary of the disk $R_{\rm out}=6$, we do not apply an outer damping zone. However, when $R<1.1R_{\rm in}$, we prescribe a simple damping zone for the radial and vertical components of the momentum ($\rho v_R$, $\rho v_z$). In this disk innermost region, we also add a band of resitivity such that $\eta_{\rm O}=H^2(R_{\rm in})\Omega_K(R_{\rm in})$ to avoid artifical magnetic coupling between the inner boundary condition and the computational domain.

\begin{table}[!ht]
\begin{center}
\begin{tabular}{cccc}
\hline
name & $M_p$ & $\beta$ & $N_{\rm orb}$ \\
\hline
\emph{$3$Mj-$\beta_3$} & $3M_{j}$ & $10^3$ & $200$ \\
\emph{$3$Mj-$\beta_4$} & $3M_{j}$ & $10^4$ & $200$ \\
\emph{Mj-$\beta_3$} & $M_{j}$ & $10^3$ & $400$ \\
\emph{Mj-$\beta_4$} & $M_{j}$ & $10^4$ & $200$ \\
\emph{Ms-$\beta_3$} & $M_{s}$ & $10^3$ & $200$ \\
\emph{Ms-$\beta_4$} & $M_{s}$ & $10^4$ & $400$ \\
\emph{$10$Me-$\beta_3$} & $10M_{\oplus}$ & $10^3$ & $200$ \\
\emph{$10$Me-$\beta_4$} & $10M_{\oplus}$ & $10^4$ & $200$ \\
\hline
\emph{$3$Mj-$\alpha_0$} & $3M_j$ & $\infty$ & $200$ \\
\emph{\rm{3D}-$\beta_3$} & $0$ & $10^3$ & $100$ \\
\emph{\rm{axi}-$\beta_3$} & -- & $10^3$ & $200$ \\
\emph{\rm{axi}-$\beta_4$} & -- & $10^4$ & $200$ \\
\\
\end{tabular}
\caption{Table that lists the simulations, characterized by their name, the planet mass and the $\beta$ parameter. Each simulation has been run during $N_{\rm orb}$ orbital periods. \label{table:runs}}
\end{center}
\end{table}

\subsection{Numerical protocol and parameter study}

Before introducing the planet, we run two axisymmetric simulations for $100$ orbits at $R=1$, for two values of the initial midplane $\beta_0$ parameter. This allows the system to launch a MHD-driven wind and reach a quasi steady-state at $R=1$. The final snapshot of these axisymmetric planet-free simulations is then extended in the azimuthal direction and used as an initial conditions for the full 3D simulations. This instant marks the $t=0$ of the simulations presented in this paper.  We then introduce the planet as described in Section~\ref{sec:method_planet}, which evolves thereafter during $N_{\rm orb}$ orbital periods. In practice, we performed several 3D simulations varying the planet mass $M_p$ and the initial $\beta_0$-plasma parameter. In Table~\ref{table:runs}, we detail these parameters that we used in eight three-dimensional MHD simulations. The names and duration of the various runs are indicated. We also show additional simulations that were used for initialization or comparison purpose, in particular the axisymmetric two-dimensional simulations, a three-dimensional planet-free simulation with $\beta_0=10^3$ and a purely hydrodynamic inviscid run. Concerning the computational cost, $475\,000$ GPU hours have been granted for the simulations presented here, which corresponds to an estimated energy consumption of about \SI{200}{MWh}, equivalent of an emission of the order of $20$ in tons of CO$_2$. 

\subsection{Disk accretion and wind launching}
\label{sec:dw_interactions}

\begin{figure}[htbp]
    \centering
    \includegraphics[width=0.5\textwidth,keepaspectratio]{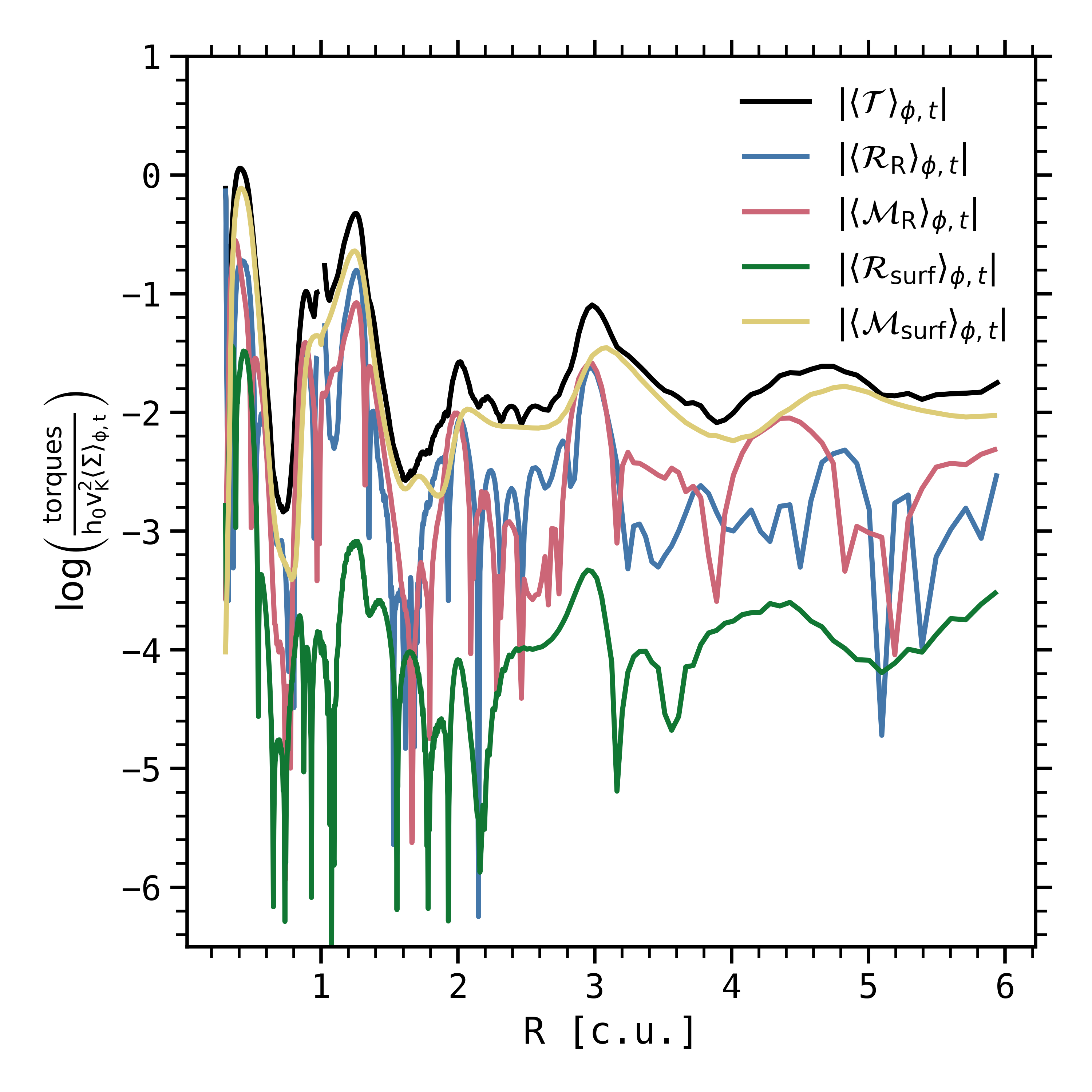}%
    \caption{ Radial profile of the four torques $\mathcal{R}_R$ (blue), $\mathcal{M}_R$ (red), $\mathcal{R}_{\rm surf}$ (green), $\mathcal{M}_{\rm surf}$ (yellow) and their sum $\mathcal{T}$ (black) normalized by $h_0 v_K^2 \langle\Sigma\rangle_{\phi,t}$, azimuthally-averaged and temporally-averaged over the last $100$ orbits for the run \emph{Mj-$\beta_3$}. We consider here only the logarithm of the absolute value of these quantities.}
    \label{fig:main_stress}
\end{figure}

We define here some important parameters that are classically used to describe disk accretion and wind launching. We first define the accretion rate as
\begin{equation}
    \dot{M}_{\rm acc}=\displaystyle -2\pi R \overline{\langle\rho v_R\rangle_{\phi}}.
    \label{eq:mdot}
\end{equation}

\noindent Using the equation of angular momentum conservation, $\dot{M}_{\rm acc}$ is linked to the transport of angular momentum by radial and surface stresses, respectively noted $\mathcal{W}_{R\phi}$ and $\mathcal{W}_{z\phi}$:
\begin{equation}
\begin{aligned}
    \displaystyle\frac{\dot{M}_{\rm acc}\Omega_K}{4\pi} &= \frac{1}{R}\partial_R\left(R^2\overline{\langle\mathcal{W}_{R\phi}\rangle_\phi}\right) + R\left[\langle\mathcal{W}_{z\phi}\rangle_\phi\right]_{-}^{+} - \overline{\langle\Gamma_{p\rightarrow d}\rangle_\phi}, \\
    &= \underbrace{\mathcal{R}_R + \mathcal{M}_R + \mathcal{R}_{\rm surf} + \mathcal{M}_{\rm surf}}_{\mathcal{T}} + \mathcal{P},
    \label{eq:angular_momentum}
\end{aligned}
\end{equation}

\noindent with $\mathcal{W}_{X\phi}=\rho v_X(v_\phi-v_K)-B_X B_\phi$ the $\phi$ component of the stress, and $\mathcal{P}=-\overline{\langle\Gamma_{p\rightarrow d}\rangle_\phi}$. $\Gamma_{p\rightarrow d}=-\rho\left(\vec{r_p}\times\vec{\nabla}\Phi_p\right).\vec{e_z}$ is the torque per unit of volume exerted by the planet on the gas, with $\Phi_p$ defined in Eq.(\ref{eq:planet_potential}). In the r.h.s. of Eq.~(\ref{eq:angular_momentum}), we have therefore four terms that can contribute to the total planet-free torque $\mathcal{T}$, and therefore to $\dot{M}_{\rm acc}$: the radial Reynolds ($\mathcal{R}_R$) and Maxwell ($\mathcal{M}_R$) torques, and the surface Reynolds ($\mathcal{R}_{\rm surf}$) and Maxwell ($\mathcal{M}_{\rm surf}$) torques. Figure~\ref{fig:main_stress} shows the contribution of the absolute value of these four torques in the accretion, as a function of the distance to the star, for the case \emph{Mj-$\beta_3$}, that is for the largest magnetization and the Jupiter-mass planet. The dominant term is $\mathcal{M}_{\rm surf}$ (yellow curve), which means that the vertical extraction of angular momentum at the disk surface seems to be the main driver of accretion, in particular when $\beta_0=10^3$. On the contrary, the surface Reynolds torque $\mathcal{R}_{\rm surf}$ (green curve) has little impact on the accretion mechanism. Concerning the radial torques ($\mathcal{R}_R$ in blue and $\mathcal{M}_R$ in red), they have a similar influence on the accretion, and can be comparable to $\mathcal{M}_{\rm surf}$ if combined. In summary, $\mathcal{M}_R\simeq\mathcal{R}_R$, $\mathcal{M}_{\rm surf}\gg\mathcal{R}_{\rm surf}$ and $\mathcal{M}_{\rm surf}+\mathcal{R}_{\rm surf}\gtrsim\mathcal{M}_R+\mathcal{R}_R$ in general, except for strong negative gradients in the gas density, for example near the gaps' inner edge. Although there is a variability in the radial profile of the four torques, the torques per unit of surface density are relatively constant radially on average. Due to the importance of the $\mathcal{M}_{\rm surf}$ term, we define the dimensionless parameter $\upsilon$ as in \citet{Lesur2021}:
\begin{equation}
    \upsilon = -\displaystyle\frac{\left[\langle B_\phi B_z\rangle_\phi\right]_{-}^{+}}{\langle\Sigma\rangle_\phi\Omega_K^2 H}= \displaystyle\frac{\mathcal{M}_{\rm surf}}{h_0 v_K^2\langle\Sigma\rangle_\phi},
    \label{eq:upsilon}
\end{equation}

\noindent which, when positive, gives an idea of the vertical evacuation of angular momentum from the disk by the magnetic torque at the disk surface. In practice with this definition, $\upsilon$ is positive when the wind extracts angular momentum from the disk. We note that for $\beta_0=10^4$, that is at lower magnetization, assuming that $\mathcal{R}_{\rm surf}$ is still negligible, $\mathcal{R}_R$ is actually the dominant term in the internal parts of the disk ($R<3$, where $\mathcal{R}_R/\mathcal{M}_{\rm surf}\leq10$), whereas $\mathcal{M}_{\rm surf}$ takes over in the external parts of the disk. It is safe to neglect $\mathcal{R}_{\rm surf}$ here as it is proportional to the density, which is small at the disk surface $z=4H$. We also define the ejection efficiency dimensionless parameter $\xi$ as:
\begin{equation}
    \xi=\displaystyle\frac{2\pi R^2\left[\langle\rho v_z\rangle_\phi\right]_{-}^{+}}{\dot{M}_{\rm acc}},
\end{equation}

\noindent which quantifies the fraction of the accreted material that is ejected. Another important quantity related to the radial transport efficiency is the $\alpha_\nu$ \citep[][]{Shakura1973} parameter:
\begin{equation}
    \alpha_\nu = \displaystyle\frac{\overline{\langle\mathcal{W}_{R\phi}\rangle_\phi}}{\overline{\langle P \rangle_\phi}} = \alpha_\nu^R+\alpha_\nu^M.
    \label{eq:alpha_tot}
\end{equation}

\noindent where $\alpha_\nu^R$ and $\alpha_\nu^M$ are respectively related to the Reynolds and Maxwell components of the radial stress. We also define a dimensionless parameter $\alpha_{\rm dw}$, similar to the one introduced in \citet{Tabone2022}, and which we can divide in Reynolds ($\alpha_{\rm dw}^R$) and Maxwell ($\alpha_{\rm dw}^M$) contributions:
\begin{equation}
    \alpha_{\rm dw} = \displaystyle\frac{R\left[\langle\mathcal{W}_{z\phi}\rangle_\phi\right]_{-}^{+}}{\overline{\langle P \rangle_\phi}} = \alpha_{\rm dw}^R+\alpha_{\rm dw}^M.
\end{equation}

\noindent We note that $\upsilon\simeq h_0 \alpha_\mathrm{dw}^M\simeq h_0\alpha_\mathrm{dw}$, using Eq.~\ref{eq:upsilon}, $\overline{\langle P \rangle_\phi}\simeq \langle\Sigma\rangle_\phi \tilde{c_s}^2$ and $\alpha_\mathrm{dw}^R\ll \alpha_\mathrm{dw}^M$.

In Appendix~\ref{sec:appendix_b2}, we address the radial transport efficiency and the level of turbulence in the 3D planet-free run \emph{\rm{3D}-$\beta_3$}, temporally averaged between $50$ and $100$ orbits. In particular, we compare $\alpha_\nu$ but also the turbulent component of $\alpha_\nu^R$ with their corresponding values in the run \emph{$10$Me-$\beta_3$}. We show that the total radial transport of angular momentum is barely affected by the presence of a low-mass planet, except close to its coorbital region. We also show that the level of turbulence is similar in both cases near the coorbital region, but $2-3$ times lower elsewhere for the low-mass planet case. This may be due to the presence of non-axisymmetric spiral features that weaken the turbulence, as in \cite{Ziampras2022}, but is beyond the scope of this study.



\section{Results}
\label{sec:3d_results}

\subsection{Gap opening}
\label{sec:GO}

\subsubsection{Overview}
\label{sec:GO_overview}

\begin{figure}[htbp]
    \centering
    \includegraphics[width=0.5\textwidth,keepaspectratio]{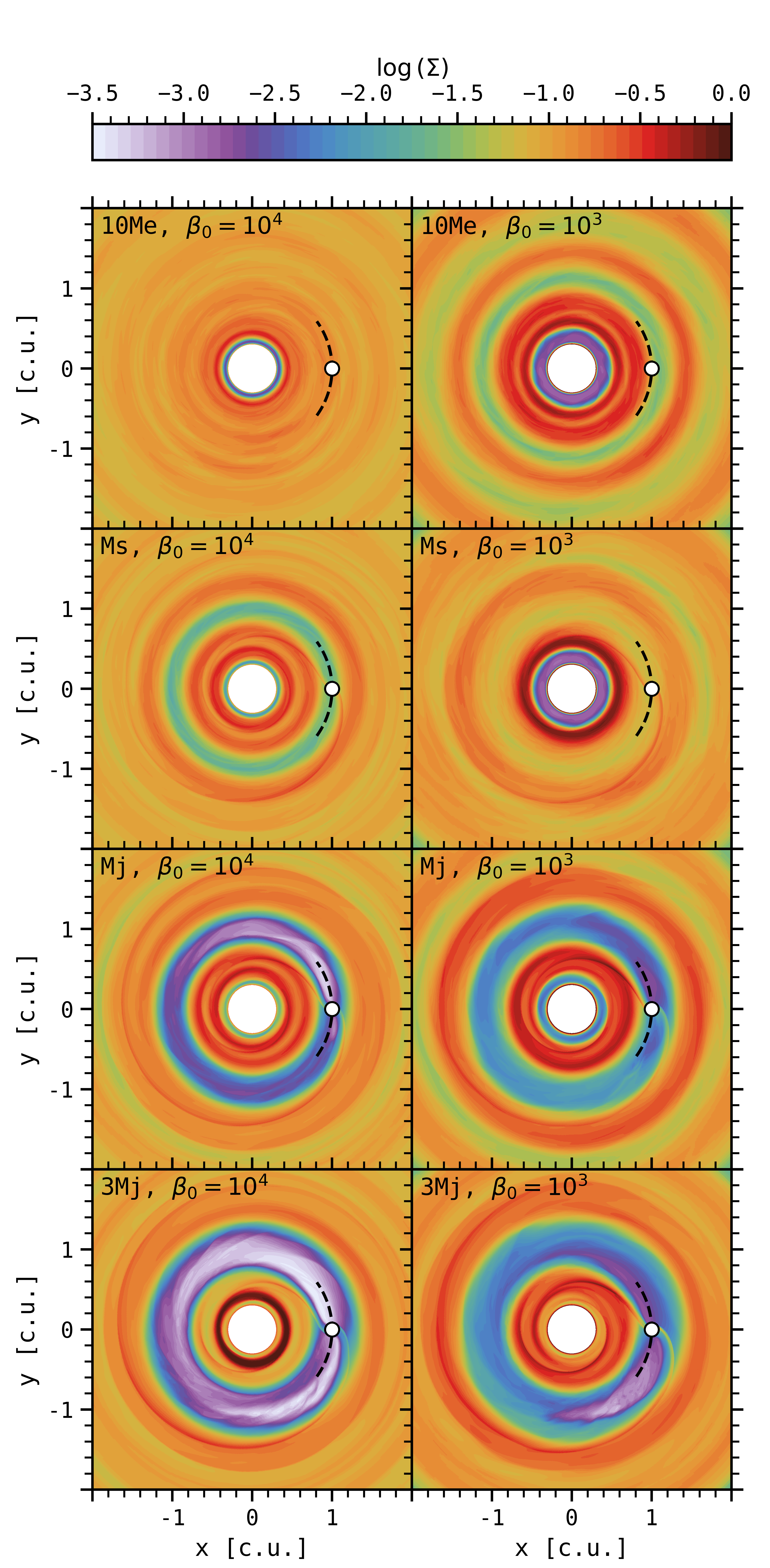}%
    \caption{Gas surface density maps for the eight MHD runs, after $200$ planet orbits. The four rows show the different values of the planet mass. The two columns show the different values of the initial $\beta_0$ parameter. The white circle marks the planet location.}
    \label{fig:main_allsigma}
\end{figure}

\begin{figure}[htbp]
    \centering
    \includegraphics[width=0.3\textwidth,keepaspectratio]{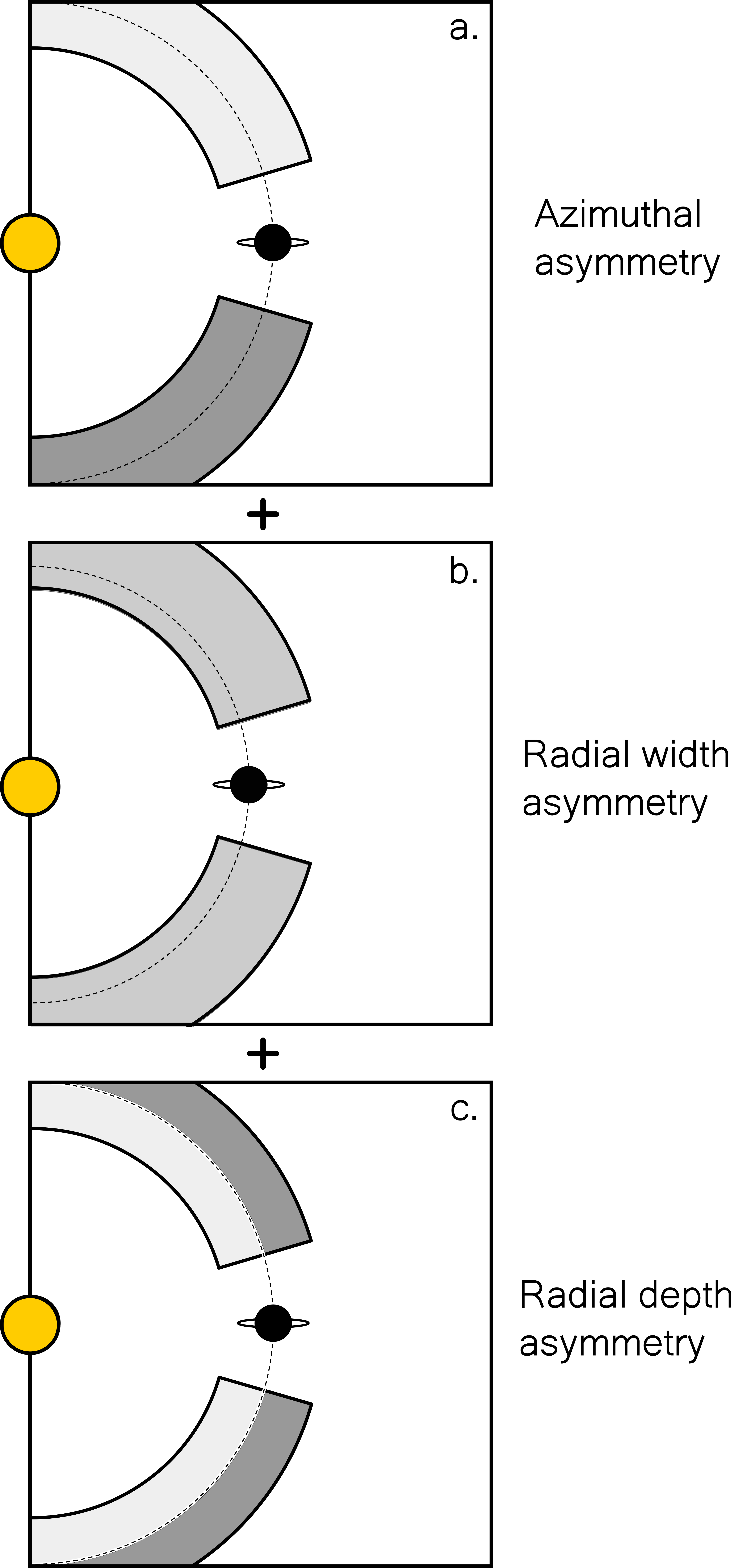}%
    \caption{Sketch of the different asymmetries of the planet gap, for massive planets and highly magnetized disks. The yellow and black circles represent respectively the central star and the planet. The horseshoe region is represented in shades of gray, with high-density zones in light gray and low-density zones in dark gray. The dashed line corresponds to the planet orbit. The global shape of the gap results from the combination of three asymmetries: a. Azimuthal azymmetry of the horseshoe region b. Radial width asymmetry, with the outer gap wider than the inner gap. c. Radial depth asymmetry, with the outer gap deeper than the inner gap.}
    \label{fig:main_asymmetries}
\end{figure}

We consider the opening of a planet gap in a wind-launching disk. The disk surface density structures after $200$ planet orbits are shown in the cartesian maps of Figure~\ref{fig:main_allsigma}, for the different planet masses (four rows), and $\beta_0$ parameters (two columns). The colormap indicates the logarithm of the gas surface density, with purple for low-density regions, and red for high-density regions. In these maps, we detect the presence of an annular planet-induced gap, whenever $q_p\geq3\times10^{-4}$ ($M_s$). In terms of $q_p/h_p^3$ parameter, this is coherent with the classical condition for non-linear effects in the flow whenever $q_p/h_p^3\lesssim1$, although this condition also depends on the level of turbulence $\alpha_\nu$ \citep[see the gap-opening criterion in][henceforth CR6]{Crida2006} and should be less true for highly magnetized disks \citep[see the extended gap-opening criterion in][]{Elbakyan2022}. The gap is wider and deeper when the planet mass is higher, with at least a factor $10$ between the $3M_j$ and the $M_s$ cases. For a given $M_p \geq M_s$, the gap is denser when $\beta_0$ is larger, with a factor of the order of $5$. For all $\beta_0$, when $M_p \geq M_j$, the horseshoe region appears clearly asymmetrical, with a difference in density in front of and behind the planet in azimuth (see panel a. of the sktech presented in Figure~\ref{fig:main_asymmetries}). As in \citet{Baruteau2011}, this kind of asymmetry in the horseshoe region could generate a positive or negative contribution in the total gravitational torque exerted by the gas onto the planet due to a static corotation torque (see Section~\ref{sec:torques}). For massive planets ($M_p \geq M_j$) and highly magnetized disk ($\beta_0=10^3$), in addition to this azimuthal asymmetry, there is also a two-fold radial asymmetry, with the outer gap wider (see panel b. of Figure~\ref{fig:main_asymmetries}) and deeper (see panel c. of Figure~\ref{fig:main_asymmetries}) than the inner gap. In practice, the planet is therefore closer to the gap's inner edge than the gap's outer edge. After $400$ orbits, for the case \emph{Mj-$\beta_3$}, the inner gap can be three times narrower and three times denser than the outer gap. The sketch in Figure~\ref{fig:main_asymmetries} isolates these three asymmetries, labeled as azimuthal asymmetry, radial width asymmetry and radial depth asymmetry.

\begin{figure*}
    \centering
    \includegraphics[width=0.99\hsize]{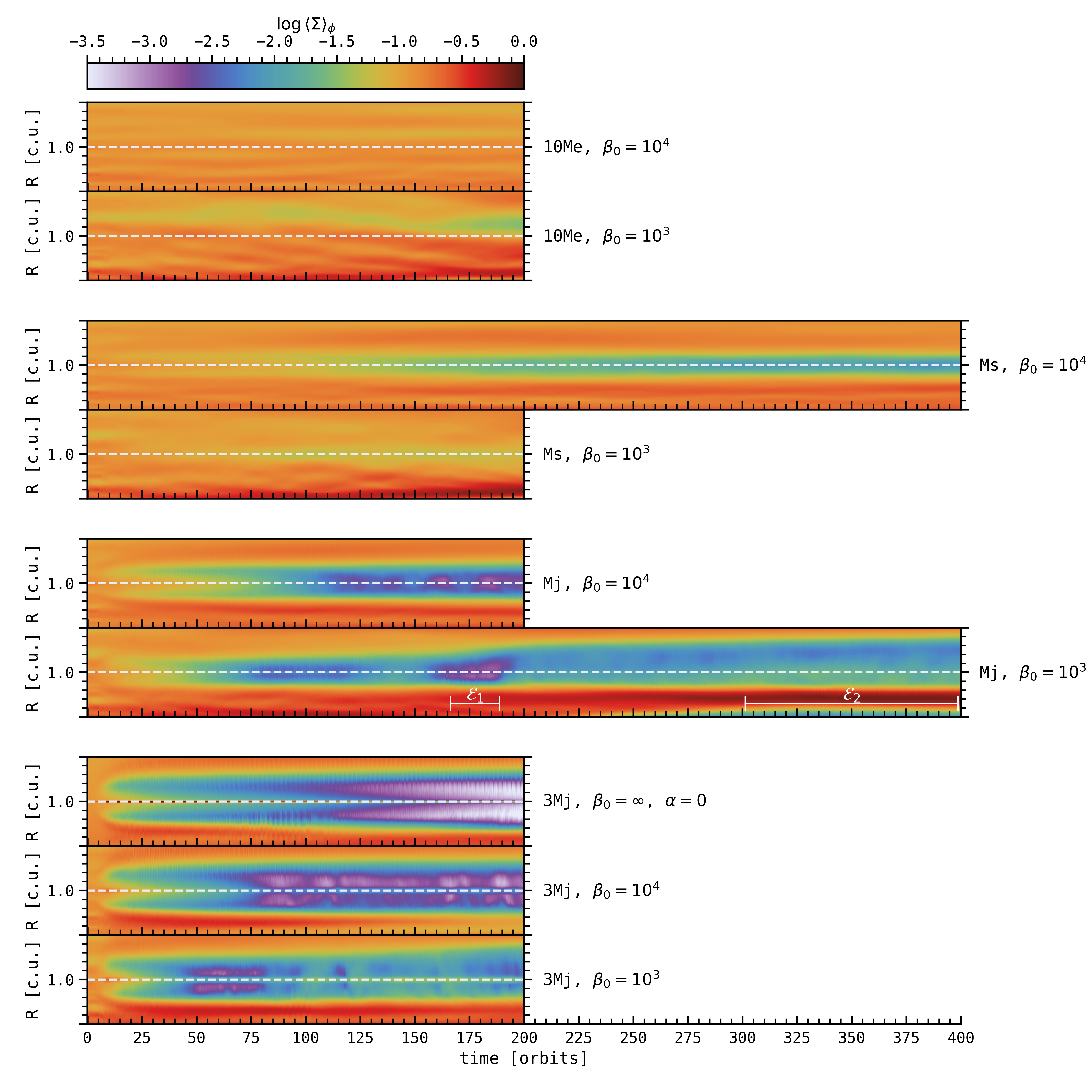}%
    \caption{Space-time diagrams showing the evolution of the logarithm of the gas surface density $\Sigma$ for the nine runs divided in four groups of different planet masses. The top plot of each group has the lowest magnetization, whereas the bottom plot has the highest. The radial extent of all the diagrams range from $0.5$ to $1.5$ code units, with the planet at $R=1$. We pinpoint as $\mathcal{E}_1$ and $\mathcal{E}_2$ two episodes in the run \emph{Mj-$\beta_3$} that will be discussed in Section~\ref{sec:HM2_accretion}.}
    \label{fig:main_spacetime}
\end{figure*}

We are able to retrieve the inner and outer planet wakes in all cases, although the fixed colorbar that we choose does not allow to disentangle for the lowest planet mass case the wakes-induced perturbations from perturbations independent of the planet presence. For all planet masses and disk magnetizations, spontaneous zonal flows emerge in the simulations, associated with concentric density rings, following a "self-organisation" process already described in a number of planet-free simulations \citep{Bethune2017,Suriano2017,Suriano2018,Suriano2019,Riols2020,Cui&Bai2021}. Comparing the planet-free case \emph{\rm{3D}-$\beta_3$} and \emph{$10$Me-$\beta_3$} in Appendix~\ref{sec:appendix_b1} confirms that the self-organized density rings and gaps are present at high magnetization regardless of the presence of a planet. Nevertheless, we show that despite its low mass, a $10M_{\oplus}$ planet is still able to impact its environnment, especially regarding the distribution of density, the accumulation of poloidal magnetic field and the strength of the wind torque. The amplitude of gas surface density perturbations $(\Sigma-\Sigma_{\rm th})/\Sigma_{\rm th}$ are larger for smaller $\beta_0$. This is particularly visible in the top row of Figure~\ref{fig:main_allsigma}, with $\sim 80\%$ in amplitude for the case \emph{$10$Me-$\beta_3$} and $\sim 20\%$ in amplitude for the case \emph{$10$Me-$\beta_4$}, compared to a simple theoretical power-law profile in $\Sigma_{\rm th} = \Sigma_0 R^{-0.5}$. These gaseous rings near $R=1$ tend to be smoothed out when the planet is able to generate sufficient perturbations. The competition between self-organisation rings and density gaps induced by massive planets can be seen in the right column of Figure~\ref{fig:main_allsigma}, with the low-density ring near the planet location and the high-density ring just outside the planet orbit. For the low-mass planet case (\emph{$10$Me-$\beta_3$}), the gas perturbations are dominated by self-organisation. For the intermediate mass case (\emph{Ms-$\beta_3$}), the competition between self-organisation and the planet tends to diminish the contrast in density. For larger planet masses (\emph{Mj-$\beta_3$}, \emph{$3$Mj-$\beta_3$}), the strong gas perturbations near the planet location are dominated by planet-driven flows. Along with these density perturbations, we detect multiple pressure bumps that are shaped at the same time by the planet and self-organisation. For example, in the run \emph{Mj-$\beta_3$}, we obtain four pressure maxima outside the planet location after $400$ orbits. The three outermost pressure maxima are probably linked to self-organisation, as we retrieve them for all planet masses, and they are less pronounced at lower magnetization.

When the planet is sufficiently massive to carve a deep annular gap around its orbit, it will form a pressure maximum at the outer edge of its gap due to the deposition of the angular momentum flux carried away by the planet's outer wake \citep[][]{Baruteau2014,Bae2016}. Dust traps, and therefore dust rings, can be naturally obtained at pressure bumps \citep[see, e.g.,][]{Perez2019,Wafflard-Fernandez&Baruteau2020}, which is of prime interest as we commonly detect annular substructures with ALMA in radio \citep[see, e.g.,][]{Huang2018,Guzman2018,Perez2019} and SPHERE in near-IR \citep[see, e.g.,][]{Avenhaus2018}. Post-processing our simulations with dust radiative transfer calculations, we would expect multiple bright and dark rings of emission, either due to the planet's outer pressure bump \citep[as in][]{Nazari2019}, or due to self-organisation \citep[as in][]{Riols2020}. Note that the emission rings induced by self-organisation are expected to be deeper and more extended in 3D than in the 2D axisymmetric models of \cite{Riols2020}. In any case, we have here two different mechanisms able to generate multiple annular substructures all at once, which could be valuable for comparison purpose (in particular concerning the kinematical signatures of such gaps in the gas).

We note that we also retrieve the formation of several small-scale vortices at the gap's outer edge corresponding to vortensity minima, probably arising because of the Rossby-Wave Instability \citep[RWI,][]{Lovelace1999}. Such vortices eventually merge in one large-scale vortex. This is particularly visible early in the run \emph{$3$Mj-$\beta_4$}. Then, the vortex gradually disappears and becomes axisymmetric. Such vortex is actually composed of numerous small-scale structures that resemble the ones obtained with elliptical instability in \citet{Lesur&Papaloizou2009}. Large-scale vortices are mostly transient in our non-ideal MHD simulations, but some non-axisymmetric structures can also appear quite late until the end of the simulation, for example at $R\simeq1.3$ for the run \emph{$10$Me-$\beta_4$}.

In Figure~\ref{fig:main_allsigma}, there is evidence that an evacuated region is generated by the inner boundary condition, especially when $\beta=10^3$. This is probably due to the fact that magnetic field lines do not cross the inner edge of the grid, and therefore tend to efficiently accumulate in that region. This accumulation of magnetic field is accompanied by a depletion of matter (see Section~\ref{sec:HM2_magnetic}), and tends to drift outward with the same mechanism that makes inner cavities \citep[as presented in][]{Martel&Lesur2022} and planet gaps (as presented here, see panel b. of Figure~\ref{fig:main_asymmetries} and Section~\ref{sec:HM2_accretion}) drift outward. Although the mechanisms behind the radial asymetries of the planet gap and the inner region are similar in their evolution, their origin is different, as they result respectively from the carving of a gap and an artifact in the inner boundary.

\subsubsection{Temporal variability}
\label{sec:GO_variability}

Figure~\ref{fig:main_allsigma} gives a good idea of the gas surface density in the disk near the planet at a given time. In order to apprehend the temporal evolution of the gas distribution, we compute in Figure~\ref{fig:main_spacetime} the space-time diagram of the logarithm of the azimuthally-averaged $\Sigma$ for the 8 non-ideal MHD runs. We also show the space-time diagram of the inviscid hydrodynamical simulation (\emph{$3$Mj-$\alpha_0$}) for comparison purpose. These diagrams confirm that a planet gap forms when $M_p \geq M_s$, and that the global shape in depth and width of this gap depends on the planet mass and the initial magnetization. For the low-mass planet cases, we notice the formation and evolution of self-organized rings and gaps, visible for example in the case \emph{$10$Me-$\beta_3$} with the under-density that appears near $R=1.2$.

The planet gap is denser when the initial magnetization increases ($\beta_0$ decreases), which is particularly visible when $M_p = 3M_j$ (bottom group of Figure~\ref{fig:main_spacetime}). A similar trend was found in simulated magnetized cavities of transition disks \citep[][their Fig. 22]{Martel&Lesur2022}, and can be explained mostly with the interdependence between the radial profiles of the accretion rate $\dot{M}_{\rm acc}$, $\beta$ and $\Sigma$, by comparing their values in the outer disk (e.g. $\beta_{\rm disk}\simeq\beta_0$) and their counterpart in the outer gap (e.g. $\beta_{\rm gap}$). Firstly, $\dot{M}_{\rm acc}$ is known to be a function of $\Sigma\beta^{-b}$ \citep[$0<b<1$, with in particular $b=0.78$ in][]{Lesur2021}. Secondly, when a gap forms, the magnetization in the gap self-regulates with $\beta_{\rm gap}\sim 10$ (see bottom panel of Figure~\ref{fig:appendix_sbgap}). As a result, in the gap, $\dot{M}_{\rm acc}\propto \Sigma_\mathrm{gap}$. If we make the assumption of a quasi steady-state, which implies that $\dot{M}_{\rm acc}$ in the gap matches $\dot{M}_{\rm acc}$ in the disk, then $\Sigma_\mathrm{gap}/\Sigma_\mathrm{disk}\propto \beta_\mathrm{disk}^{-b}\simeq\beta_0^{-b}$.


Along with the increase in density, we detect episodes of activity in the gap when increasing the magnetization and for $M_p \geq M_j$. In particular, we detect episodes of partial filling of the gap due to bursts of accretion in the gap from the outer disk. We will come back to this accretion behavior in Section~\ref{sec:HM2_accretion}. We also recover the radial asymmetries as suggested in Figure~\ref{fig:main_allsigma} for the cases \emph{$3$Mj-$\beta_3$} and \emph{Mj-$\beta_3$}, whether it be in depth or in width. This is particularly visible for the Jupiter-mass planet after $400$ orbits, with the outer gap slowly spreading outwards. We will discuss this gap dynamics in section~\ref{sec:HM2_accretion}. 

Looking again at the bottom group of Figure~\ref{fig:main_spacetime} (e.g. for a $3$ Jupiter mass planet), we evaluated how the gap density evolves with time, following the procedure of \cite{Fung2014} and \cite{Fung&Chiang2016} (see Appendix~\ref{sec:appendix_c} for more details). Using this metrics, we notice that the gap reaches its minimum density earlier when the magnetization is higher, but in that case the gap opening speed is also larger at least early on in the simulations. Qualitatively, this suggests that the accretion, and therefore mainly the wind torque (via the dominant term $\mathcal{M}_{\rm surf}$), tends to help the gap opening until $\beta$ in the gap reaches a threshold value (of the order of $1-10$) which in turn makes the density reach a quasi-equilibrium value. When the initial magnetization is larger, the wind torque is also larger, helping even more the gap to be carved and to reach earlier its quasi-equilibrium state.

Finally, we observe a strong time-variability in the gap density in magnetized models, with a larger amplitude when the magnetization increases, which would imply that the accretion varies in time, with stronger episodes as the magnetization increases (see snapshots in Figure~\ref{fig:appendix_mass_flux} of Appendix~\ref{sec:appendix_a} that illustrate such temporal variability).

\subsubsection{Gap opening criterion in a wind-launching disk}
\label{sec:GO_criterion}

The goal of this paragraph is to obtain qualitatively a criterion for gap opening in a disk where accretion is mostly driven by MHD winds via vertical extraction of angular momentum. We start by stating that a gap may form if the time it takes for a parcel of gas being accreted at velocity $v_\mathrm{acc}$ to cross the Hill sphere of the planet $\tau_{\rm acc}=R_{\rm hill}/v_\mathrm{acc}$ is longer than the time it takes for the planet to clear up the Hill sphere $\tau_{\rm p}\simeq R_p/(R_{\rm hill}\Omega)$. This last term is obtained by saying that the speed at which a fluid element is evacuated from the Hill sphere by the planet is $R_{\rm hill}\Omega$, therefore the typical timescale to clean a full ring at the planet location $R_p$ is approximately given by $\tau_{\rm p}$.

\begin{figure}[htbp]
    \centering
    \includegraphics[width=0.5\textwidth,keepaspectratio]{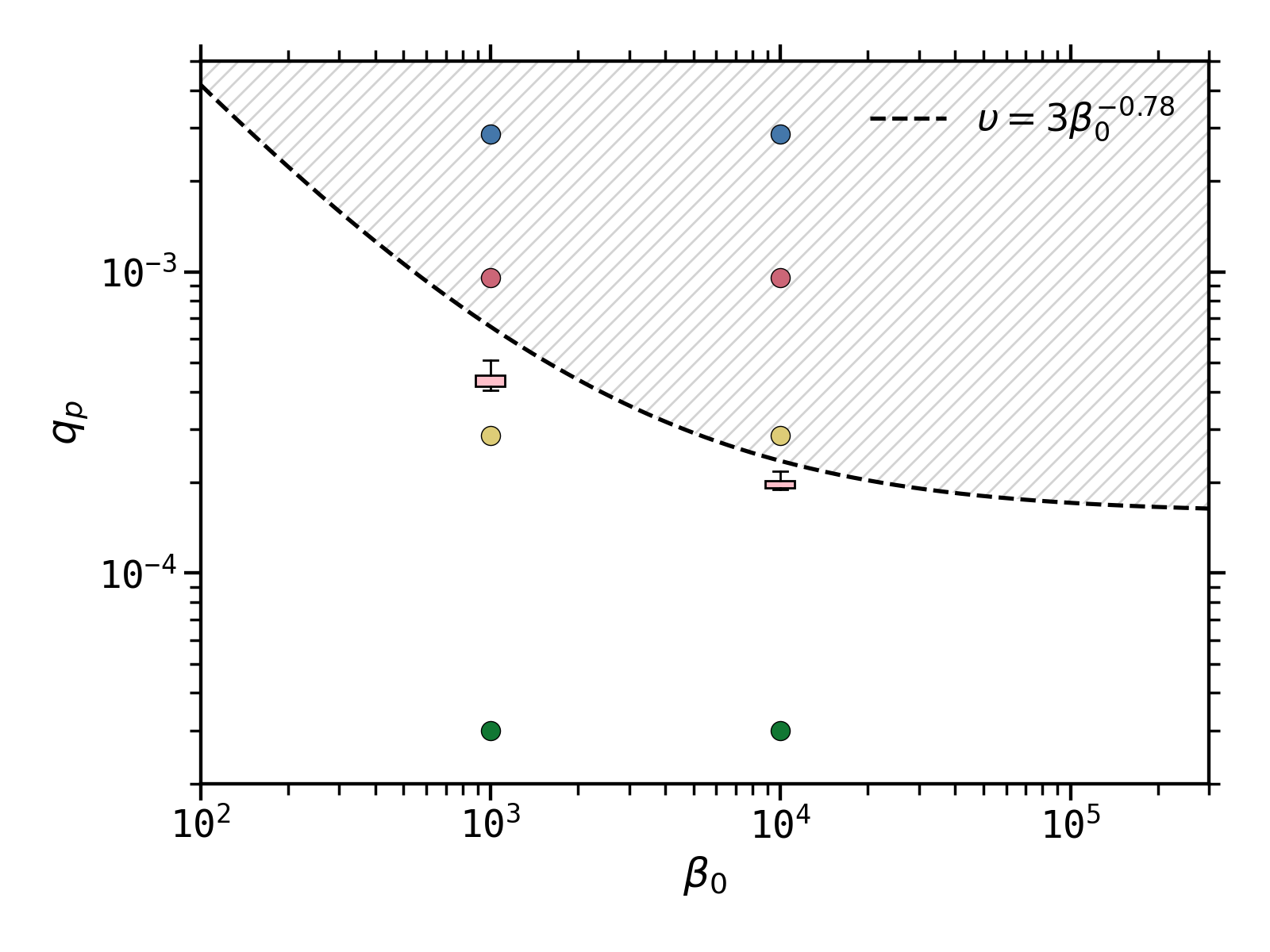}%
    \caption{Gap opening criterion with MHD winds displayed in the ($q_p$,~$\beta_0$) domain. The dashed line indicates the minimum mass to open a gap from this criterion, using a scaling law between $\upsilon$ and $\beta_0$. Planets in the hatched area are able to open a gap. To obtain the pink box plots, we estimated $\upsilon$ in our low-mass planet simulations.}
    \label{fig:main_criterion}
\end{figure}

With these two timescales, the criterion for gap opening reads
\begin{equation}
    \tau_\mathrm{p}/\tau_\mathrm{acc}<1. 
    \label{eq:crit_time}
\end{equation}

\noindent Using the equation right before Eq.~(17) in \cite{Lesur2021}, we can link the accretion speed $v_\mathrm{acc}$ to the normalized magnetic wind torque $\upsilon$. We get $v_\mathrm{acc}=2R_p\Omega h\upsilon$. The gap opening criterion then becomes
\begin{equation}
    \frac{2 R^2 h\upsilon}{R_H^2}<1.
    \label{eq:crit_from_time}
\end{equation}

\noindent Using the expression for the Hill radius in Eq.~(\ref{eq:rhill}), we eventually get
\begin{equation}
    2 h\upsilon(q_p/3)^{-2/3}<1.
    \label{eq:crit1}
\end{equation}

\begin{figure*}
    \centering
    \includegraphics[width=0.99\hsize]{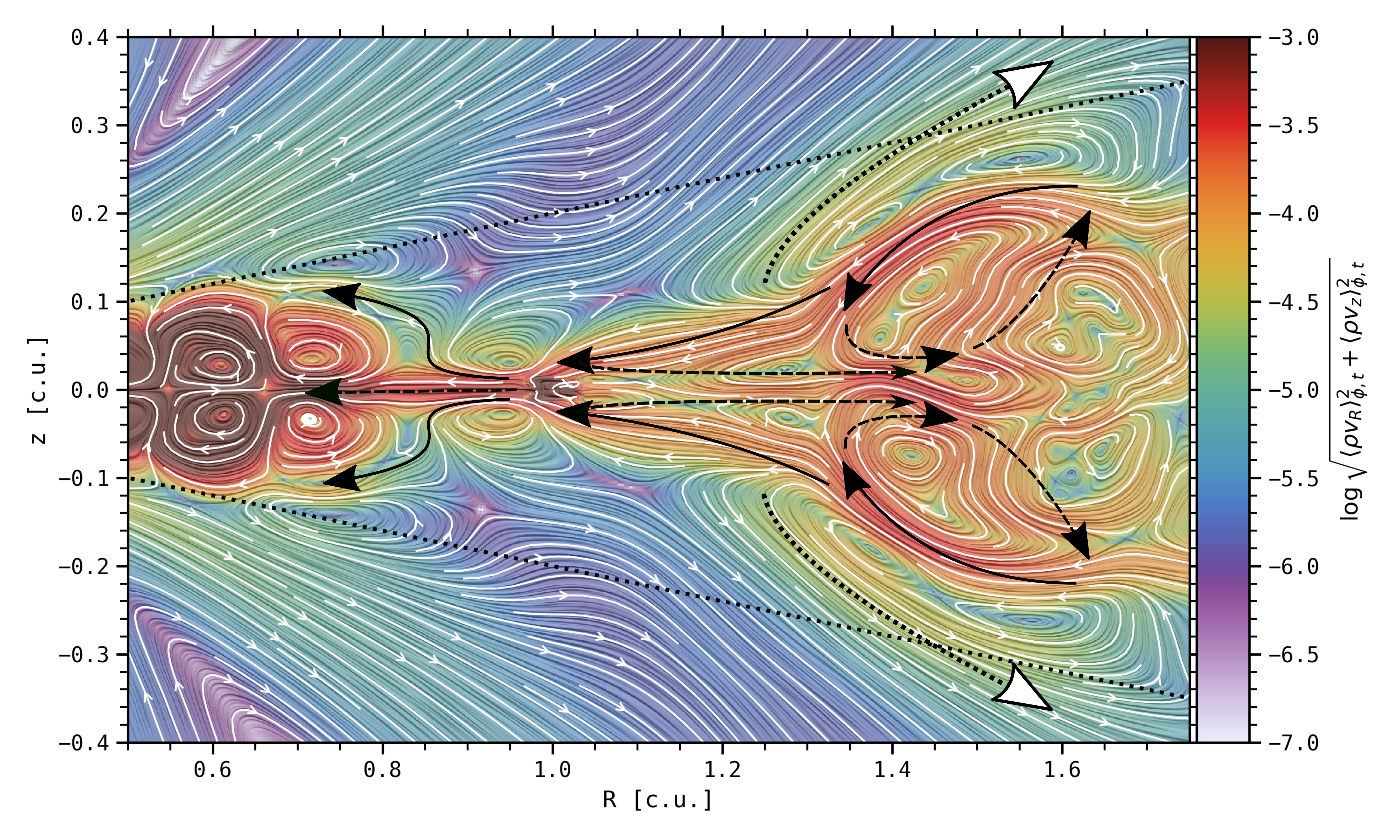}%
    \caption{Meridional flows around the planet for the run \emph{$3$Mj-$\beta_4$}, averaged in azimuth and over the last $100$ orbits. The background color represents the logarithm of the poloidal mass flux. The white streamlines and the LIC correspond to the radial and vertical components of the poloidal mass flux. Dashed black arrows, solid black arrows and dotted black and white arrows show respectively the planet-driven flows, the upper-layer accretion flows and the wind-driven flows. The upper and lower surfaces $\pm z_{\rm w}\simeq \pm 4H(R)$ are indicated by the black dotted lines.}
    \label{fig:main_3Mj_meridional_flux}
\end{figure*}

\noindent This should also be combined to another criterion which states that the disk thickness should be smaller than the Hill radius
\begin{equation}
    h(q_p/3)^{-1/3}<1.
    \label{eq:crit2}
\end{equation}

\noindent Using Eq.~(\ref{eq:crit1}) and Eq.~(\ref{eq:crit2}), we can eventually get a combined criterion for gap opening as in CR6, when accretion is dominated by MHD winds
\begin{equation}
    \frac{3h}{4(q_p/3)^{1/3}}+\frac{2 h\upsilon}{(q_p/3)^{2/3}}<1.
    \label{eq:criterion}
\end{equation}

\noindent Note that we have kept the factor $3/4$ in the original CR6 formulation for the disk thickness criterion. When accretion is dominated by viscosity, we can replace $\tau_{\rm acc}$ by $\tau_\nu = R_{\rm hill}^2/(C\nu)$, with $C$ a constant factor and rewrite the gap-opening criterion of Eq.~\ref{eq:criterion}. By considering $\mathcal{R}=R_p^2 \Omega/\nu$ the Reynolds number, Eq.~\ref{eq:crit1} would become $3C/(q_p\mathcal{R})<1$ in the fully viscous case. We therefore retrieve the same scaling $\propto 1/q_p\mathcal{R}$ as in CR6 criterion (see their Eq.~15). Note that their factor $50$ comes from a fitting procedure. Coming back to the MHD wind-driven accretion disk, we can estimate the critical $q_p=q_p^{C}$ to open a gap in a MHD wind-driven accretion disk, with $h>0$ and $\upsilon>0$:
\begin{equation}
    q_p^{C} = \frac{192(h\upsilon)^3}{\left(\displaystyle\sqrt{\frac{9h^2}{16}+8h\upsilon}-\frac{3h}{4}\right)^3}.
    \label{eq:mass_min}
\end{equation}

In order to make this expression useful, one must estimate the wind torque coefficient $\upsilon$. Here, we propose two approaches: first using the self-similar solutions from \cite{Lesur2021} with $\tilde{A}_{\rm m}=1$, giving $\upsilon=3\beta_0^{-0.78}$. In this case our gap openning criterion becomes essentially a function of $\beta_0$ and $h$. The second approach is to measure $\upsilon$ in our simulations with a $10M_{\oplus}$ planet, in the radial interval $\left[R_p - 2R^{\rm jup}_{\rm hill},R_p + 2R^{\rm jup}_{\rm hill}\right]$ and assuming that the torque obtained in this case is only barely affected by the planet.

We plot in Figure~\ref{fig:main_criterion} the gap opening criterion obtained from Eq.~(\ref{eq:criterion}) using these two approaches and assuming an aspect ratio $h=h_0=0.05$. The hatched area above the dashed line corresponds to Eq.~(\ref{eq:criterion}) with the self similar scaling for $\upsilon$ \citep{Lesur2021} while the pink box plots indicate the values critical $q_p^C$ estimated from $\upsilon$ measured in our simulations. The $8$ colored dots correspond to the $8$ ($q_p$,~$\beta_0$) pairs in our  non-ideal MHD simulations, with the $3M_{j}$, $M_{j}$, $M_{s}$ and $10M_{\oplus}$ cases respectively in blue, red, yellow and green. 
We find that our expression correctly predicts gap formation where we find them (see figure~\ref{fig:main_allsigma}). For a given planet mass, this plot confirms that it is harder for a planet to carve a gap at higher initial magnetization (decreasing $\beta_0$). More precisely, the minimum mass for a planet to open a gap lies between a mass of Saturn and a mass of Jupiter for $\beta_0=10^3$, whereas it is inferior to a mass of Saturn for $\beta_0=10^4$.

For this estimation of a gap opening criterion in a wind-launching disk, it should be mentioned that we did not consider the self-organized structures, and neglected the mechanism of magnetic accumulation in the gaps. Because $\beta$ decreases also in the gap, $\upsilon$ tends to increase as described by the self-similar scaling law, and the carving of the gap will be easier. We mentioned this runaway gap opening in the last paragraph of Appendix~\ref{sec:appendix_c} (see, e.g., the dashed and solid blue lines in the top panel of Figure~\ref{fig:appendix_sbgap}).

\subsection{Case study: high planet mass and high magnetization runs}
\label{sec:HM2}

We focus now mainly on the simulations with $M_p \geq M_j$ and $\beta_0=10^3$.

\begin{figure*}
    \centering
    \includegraphics[width=0.99\hsize]{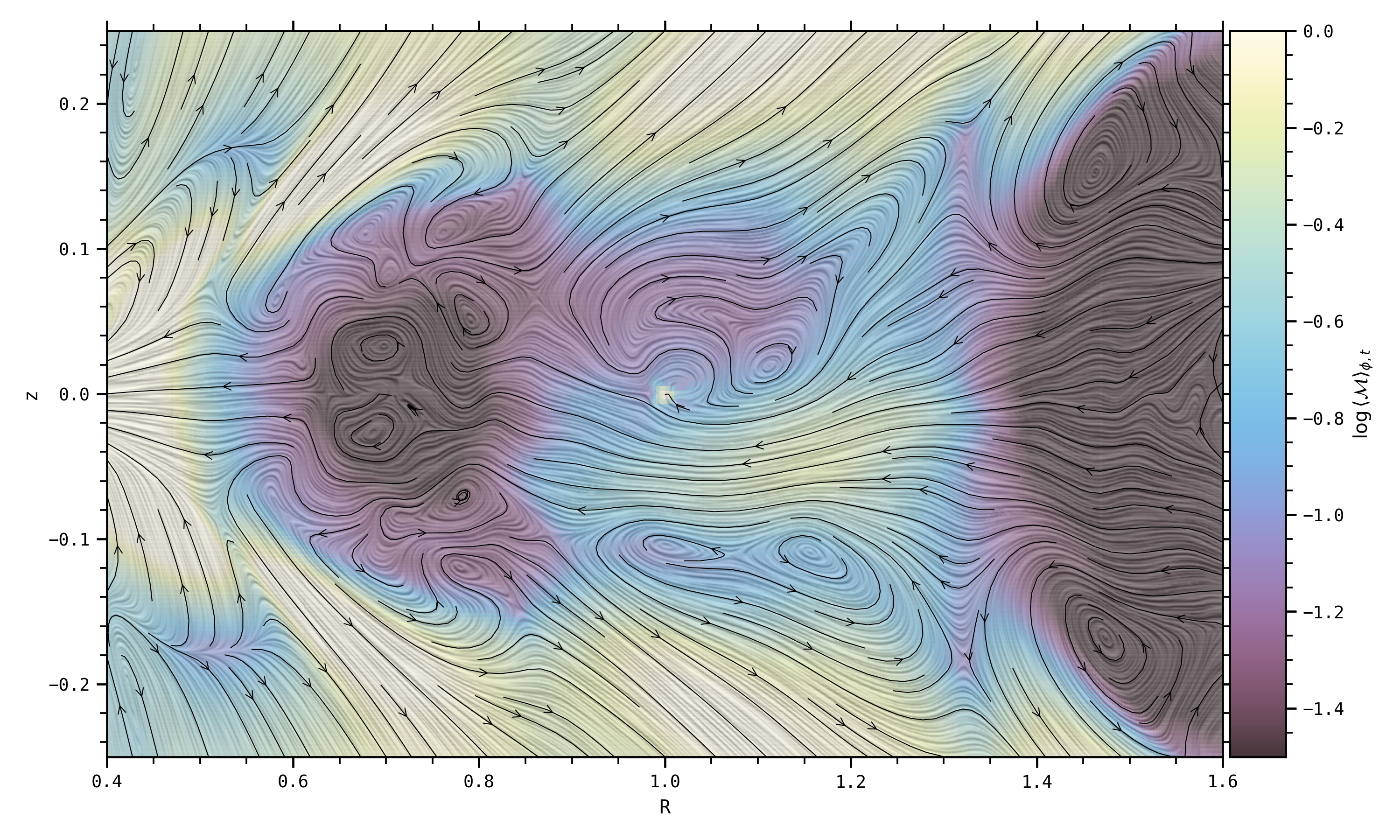}%
    \caption{2D map of the Mach number, defined as $\sqrt{\langle\rho v_R\rangle_{\phi,t}^2+\langle\rho v_z\rangle_{\phi,t}^2}/\langle\rho\rangle_{\phi,t}/\langle c_s \rangle_{\phi,t}$, in the ($R$-$z$) plane for the run \emph{Mj-$\beta_3$}, averaged in azimuth and over the last $100$ orbits. The background color represents the logarithm of the Mach number. The black streamlines and the LIC correspond to the radial and vertical components of the poloidal Mach number. The nearly-sonic top-down non-symmetrical streamer is visible in white, in the gap and under the planet.}
    \label{fig:main_Mj_mach}
\end{figure*}

\subsubsection{Meridional flows: accretion layers and sonic streamers}
\label{sec:HM2_MF}

We discuss here the behavior of the flow in the vicinity of a planet in a fixed circular orbit. In the next paragraphs, we focus on the mass flux instead of the velocity, as mass flux brings a more precise view of the gas bulk motion. In Figure~\ref{fig:main_3Mj_meridional_flux}, we show in background color the magnitude of the poloidal mass flux $\langle\rho v_p\rangle_{\phi,t}$ in log scale, averaged azimuthally and temporally over the last $100$ orbits for the run \emph{$3$Mj-$\beta_4$}. The texture in LIC (line integral convolution) and the white arrows indicate the direction of the radial ($\langle\rho v_R\rangle_{\phi,t}$) and vertical ($\langle\rho v_z\rangle_{\phi,t}$) components of this mass flux. Following the nomenclature in \cite{Fung&Chiang2016}, the black arrows help to schematically classify the different flows in the (R-z) plane in three categories:

\begin{itemize}
    \item \textbf{Planet-driven flows} (dashed black arrows) are probably linked to the planet's repulsive Lindblad torques.
    \item \textbf{Upper layer accretion flows} (solid black arrows) are driven by the accretion of material from the surface layers of the outer disk to the planet poles and from the inner gap to the surface layers of the inner disk.
    \item \textbf{Wind-driven flows} (dotted black and white arrows) act to carry material away from the lower altitudes of the disk.
\end{itemize}

\noindent These flows eventually collide, driving the merged flow upwards and resulting in a large-scale meridional circulation in the outer disk \citep[as in][]{Fung&Chiang2016}. We also observe a small-scale localized meridional circulation close to the planet, with bigger loops in the inner gap when $R \in [0.9,1]$ and $|z| \in [0.0,0.1]$ than in the outer gap when $R \in [1,1.1]$ and $|z| \in [0.0,0.02]$. This asymmetry is due to the two accretion layers in $R \in [1.0,1.3]$ and $|z|<0.1$. On the one hand they carry material from the outer disk onto the poles of the planet, and on the other hand they pinch and confine the localized outer meridional circulation even closer to the planet by counteracting the planet-driven flows. If we increase the magnetization, the mass flux is globally increased in the accretion layers compared to the case $\beta_0=10^4$. This behavior is the result of an enhanced $\mathcal{M}_{\rm surf}$ component (magnetic torque at the disk's surface) and to a lesser extent to an enhanced radial torque ($\mathcal{R}_R+\mathcal{M}_R$) leading overall to a more efficient accretion. In addition, we notice that the accretion layers are more collimated towards the midplane, thus carrying denser material compared to the lower magnetization case. This behavior prevents even more the planet-driven flows to counterbalance accretion. In that high magnetization case, accretion onto the planet -- as well as planet growth, should therefore be more efficient because of this denser inflow coming from lower altitudes toward the planet poles, as observed in \citet{Gressel2013}.

\begin{figure*}
    \centering
    \includegraphics[width=0.99\hsize]{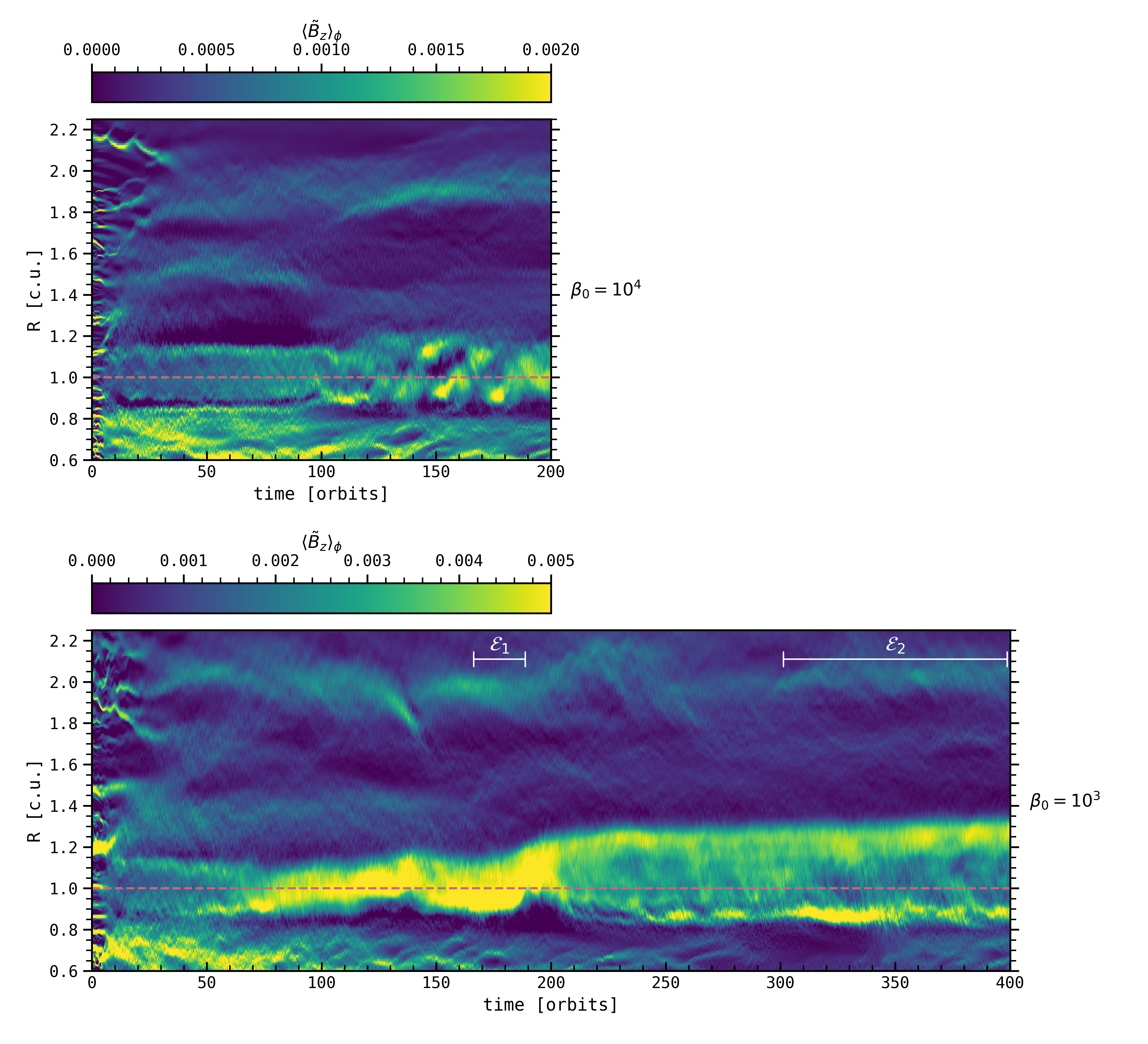}%
    \caption{Space-time diagram of the vertical component of the magnetic field in the midplane $\langle \tilde{B}_z \rangle_\phi$, for the runs \emph{Mj-$\beta_4$} (top panel) and \emph{Mj-$\beta_3$} (bottom panel). The horizontal dashed lines indicate the location of the planet. The two episodes $\mathcal{E}_1$ and $\mathcal{E}_2$ are indicated in white.}
    \label{fig:main_Mj_spacetime_Bz_vm}
\end{figure*}

When $M_p \geq M_j$ and $\beta_0=10^3$, in addition to the structures observed in Figure~\ref{fig:main_3Mj_meridional_flux}, we note the presence of a streamer at $z<0$ which accretes even more material towards the inner disk. In Figure~\ref{fig:main_Mj_mach}, we show that this top-down non-symmetrical flow (for $z \in [-0.1,0]$) has nearly-sonic velocities (of the order of $60\%$ of the sound speed) in the gap. 
Several other studies have pointed out that sonic accretion was expected in strongly magnetized disks, as is the case in our planet gaps \citep[][]{Combet&Ferreira2008,Martel&Lesur2022}. This is also coherent with the fact that the mass flux $\langle\rho v_p\rangle_{\phi,t}$ is approximately radially constant (see Figure~\ref{fig:main_3Mj_meridional_flux}), whether it be in the outer disk, the planet gap or the inner disk, in spite of a deep depletion of gas in the gap. Because the gap density is low, the velocity has necessarily to increase to keep a constant mass flux, reaching a nearly sonic speed in the gap and a $\beta$ much smaller than $\beta_0$ (reaching locally values close to unity, see e.g. the bottom panel in Figure~\ref{fig:appendix_sbgap}). Note that this approach is time-averaged and the large-scale gas motion is actually extremely variable in time, with complex redistribution of matter in the gap and evacuation in the wind (see Appendix~\ref{sec:appendix_a}).

\begin{figure*}
    \centering
    \includegraphics[width=0.99\hsize]{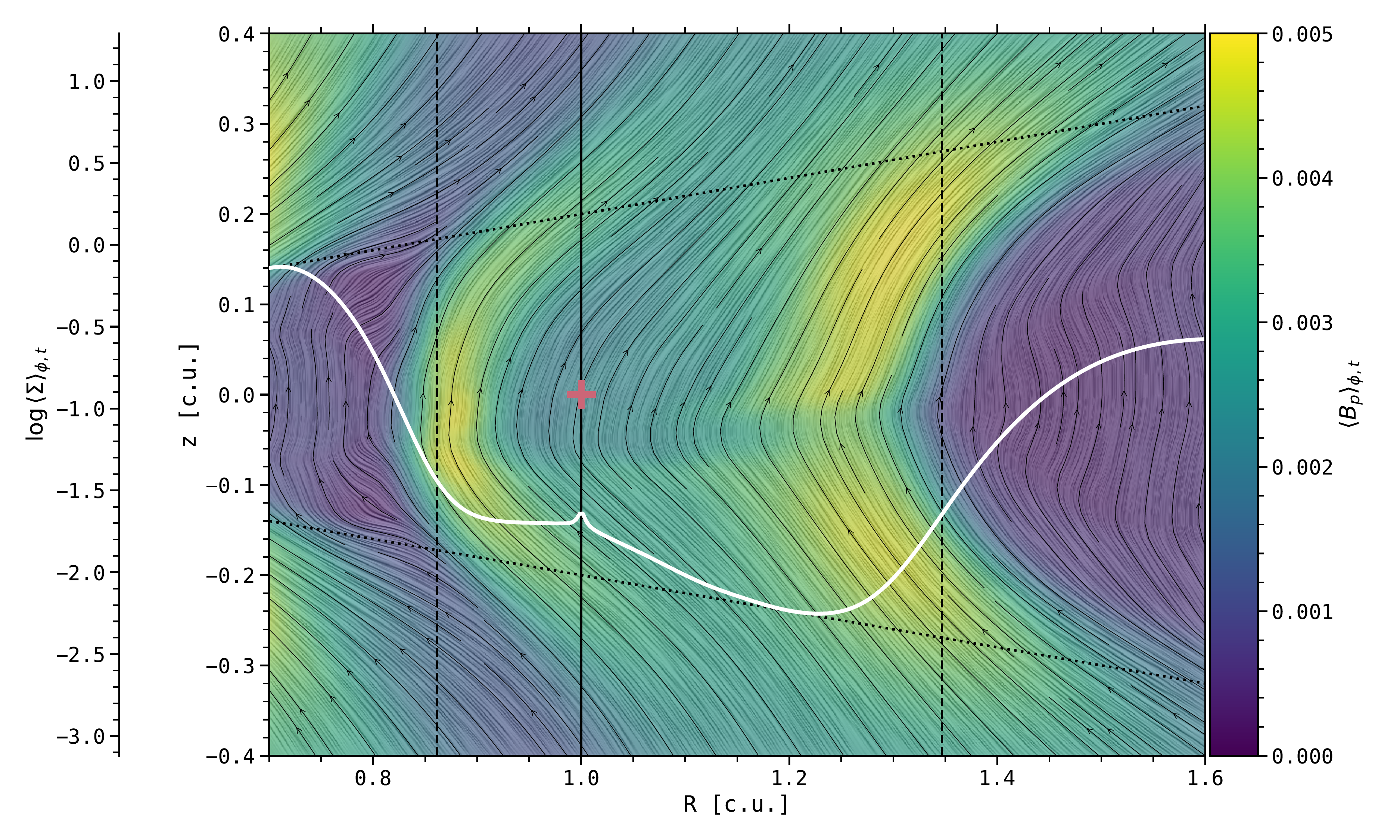}%
    \caption{Poloidal magnetic field in the gap for the run \emph{Mj-$\beta_3$}, averaged in azimuth and over the last $100$ orbits (episode $\mathcal{E}_2$). The background color represents the logarithm of the poloidal magnetic field. The black streamlines and the LIC correspond to the radial and vertical components of the poloidal magnetic field. The white curves represents the radial profile of $\log(\Sigma)$, also averaged in azimuth and over the last $100$ orbits. The red cross and solid black line pinpoint the location of the planet. Vertical black dashed lines delineate the gap edges at $R\simeq R_p-2R_{\rm hill}$ and $R\simeq R_p+5R_{\rm hill}$. The upper and lower surfaces $\pm z_{\rm w}\simeq \pm 4H(R)$ are indicated by the black dotted lines.}
    \label{fig:main_Mj_bpbprho}
\end{figure*}

In conclusion, we retrieve in our MHD simulations a large-scale meridional circulation of the gas, as in purely hydrodynamical 3D simulations \citep[][]{Szulagyi2014, Morbidelli2014, Fung&Chiang2016, Teague2019,Szulagyi2022}, though less clearly symmetrical between outer disk region and inner disk region due to the different dynamical processes at stake. The gas in the outer disk flows from the upper accretion layers into the gap and eventually falls towards the midplane. The planet is able to maintain a sustainable gap via its tidal torque, which tends to empty the planet's coorbital region. Therefore, part of the accretion flow is advected inwards in the inner gap, whereas the rest is pushed back in the outer disk. At the gap's outer edge, the gas still at low altitude is then progressively driven upwards to reach again the surface accretion layers. Note that this meridional circulation is not a closed loop. On the one hand, the accretion occurs from the disk's surface layers, and inevitably transports gas material inwards, regardless of the planet presence. On the other hand, part of the gas is evacuated vertically in the wind at the gap's outer edge during its descent towards the midplane. We will see in the next sections that this gas dynamics in the gap comes from the distribution and dynamics of magnetic field lines.


\subsubsection{Magnetic field}
\label{sec:HM2_magnetic}

We focus here on the large-scale magnetic field threading the disk and its long term behavior. We consider first the temporal evolution of the vertical component of the azimuthally-averaged magnetic field in the midplane $\langle \tilde{B}_z \rangle_\phi$. Figure~\ref{fig:main_Mj_spacetime_Bz_vm} shows the space-time diagram of this quantity for the runs \emph{Mj-$\beta_4$} (top panel) and \emph{Mj-$\beta_3$} (bottom panel), to be compared with the corresponding space-time diagrams of $\log\langle\Sigma\rangle_\phi$ in Figure~\ref{fig:main_spacetime}. In both cases, during the first orbits, several small-scale structures are visible, each one corresponding to the self-organisation of vertical magnetic field \citep{Riols&Lesur2019} in the axisymmetric 2.5D simulation used for the initial condition. In 3D, they tend to quickly merge into more extended and diffuse structures, see for example at $R\simeq2$ in both cases. We note that this large zonal field is associated to a dark ring of matter (i.e. a gap induced by self-organisation). The anti-correlation between density and magnetic fluctuations in gaps formed in planet-free disks have been reported in various works \citep[see, e.g.,][]{Bethune2017,Suriano2018,Riols&Lesur2019}. 

At high magnetization (bottom panel of Figure~\ref{fig:main_Mj_spacetime_Bz_vm}), as the planet is growing, the gap is formed and reaches its minimum density (on average) after $\simeq 50$ planet orbits, as indicated in Section~\ref{sec:GO_variability}. After the planet gap has opened, it becomes a prime area of magnetic field accumulation between $75$ and $200$ orbits, with $\langle \tilde{B}_z \rangle_\phi$ reaching its highest value. If we focus on $\log\langle\Sigma\rangle_\phi$, this strong magnetic accumulation is concomitant to a strong episode of gas depletion and a fast outward drift of the outer gap (episode noted $\mathcal{E}_1$ in Section~\ref{sec:HM2_accretion}). This $\langle \tilde{B}_z \rangle_\phi$ accumulation at the gap center has been observed for low-mass planets in various ideal \citep[][]{Papaloizou2004,Zhu2013,Carballido2017} and non-ideal \citep[][]{Keith&Wardle2015} MHD simulations. Subsequently, after $200$ orbits, the magnetic field is divided in two regions of accumulation, one at the transition between the inner disk and the inner gap (near $R\simeq0.85$), and the other one at the transition between the outer gap and the outer disk (near $R\simeq1.3$). \citet{Carballido2017} obtained as well for their highest planet masses (a few $M_j$) a double accumulation for the vertical magnetic field (as well as $\alpha_\nu$), larger close to the gap edges than at the gap center. In addition, they argued that the radial band of large $\langle \tilde{B}_z \rangle_\phi$ is smaller than the band of low-density (i.e. the density gap), which is also what we retrieve in our simulations: $\simeq0.45$ code units for $\langle \tilde{B}_z \rangle_\phi$ at $400$ orbits instead of $\simeq0.55$ code units for $\langle\Sigma\rangle_\phi$, corresponding to a difference of \SI{1}{au} for a planet orbiting at \SI{10}{au}. Concerning the temporal variability of magnetic field accumulation, the two accumulation regions in Figure~\ref{fig:main_Mj_spacetime_Bz_vm} do not exhibit a constant $\langle \tilde{B}_z \rangle_\phi$ in time, with episodes of strong magnetic field accumulation (e.g. between $310$ and $350$ orbits) alternating with episodes of weaker $\langle \tilde{B}_z \rangle_\phi$ accumulation (e.g. between $285$ and $305$ orbits). This behavior may be coupled to the sporadic accretion of matter illustrated in Appendix~\ref{sec:appendix_a}.

At lower magnetization (top panel of Figure~\ref{fig:main_Mj_spacetime_Bz_vm}), the behavior seems a little bit more complex. We first witness a phase of $\langle \tilde{B}_z \rangle_\phi$ accumulation at the gap center ($<100$ orbits), though less efficient than in the case of stronger magnetization (see maximum value of the colorbars between the two panels). This episode is then followed by an oscillation of the magnetic field accumulation between the inner gap edge and the outer gap edge, which could be due to a complex behavior of matter in the planet's horseshoe region.

Coming back to the two-fold accumulation of the magnetic field, we represent in Figure~\ref{fig:main_Mj_bpbprho} the distribution and topology of the time-average poloidal magnetic field and the disk surface density for the run \emph{Mj-$\beta_3$}. 
We retrieve the double accumulation of the poloidal magnetic field at the transitions between the inner disk and the inner gap and between the outer gap and the outer disk, as presented in \citet{Carballido2017}. The magnetic field lines are nearly vertical in the outer disk. We however detect the presence of kinks in the $\langle B_p \rangle_{\phi,300-400}$ lines, as described in \citet{Lesur2021,Martel&Lesur2022}. Such structures trace the two disk's upper accretion layers that bend magnetic field lines as material is carried inward in the gap. If we now focus on the gap, the magnetic field lines are pinched in a plane parallel to the midplane but such that $z<0$. This localisation actually matches the sonic accretion streamer, indicating here again that poloidal magnetic fields lines are bent by the streamer. 

\begin{figure}[htbp]
    \centering
    \includegraphics[width=0.5\textwidth,keepaspectratio]{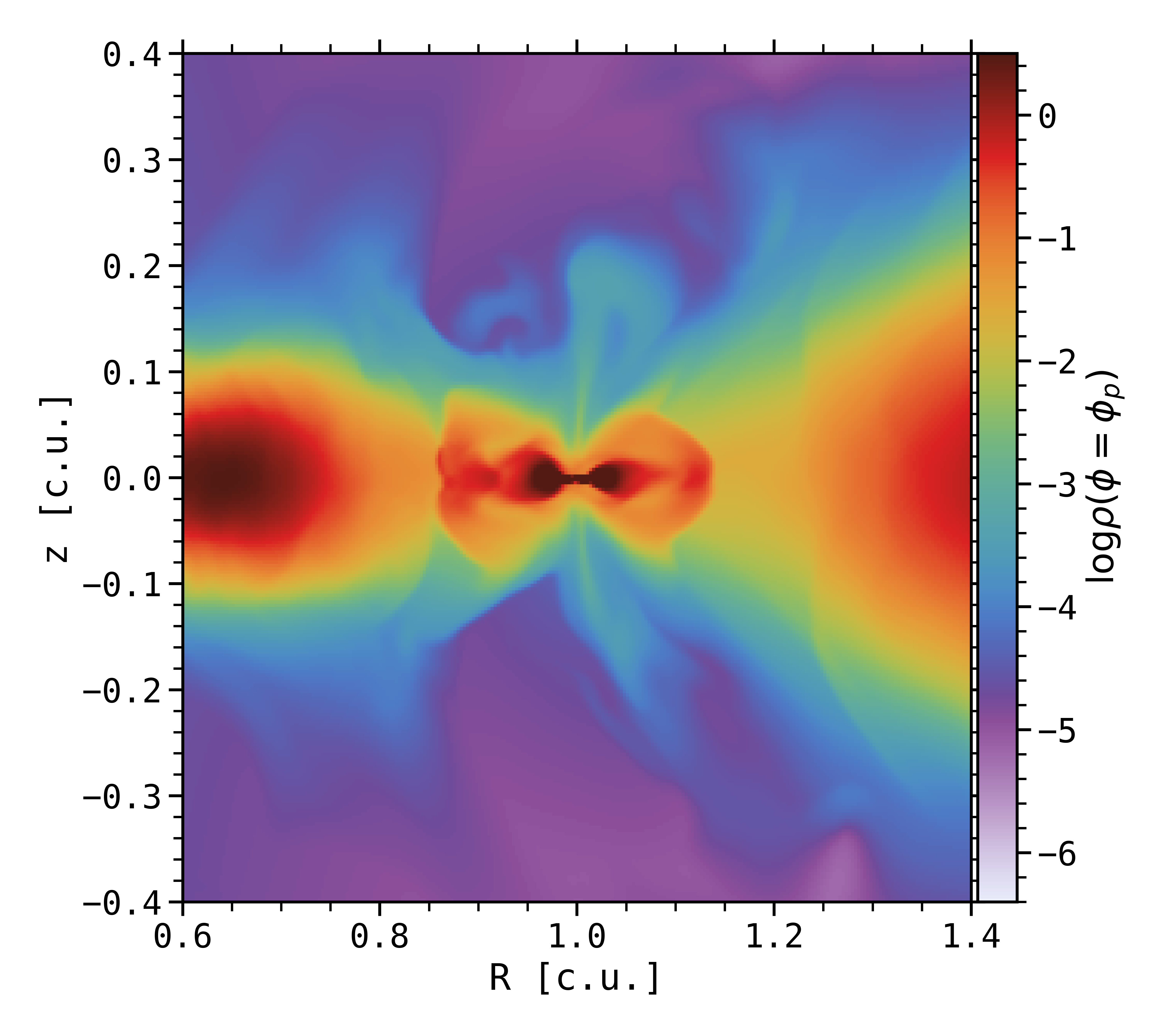}%
    \caption{Azimuthal cut of the $\log{(\Sigma)}$, at the azimuth of the planet $\phi_p$, at $\simeq114$ orbits for \emph{$3$Mj-$\beta_4$}. A protoplanetary jet is clearly visible, launching from the planet poles.}
    \label{fig:main_jet}
\end{figure}

Magnetic field accumulation in the gap is so strong that it could be responsible for sporadic protoplanetary jets. We demonstrate this in Figure~\ref{fig:main_jet} with an azimuthal cut of $\log{(\Sigma)}$ near the location of the planet, at $\simeq114$ orbits for the run \emph{$3$Mj-$\beta_4$}. This zoom near the planet shows several interesting structures : the gap, the density maxima on both gap edges, a circumplanetary disk (CPD) and an outflow from the planet poles. We can characterize the CPD by fitting a power law to the azimuthal velocity $v_\phi-v_K$ in the two regions close to $R=1$ in the Hill sphere. We obtain $0.038(R-1)^{-0.459}$ and $0.034(R-1)^{-0.453}$ respectively for the outer CPD and the inner CPD, while with the planet mass we have $\sqrt{q_p}\simeq0.032$, confirming that the CPD is rotating at the expected Keplerian velocity around the planet. Turning now to the meriodional flows, we find that, in addition to the circulation presented in Section~\ref{sec:HM2_MF}, we get an outflow from the planet poles ($R\simeq1$, $|z|<0.25$). At higher altitudes, the outflow is slightly bent in the disk wind direction where the outflow material eventually falls back towards the planet. We note however that this outflow is only occasionally present, as it is not visible in time-average plots. Outflows from CPDs have already been observed in 3D non-isothermal MHD simulations \citep{Gressel2013} where it was suggested that they could be magnetocentrifugally driven. For the sake of conciseness, we however postpone the detailed investigation of these outflows to a future study.



Overall, these results indicate that the poloidal magnetic field is very dynamical and strongly affected by the planet, with a significant accumulation in the gap. This in turn impacts the strength of magnetic torques and of the accretion flow, which we explore next.

\subsubsection{Accretion rate, wind torque and gap asymmetries}
\label{sec:HM2_accretion}

In this section, we focus on the case \emph{Mj-$\beta_3$}. We look at the link between the accretion rate $\dot{M}_{\rm acc}$ defined in Eq.~(\ref{eq:mdot}), the specific wind torques via $\mathcal{M}_{\rm surf}$ and the $\upsilon$ parameter defined in Eq.~(\ref{eq:upsilon}), and the radial asymmetries of the planet gap. In particular, we focus on two specific moments in the simulation: the first one, noted $\mathcal{E}_1$, corresponds to the strong episode of gas depletion between $165$ and $200$ orbits, the second one, $\mathcal{E}_2$, corresponds to the last $100$ orbits of the simulation, after the magnetic field has been divided in two regions of accumulation. During $\mathcal{E}_1$, the gap in the density lies approximately between $R_p - 2R_{\rm hill}$ and $R_p + 3R_{\rm hill}$, whereas during $\mathcal{E}_2$, the outer gap edge is further out in the disk, near $R_p + 5R_{\rm hill}$. In the graphs of this section, we will highlight these three characteristic radii with vertical red dashed lines.  

\begin{figure}[htbp]
    \centering
    \includegraphics[width=0.5\textwidth,keepaspectratio]{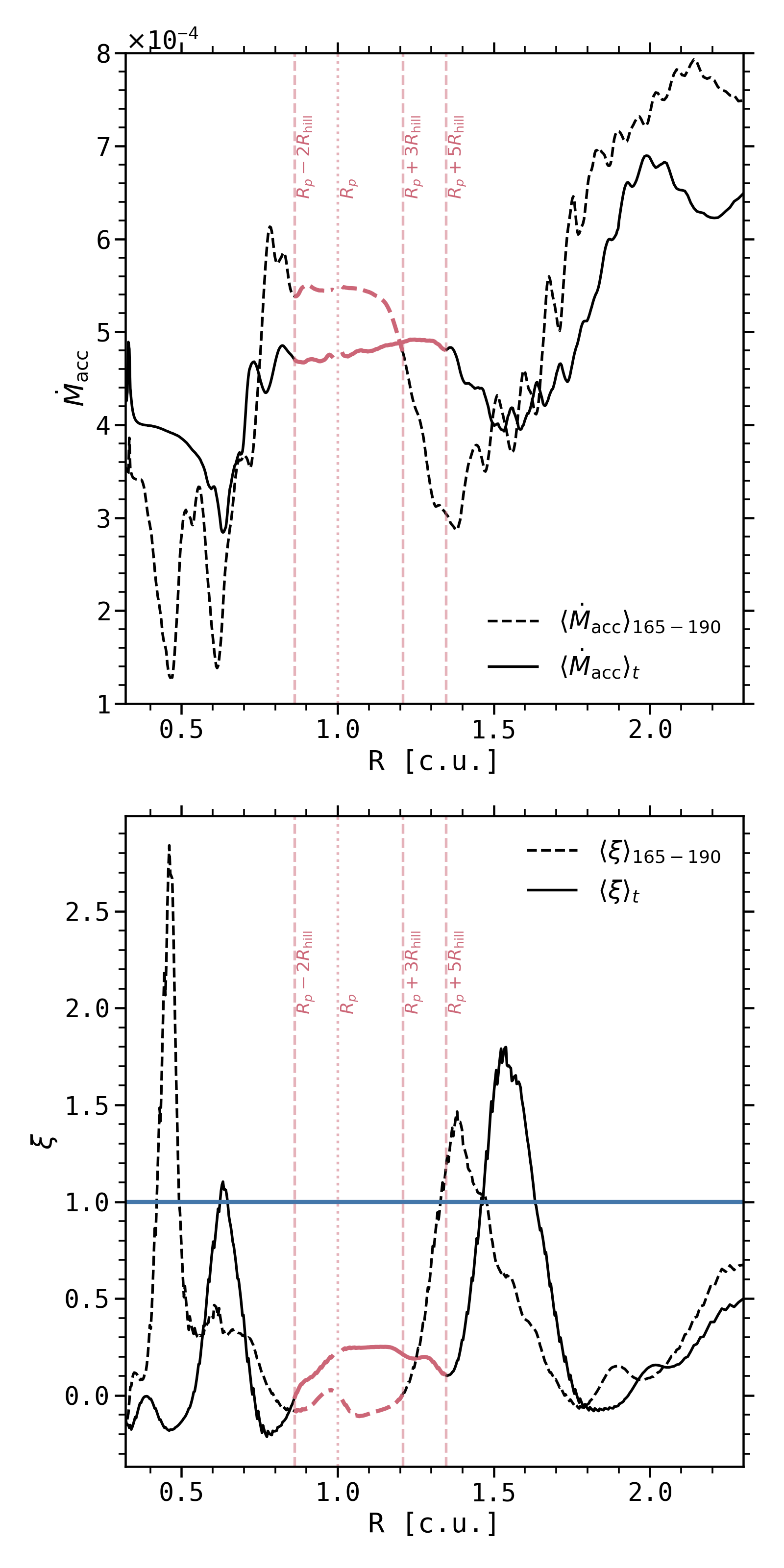}%
    \caption{Radial profile of the accretion rate $\dot{M}_{\rm acc}$ (top panel) and the ejection parameter $\xi$ (bottom panel), averaged azimuthally and temporally over the last $100$ orbits (solid line) in the run \emph{Mj-$\beta_3$} for $\mathcal{E}_2$ and over $25$ orbits during the episode $\mathcal{E}_1$ of gas depletion and fast outward drift of the outer gap (dashed line). We removed the contribution close to the planet and its CPD in the evaluation of $\dot{M}_{\rm acc}$ and $\xi$.}
    \label{fig:main_macc_xi}
\end{figure}

Figure~\ref{fig:main_macc_xi} displays the radial profiles of $\dot{M}_{\rm acc}$ (top panel) and the ejection efficiency $\xi$ (bottom panel) for $\mathcal{E}_1$ (dashed line) and $\mathcal{E}_2$ (solid line). For these two episodes, the curves of $\dot{M}_{\rm acc}$ and $\xi$ are red in the gap and black beyond the gap. As a general remark, we find that the accretion is stronger on average in the gap than in its vicinity (e.g. at $R\simeq0.7$ and $R\simeq1.3-1.5$), and even stronger at higher magnetization. In addition, the ejection efficiency $\xi$ is very small in the gap\footnote{We note that beyond the gap on both sides, the ejection in the wind is particularly efficient ($\xi>1$) where the accretion rate is minimum} (bottom panel of Figure~\ref{fig:main_macc_xi}), indicating that almost all of the matter that enters the outer gap eventually leaves via the inner gap (in other words, the gas "leak" due to the outflow is negligible).  The step in $\dot{M}_{\rm acc}$ and the fact that $\xi\ll 1$ in the gap are crucial to interpret the radial width asymmetry of the gap: on average, the gap progressively erodes the outer disk, while material tends to accumulate at the inner gap edge. The outward drift of the gap is therefore mainly due to a slight mismatch of the accretion rate in the gap and in the outer/inner disks. Such phenomenon has already been reported in MHD-driven disks, with the outward drift of the cavity in transition disks \citep[][]{Martel&Lesur2022}. If we compare the two episodes, $\dot{M}_{\rm acc}$ is larger in the gap during $\mathcal{E}_1$ than during $\mathcal{E}_2$. Actually, the accretion rate reaches its highest value in the gap during $\mathcal{E}_1$, which is related to the saturation of the $B_z$ everywhere in the gap during this epoch (see Figure~\ref{fig:main_Mj_spacetime_Bz_vm}). Moreover, the step of $\dot{M}_{\rm acc}$ in the gap and out of the gap is bigger during $\mathcal{E}_1$, with a steeper gradient (see in particular between $R=R_p + 3R_{\rm hill}$ and $R\simeq1.35$ for $\mathcal{E}_1$ and between $R=R_p + 5R_{\rm hill}$ and $R\simeq1.5$ for $\mathcal{E}_2$). We therefore expect a faster outward drift of the outer gap during $\mathcal{E}_1$ than during $\mathcal{E}_2$, which is indeed what we observe in the space-time diagram of $\log\langle\Sigma\rangle_\phi$ in the sixth panel of Figure~\ref{fig:main_spacetime}.

\begin{figure}[htbp]
    \centering
    \includegraphics[width=0.5\textwidth,keepaspectratio]{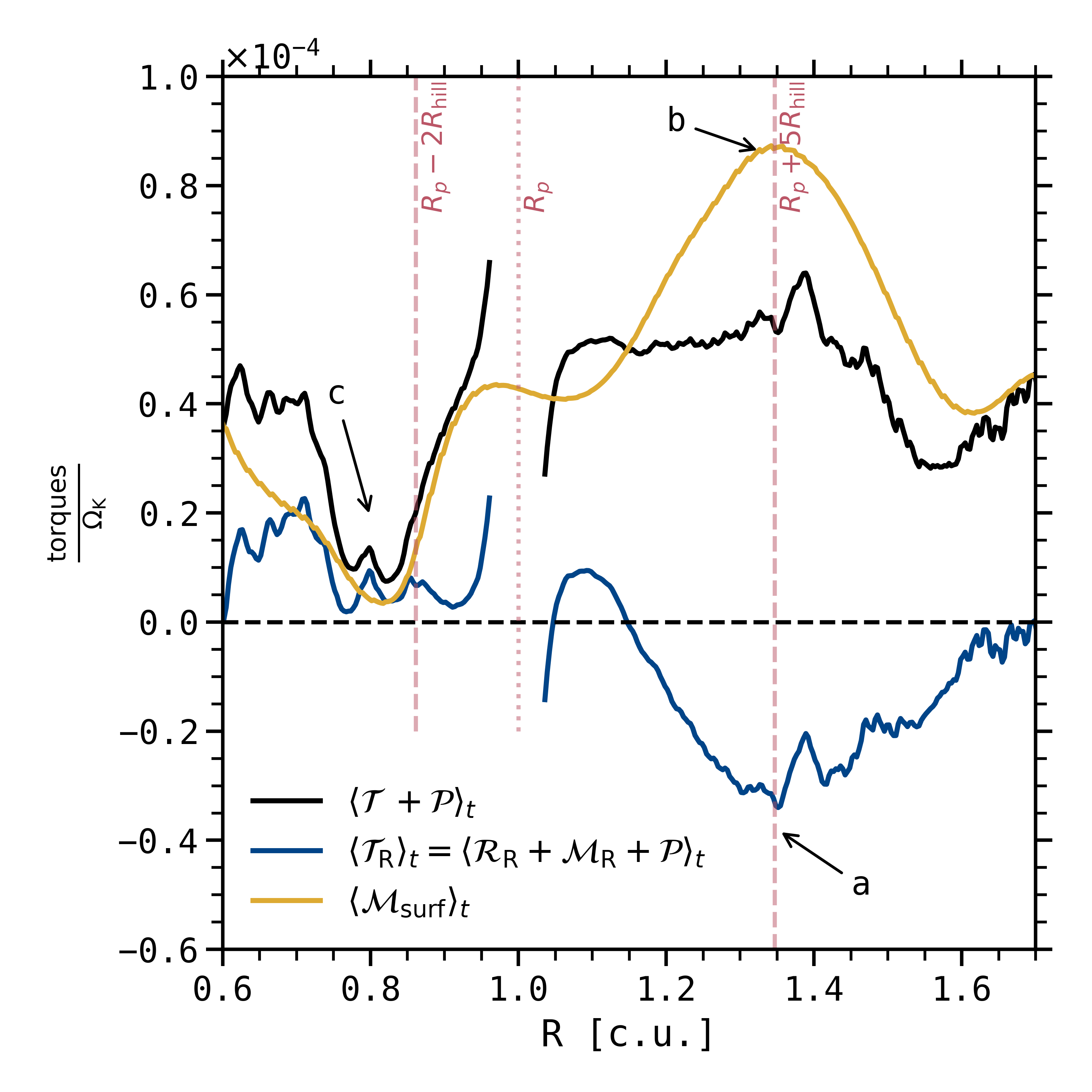}%
    \caption{Radial profile of the torques exerted by the wind ($\mathcal{M}_{\rm surf}$, yellow curve) and the sum of the radial torques and planet torques ($\mathcal{R}_R$, $\mathcal{M}_R$ and $\mathcal{P}$, blue curve), averaged over the episode $\mathcal{E}_2$ in the run \emph{Mj-$\beta_3$}. The black curve displays the sum of all torques enumerated in Eq.~(\ref{eq:angular_momentum}). We removed the contribution close to the planet and its CPD in the evaluation of the planet and radial torques. We add three labels to identify different angular momentum redistribution phenomena that we believe are essential in apprehending the gap asymmetries: \textbf{a}. Angular momentum deposition by the outer planet wake at the gap's outer edge. \textbf{b}. Angular momentum evacuation in the wind. \textbf{c}. Inefficient angular momentum extraction by the inner planet wake at the gap's inner edge, and small value of the wind torque.}
    \label{fig:main_torques_pw}
\end{figure}

Furthermore, $\dot{M}_{\rm acc}$ is approximately constant in the gap on average for both episodes, which is key for the interpretation of the gap asymmetry. If we consider the torques exerted on the gas in the gap, we have two main contributions. On one hand, the tidal torque $\mathcal{P}$ is due to the presence of the planet and deposits angular momentum in the flow in the form of spiral wakes. These wakes, in turn, transport angular momentum in the radial fluid torques $\mathcal{R}_R + \mathcal{M}_R$. We therefore gather these three torques into a planet-induced torque $\mathcal{T}_R = \mathcal{R}_R + \mathcal{M}_R + \mathcal{P}$. On the other hand, the magnetic wind torque $\mathcal{M}_{\rm surf}$ is the main driver of accretion (see Figure~\ref{fig:main_stress}). In Figure~\ref{fig:main_torques_pw}, we show the radial profiles in the gap of $\mathcal{T}_R/\Omega_K$ (blue curve), $\mathcal{M}_{\rm surf}/\Omega_K$ (yellow curve), and the total torque exerted on the gas (black curve) as expressed in Eq.~(\ref{eq:angular_momentum}), averaged over the last $100$ orbits (episode $\mathcal{E}_2$). We note that the total torque $(\mathcal{T}+\mathcal{P})/\Omega_K$ is close to the accretion rate $\dot{M}_{\rm acc}/4\pi$ displayed in the top panel of Figure~\ref{fig:main_macc_xi}, with a relative difference of $\simeq22\%$ on average in the gap. In particular, we retrieve that the total torque is approximately constant in the gap, which means that the torques $\mathcal{T}_R/\Omega_K$ and $\mathcal{M}_{\rm surf}/\Omega_K$ that contribute to the accretion somehow counterbalance each other. We also retrieve a small step in the total torque at the gap's outer edge, necessary to understand the gradual erosion of the outer disk by the gap. Regarding $\mathcal{T}_R$, we can divide it in $\mathcal{T}_{R-}$ and $\mathcal{T}_{R+}$, respectively for the planet-related torques in the inner gap and the outer gap. $\mathcal{T}_{R+}$ is negative and tends to push material outward from the planet, with angular momentum that is deposited by the planet's outer wake. However, $\mathcal{T}_{R-}$ is quite small and positive, and therefore seems inefficient at extracting angular momentum via the planet's inner wake.

We can interpret the redistribution of angular momentum in the gap as follows. Near the gap outer edge, the planet-induced torque deposits an excess of angular momentum ($\mathcal{T}_{R+}$ is negative and minimum, see label \textbf{a} in Figure~\ref{fig:main_torques_pw}) that is efficiently transferred to the wind and eventually vertically evacuated at the disk surface ($\mathcal{M}_{\rm surf}$ positive and maximum, see label \textbf{b} in Figure~\ref{fig:main_torques_pw}). Near the gap inner edge, the planet does not extract efficiently angular momentum from the gas ($\mathcal{T}_{R-}$ positive but close to zero), and the wind torque is also negligible at this location ($\mathcal{M}_{\rm surf}\gtrsim0$, see label \textbf{c} in Figure~\ref{fig:main_torques_pw}). The end result is an apparent aborted wind in the inner gap regions, and an enhanced wind in the outer gap region.

\begin{figure}[htbp]
    \centering
    \includegraphics[width=0.5\textwidth,keepaspectratio]{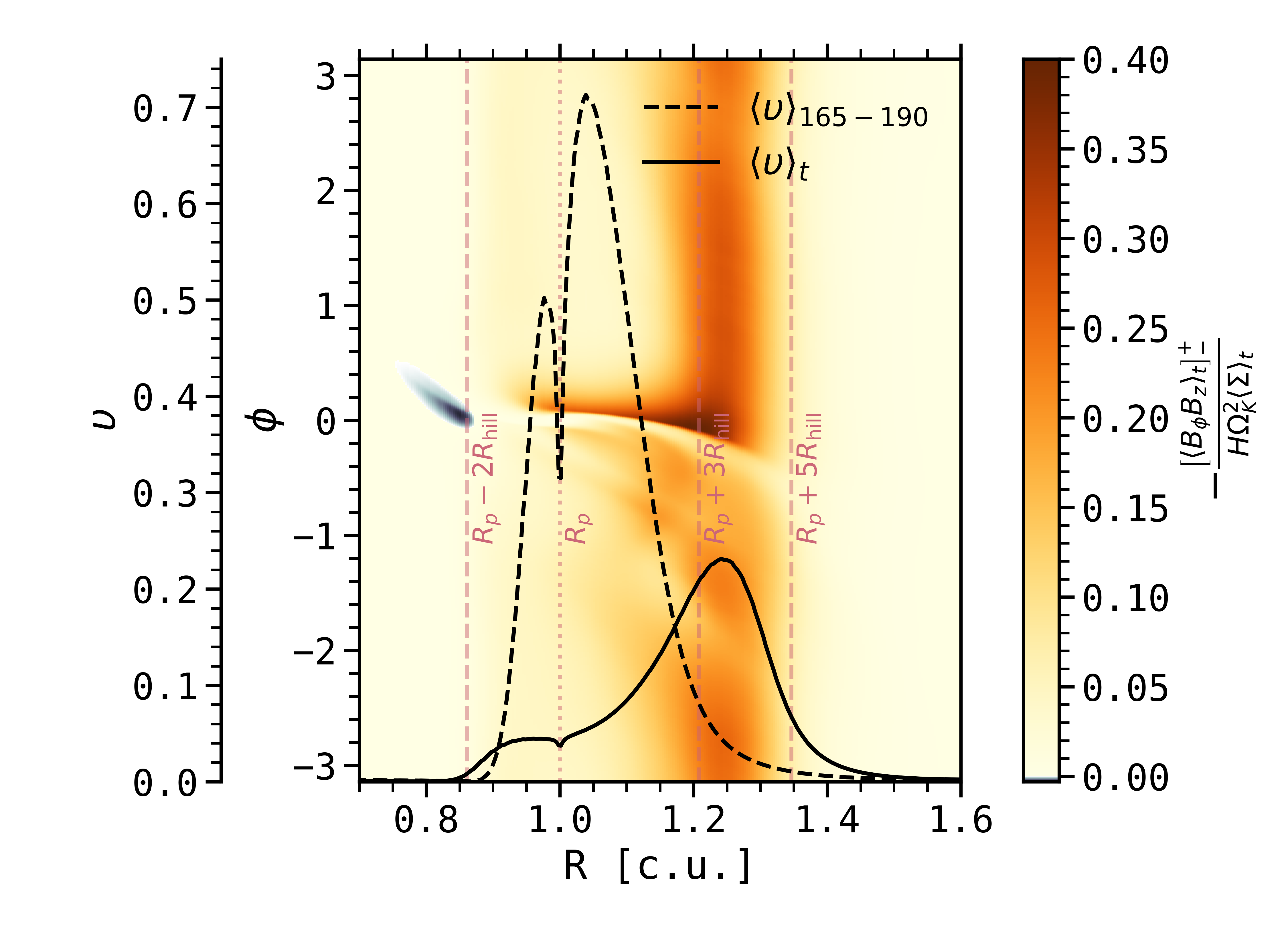}%
    \caption{2D map of $\upsilon$, i.e. the magnetic torque per unit of surface density, at the disk surface in ($R$-$\phi$) coordinates, averaged over the episode $\mathcal{E}_2$ in the run \emph{Mj-$\beta_3$}. We highlight in dark blue the region where the magnetic torque is negative, which corresponds here to a slight deposition of angular momentum. The solid and dashed black lines show respectively $\upsilon$ during $\mathcal{E}_2$ and $\mathcal{E}_1$.}
    \label{fig:main_upsilon}
\end{figure}

Due to the predominance of the wind torque in the accretion, we focus in Figure~\ref{fig:main_upsilon} on the evolution in the gap of the $\upsilon$ parameter, which is a dimensionless measure of $\mathcal{M}_{\rm surf}$ (Eq.~\ref{eq:upsilon}). This plot shows the 2D map of $\upsilon$ in ($R$-$\phi$) coordinates for $\mathcal{E}_2$, as well as its azimuthally-averaged version (solid black line). We also show the azimuthally-averaged $\upsilon$ for $\mathcal{E}_1$ (dashed black line). During both episodes, $\upsilon$ is larger in the outer gap than in the inner gap, with an asymmetry of the angular momentum extraction $\Delta \upsilon\simeq0.2$ in both cases. 
In the ($R$-$\phi$) plane, we detect a clear azimuthal asymmetry of $\upsilon$, with a larger value at $\phi>0$. The wind torque seems affected by the outer planet's spiral wake, with an enhanced $\upsilon$ along the wake towards the inner gap near the CPD, from $R=1.25$ to $R=0.95$. This increased $\upsilon$ is also accompanied by a localized enhancement of the magnetization near the outer wake, which would suggest that magnetic field accumulates efficiently from the outer wake to the CPD. This scenario of magnetic accumulation in the CPD seems to be confirmed by the launching of a collimated protoplanetary jet (see \ref{sec:HM2_magnetic}).

Near the gap's inner edge, the wind torque is particularly small, and can even be negative close to the planet’s azimuth. This region of negative $\left[\langle B_\phi B_z\rangle_t\right]_{-}^{+}$ is represented in Figure~\ref{fig:main_upsilon} with the dark blue area near $R=0.8$, and can be related to label \textbf{c} in Figure~\ref{fig:main_torques_pw} at that same radial location. At this location, the wind is actually deflected due to the planet-induced deflection of the poloidal magnetic field lines at the disk surface towards the midplane, leading to a rearranging of $\left[\langle B_\phi B_z\rangle_t\right]_{-}^{+}$ (see in Figure~\ref{fig:main_Mj_bpbprho} the LIC texture and magnetic field lines in black that are nearly tangential to the surface layers at $\pm z_{\rm w}\simeq \pm 4H(R)$, also near $R=0.8$). We have checked that this deflection is all the more pronounced as the planet is massive. We can therefore have a strong magnetic accumulation both near the inner and outer gap edges as indicated in Figures~\ref{fig:main_Mj_spacetime_Bz_vm} and \ref{fig:main_Mj_bpbprho}, and yet have an apparent inefficient (efficient) wind with a small (large) fraction of mass flux escaping from the inner (outer) gap edge, as suggested in Figure~\ref{fig:main_3Mj_meridional_flux} (see in particular the wind-driven flows pinpointed by the dotted black and white arrows).

We would like to draw the attention on the fact that the one-dimensional estimation of the velocity structure of the wind (and more generally the gas flow) can be misleading, as it actually results from a complex three-dimensional dynamics. In a theoretical point of view, the radial evolution of 1D or 2D quantities like $\mathcal{M}_{\rm surf}$ or $\left[\langle B_\phi B_z\rangle_t\right]_{-}^{+}$ could suggest that there is no wind near the gap's inner edge (the wind torque is almost zero), while the 3D dynamics indicates in reality that this behavior is due to the local deflection of the wind downwards and then upwards (with an N-shaped structure of the wind for the most massive planets, see e.g. the intersection of the upper and lower surfaces $\pm z_{\rm w}\simeq \pm 4H(R)$ with the white streamlines in Figure~\ref{fig:main_3Mj_meridional_flux}). Similarly in an observational point of view, for such a wind we would expect a predominant role of the emitting surface on the kinematic signature of the wind. At high altitude, e.g. for an optically thick tracer such as CO, one would expect to measure an outflow at all radii. At lower altitude (but still far enough from the disk bulk), e.g. for an optically thinner tracer such that $^{13}$CO, one would expect to measure an outflow signature above the outer gap and an inflow signature above the inner gap due to the deflection of the wind in this region. At even lower altitude, e.g. for an even optically thinner observational tracer like C$^{18}$O, one would this time expect to be sensitive to the nearly sonic accretion flows, from the outer disk's surface layers to the planet poles and from the inner gap to the inner disk's surface layers.

\begin{figure*}
    \centering
    \includegraphics[width=0.99\hsize]{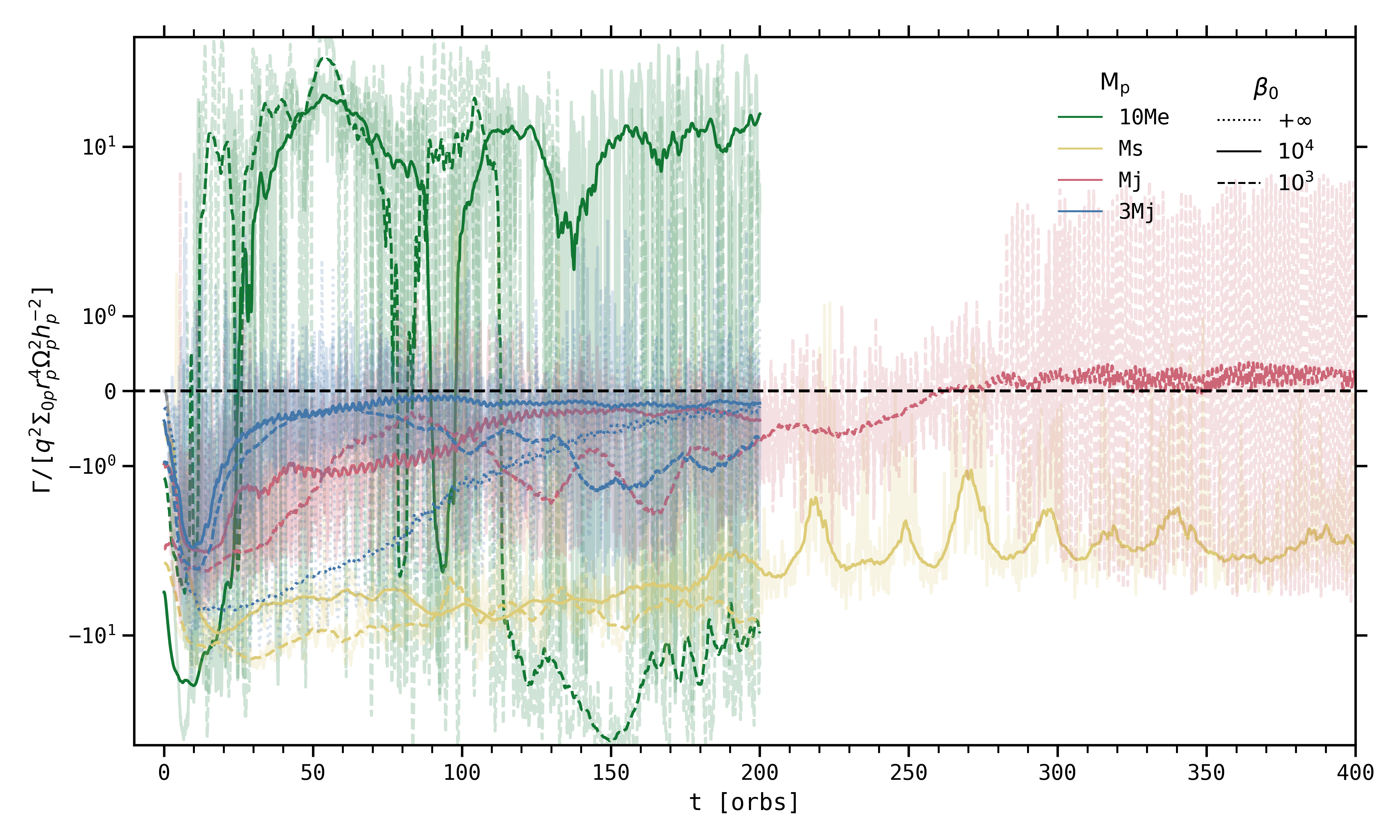}%
    \caption{Temporal evolution of the gravitational torques exerted by the gas onto the planets for the nine runs. The torques are normalized by $\Gamma_0=q_p^2\Sigma_0 R_p^4\Omega_K(R_p)^2 h_0^{-2}$. The different colors show different planet masses: $10M_{\oplus}$ (green), $M_s$ (yellow), $M_j$ (red) and $3M_j$ (blue). The different line-styles indicate the initial magnetization: $\beta_0=+\infty$ (inviscid hydrodynamical run, dotted line), $\beta_0=10^4$ (solid line) and $\beta_0=10^3$ (dashed line). The transparent variable curves show the instantaneous torques, on top of which are displayed the torques with a moving average of $10$ orbits.}
    \label{fig:main_torques}
\end{figure*}

To sum up, one of the key point here is the connection between the magnetic field accumulation, the wind torque and the planet-related torques. If $B_p$ accumulates at the gap edges, we would expect a strong wind torque at these locations. Because the outer planet wake deposit angular momentum near the outer gap edge and because magnetic field lines are deflected by the planet at a certain altitude above the inner gap edge, our estimate of the wind torque is larger in the outer gap than in the inner gap.
We next explore how the erosion of the outer disk by the gap affects the effective torque exerted by a magnetized disk onto massive planets.

\subsection{Gravitational torques}
\label{sec:torques}

We discuss here the evaluation of the gravitational torque $\Gamma$ exerted by the gas onto the different planets in all 3D simulations, and focus more specifically on high mass planets. Such torque is defined in {\fontfamily{cmtt}\selectfont IDEFIX} as
\begin{equation}
    \Gamma=-\displaystyle \sum_{i,j,k}^{N_R,N_\phi,N_z}\Gamma^{i,j,k}_{p\rightarrow d}V^{i,j,k}_{\rm cell},
    \label{eq:gamma_planet}
\end{equation}

\noindent with $\Gamma^{i,j,k}_{p\rightarrow d}$ the torque per unit of volume defined in Section~\ref{sec:dw_interactions} and evaluated for the cell $i,j,k$ of volume $V^{i,j,k}_{\rm cell}$. The estimate of this quantity is valuable for disk/planet interactions modeling, because it has a direct impact on how the planet would migrate if its orbital evolution were taken into account. In this case indeed, numerically, the value of the instantaneous torque is used to update the orbital parameters of the planet. Figure~\ref{fig:main_torques} displays the temporal evolution of $\Gamma$ for our $8$ non-ideal MHD simulations and the inviscid hydrodynamical simulation \emph{$3$Mj-$\alpha_0$}, with a moving average of $10$ orbits and normalized by $\Gamma_0=q_p^2\Sigma_0 R_p^4\Omega_K(R_p)^2 h_0^{-2}$. Each color corresponds to a planet mass $M_p$, and the line-style indicates the value of $\beta_0$.

We first notice that $\Gamma$ is negative for the three most massive planets, with a smaller $|\Gamma|$ for larger planet masses. It suggests a slower inward migration for more massive planets. We observe indeed a factor $\simeq10$ for $|\Gamma|$ between the cases $3M_j$ and $M_s$. This is in agreement with the models of interactions between a gap-opening planet and a viscous disk: the amount of gas inside this gap is larger if the planet is not massive enough to empty its coorbital region \citep[see, e.g.,][]{Baruteau&Masset2013}. This excess of gas boosts the so-called Lindblad torque, which is an essential (usually) negative component in the total torque. A more negative torque for a less massive planet is indeed what we observe in Figure~\ref{fig:main_torques}.

Second, at fixed $M_p$ (except for the least massive planet $10M_{\oplus}$), we show that $\beta_0$ seems to have little influence on the value of the torque, at least on the first $200$ orbits. Still, it would seem that the absolute value of the torque ends up being slightly larger when the magnetization increases (i.e. when $\beta_0$ decreases), which is especially visible between $\simeq100$ and $200$ orbits. It means that the migration is faster when the magnetization is higher. It is consistent, because a larger magnetization leads to a denser gap (see in particular Section~\ref{sec:GO_variability} and Figure~\ref{fig:main_spacetime}), and therefore an amplified Lindblad torque and a faster migration. However, for the longer run \emph{Mj-$\beta_3$}, we expect an additional mechanism to occur. Because the outer gap widens and drifts outwards (i.e. the planet is closer to the inner gap edge), we expect the external Lindblad torque to weaken and decrease compared to the internal Lindblad torque, leading to a less and less negative total torque. We checked that the total outer torque $|\Gamma_+|$ decreases more than the total inner torque $|\Gamma_-|$. We also noticed that the fluctuations of $|\Gamma_+|$ and $|\Gamma_-|$ are at first equivalent until $|\Gamma_-|$ overcomes $|\Gamma_+|$ near $\simeq280$ orbits. At that time, the inner torque starts to strongly fluctuate, with fluctuations $\simeq3$ times larger with respect to the outer torque. These fluctuations may be due to the formation of a thin density ring (corresponding to a pressure maximum) confined between the planet gap and a cavity that develops in the inner grid edge, probably linked to an artifact in the inner boundary (see end of Section~\ref{sec:GO_overview}). This ring is unstable to the RWI as a vortex develops, and becomes particularly strong after $\simeq330$ orbits. In practice, the frequency of the planet torque fluctuations may be linked to the periodic transit of this vortex in the vicinity of the inner planet wake. Due to the weakening of the external torque, we expect an increasingly slower inward migration, and even a potential reverse migration, as shown in Figure~\ref{fig:main_torques} by the sign reversal of $\Gamma$ at $270$ orbits. This phenomenon could also be obtained by the action of a positive component of the corotation torque exerted by the MHD wind, as proposed by some studies which have added hydrodynamical prescriptions of the impact of these winds on the migration of massive planets \citep[][]{Kimmig2020}. The main difference here is that the outward migration they observe is due to a dynamical corotation torque linked to the inward accretion flow and leading to a type-III runaway migration. Even without mentioning the dynamical corotation torque, it could be challenging to study migration of such planets by 3D non-ideal MHD simulations, because the dynamical evolution of the gap is complex and the characteristic time-scale for migration reversal can be quite long ($\gtrsim300$ orbits for the run \emph{Mj-$\beta_3$}).

Third, we notice that the total torque exerted by the gas on the least massive planet ($10M_{\oplus}$) seems stochastic during the simulation (for the two magnetization cases), which makes it difficult to guess the direction of migration (i.e. the response of the planet to the total torque). This is similar to what has been observed by \citet{Nelson&Papaloizou2004,Uribe2011}, with strong fluctuations of the total torque experienced by low-mass planets on an orbital time-scale, oscillating between positive and negative values. They suggested that the orbital evolution of such planet would proceed as a random walk. For the disk models without and with a low-mass planet, we show in Appendix~\ref{sec:appendix_b2} that there is a low level of turbulence in terms of angular momentum transport (i.e., accretion is mostly wind-driven). However, we also show in Appendix~\ref{sec:appendix_b3} that this turbulence plays a non-negligible role in the stochastic nature of the planet torques. This relation between the turbulence intensity and the planet torque dispersion is not linear, because the level of turbulence increases faster than the torque dispersion. As a first experiment, we activated planet migration after the first $\simeq50$ orbits of the run \emph{$10$Me-$\beta_4$}, for an initial disk density of $\Sigma_0=1.25\times10^{-4}$ in code units, which corresponds to $\geq$ \SI{10}{g.cm^{-2}} at \SI{10}{au} for a solar-mass star. The migration was found to be slow and directed outwards during at least $10$ planet orbits, at a speed of $\simeq3.5\times10^{-3}~$\SI{}{au} per orbit. Outward planet migration in wind-driven accretion disks could thus help the survival of planets at large radii that are often invoked to interpret substructures \citep[][]{Zhang2018,Isella2018,Lodato2019,Baruteau2019} and gas kinematics \citep[][]{Pinte2018,Pinte2019,Pinte2020,Izquierdo2021,Izquierdo2022} in ALMA disks. 

\begin{figure}[htbp]
    \centering
    \includegraphics[width=0.5\textwidth,keepaspectratio]{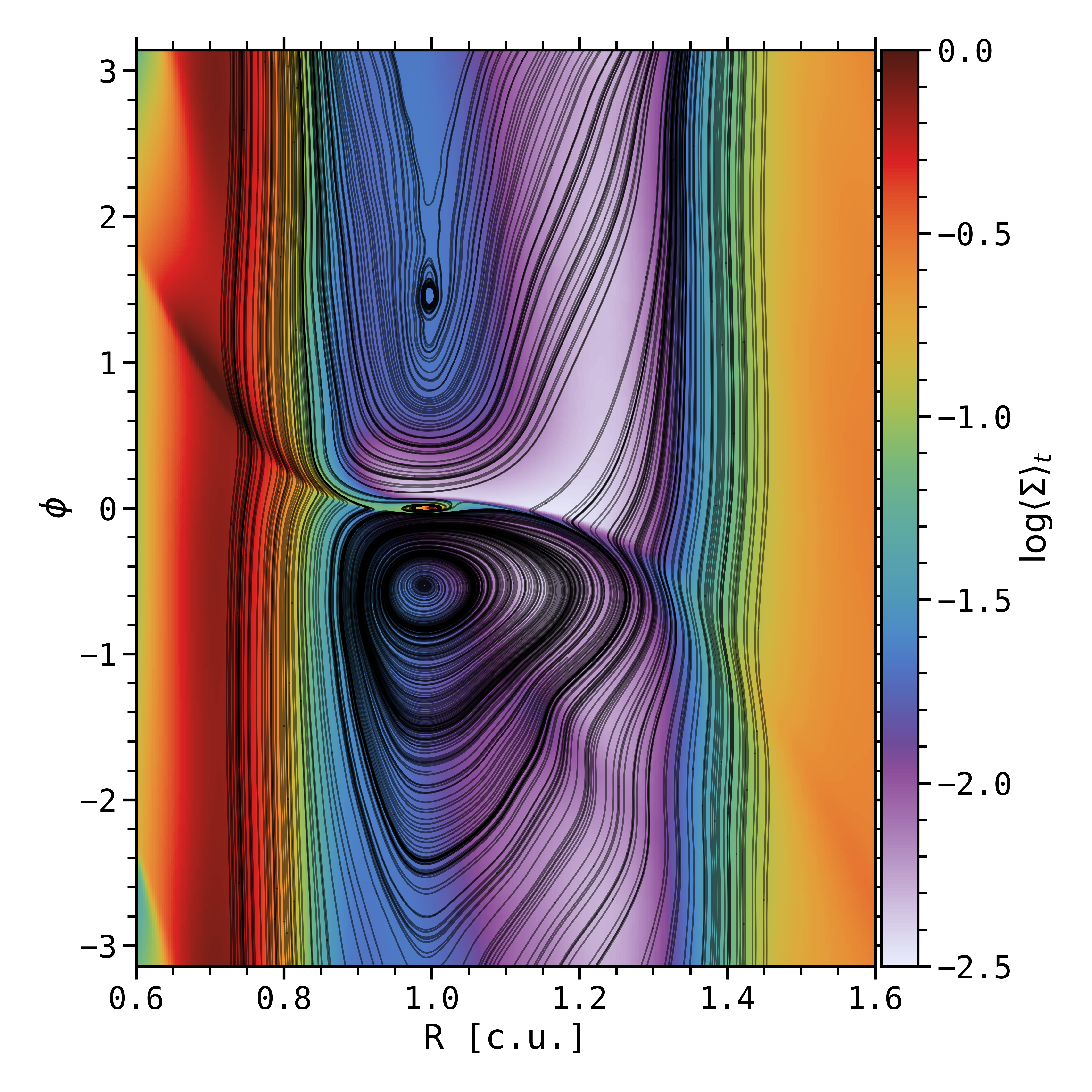}%
    \caption{Gas surface density $\log{(\Sigma)}$, averaged over the last $100$ orbits, for \emph{Mj-$\beta_3$}. The streamlines in the midplane are shown in black.}
    \label{fig:main_streamlines}
\end{figure}

When there is an episode of strong accretion, \citet{Kimmig2020} expect a mass excess in front of the planet because the gas performs an inward U-turn and then is accreted in the inner disk before it can perform an outward U-turn. In other words, it is easier for the gas to cross the horsehoe region in front of the planet than behind due to the inward accretion flow (see their Figure~11). This strong azimuthal asymmetry of the horseshoe U-turns (there are more inward U-turns than outward U-turns) would be associated with a positive dynamical corotation torque, and therefore an outward migration. However, the azimuthal asymmetry in our non-ideal MHD runs is not the same depending on time and on the simulation, with sometimes an under-density in front, sometimes behind the planet in azimuth. This could be due to the joint influence of magnetic field accumulation and accretion sporadicity, either leading to a complex redistribution of matter inside the horsehoe region or an evacuation in the wind (see Appendix~\ref{sec:appendix_a}). Besides, by looking at the streamlines in Figure~\ref{fig:main_streamlines} for the run \emph{Mj-$\beta_3$} averaged over the last $100$ orbits, we remark that the horseshoe region shrinks to a region of limited azimuthal extent, as a result of the fast radial accretion flow driven by the MHD wind. This region is centered around a libration point closer to the planet in azimuth than the Lagrange point L5. This is consistent with \citet{Peplinski2008b,Peplinski2008c} who observe such "tadpole-like" region around generalized Lagrange points G4 and G5 if the migration is respectively inward and outward. For slowly migrating planets, G4 and G5 correspond to L4 and L5 and the horseshoe region is fairly symmetric on the $2\pi$ angle of the corotation region. On the contrary, if planet migration is fast, the libration points move azimuthally closer to the planet as the horseshoe region shrinks. In the planet's reference frame, because in our case the libration point is G5$\neq$L5, the inward accretion flow that crosses the planet gap on average over the last $100$ orbits is equivalent to a fast outward migration, regarding the shape of the horseshoe region. It would therefore be worthwhile to determine how the migration of Jupiter-mass planets with such inward nearly-sonic accretion streamer inside the gap differs from the classical picture of type-II migration in viscous disks. In this latter case, we indeed expect a slow migration proportional to the disk viscosity, with a negligible role of gap-crossing flows \citep[][]{Robert2018}.

As mentioned in Section~\ref{sec:GO_overview}, for the most massive planets ($M_p\geq M_j$) and the low-magnetization case ($\beta_0=10^4$), a strong vortex appears at the gap's outer edge, before decaying after $\simeq100$ orbits. It would be interesting to compare planet migration with the two-phase vortex-driven migration presented in \citet{Lega2021,Lega2022}. Note that in all our simulations, small-scale non-axisymmetric structures seem to develop everywhere in the disk and could interact with the planet during its migration. Note also that the torque calculated here is the one exerted by the gas on a fixed planet, which does not take into account the coupling between $\Gamma$ and the migration. This link between migration and torque operates in what is called the dynamical corotation torque, which tends to amplify (positive feedback) or attenuate (negative feedback) the migration according to the physical properties of the disk and the coorbital region \citep[][]{Masset2008,Paardekooper2014}. The next step is therefore to confirm the first results presented in Figure~\ref{fig:main_torques} by letting the planet evolve radially according to $\Gamma$, and thus to study the impact of MHD winds on planetary migration. Even if the gaps seem to open in less than $100$ orbits for the most massive planets whatever the magnetization (see Section~\ref{sec:GO_variability}), the strong variability in the accretion and the secular evolution of the gap (outer gap's outward drift) can change the planet response: for the case \emph{Mj-$\beta_3$}, the torque seems to reach a steady-state positive value solely after $\simeq300$ planet orbits, which is challenging in terms of computational costs. It also raises the question of the typical time-scale at which to activate planet migration.

\section{Summary and Discussion} 
\label{sec:conclusion}

\subsection{Summary} 
\label{sec:summary}

In this study, we performed with the {\fontfamily{cmtt}\selectfont IDEFIX} code 3D non-ideal MHD simulations of a planet in a fixed circular orbit, interacting with a wind-launching protoplanetary disk. We focused in particular on the ability of massive planets to open a gap in the presence of a vertical magnetic field. One of the challenge of this work is to achieve a balance between the large-scale study of a disk with a wind and dynamical processes localized around a planet, as well as to isolate the multiple phenomena that emerge all at once in the simulations. 

We find that gap opening occurs, with density gaps induced by self-organisation everywhere in the disk, especially at higher magnetization, and induced by the planet if the planet mass is high enough (Section~\ref{sec:GO}). When comparing gap opening in our MHD simulations, we notice that the planet gap is in a counter-intuitive way denser when the initial magnetization increases (i.e. when $\beta_0$ decreases), that we interpret as a consequence of the strong dependence of the mass accretion rate with the disk magnetization. We derived a gap opening criterion when accretion is dominated by MHD winds, which depends on the planet-to-primary mass ratio $q_p$, the aspect ratio $h$ and the normalized magnetic wind torque $\upsilon$ (Section~\ref{sec:GO_criterion}). 
Besides, the combination of upper-layer accretion flows, planet-driven flows and wind-driven flows drives a large-scale gas motion that takes the form of a meridional circulation, especially in the outer disk   (Section~\ref{sec:HM2_MF}). The accretion of material comes from the surface layers of the outer disk to the inner disk. More specifically, two accreting surface layers fall directly from the outer disk towards the planet poles. We discard here the accretion onto the planet, but this should have a strong impact on the planet's internal and orbital evolutions, as presented in \citet{Gressel2013}.

Magnetic field tends to efficiently accumulate in the planet gap, which is coherent with \cite{Zhu2013}, \cite{Keith&Wardle2015} and \cite{Carballido2017}. Depending on the initial magnetization and the planet mass, it is found to accumulate firstly at the gap center, and later on at both gap edges. Such magnetic accumulation could be at the origin of complex effects, like the launching of a protoplanetary jet, magnetic reconnection and magnetic braking. We observe a well-known anti-correlation between the density and magnetic field, with for example for the run \emph{Mj-$\beta_3$} an episode (episode $\mathcal{E}_1$) of strong gas depletion in the gap associated with an efficient magnetic field accumulation. The radial extent of the density gap is larger than the magnetic accumulation region, as observed in other works (Section~\ref{sec:HM2_magnetic}).

When $\beta_0=10^3$, i.e. when the amplitude of magnetic field is large in the gap, we have an efficient angular momentum removal by the wind torque, corresponding to a high equivalent $\alpha_{\rm dw}\simeq \alpha_{\rm dw}^M \simeq 5$ in the gap for the run \emph{Mj-$\beta_3$} averaged over the last $100$ orbits (episode $\mathcal{E}_2$). We note that the radial stresses that act on the gas are also larger on average when the magnetization increases, leading to an equivalent Shakura $\&$ Sunyaev $\alpha_\nu\lesssim10^{-2}$ in the cases $\beta_0=10^4$, and $\alpha_\nu>10^{-2}$ when $\beta_0=10^3$ in most of the disk, reaching $\simeq1$ for the deepest gaps. 

The torque exerted by the planet onto the gas (planet-induced torque) mainly deposits angular momentum near the gap's outer edge. On the other hand, the magnetic wind torque is almost always extracting angular momentum from the disk and has therefore almost the same sign on both sides of the planet. The only exception is close to the gap's inner edge, and all the more so when the planet mass is larger. In that case, the outflow and its poloidal magnetic field lines are deflected by the planet's gravitational field at the disk surface, leading to a decrease of the wind torque which can even become negative. Note that this deflection could be interpreted as a local inflow signature (this is a purely geometrical effect as deflected streamlines eventually escape the disc). For low-mass planets, the planet-induced torque is relatively small, as the planet hardly generates a gap. The wind torque is therefore dominant and approximately constant in the gap, whatever the initial magnetization. 
For high mass planets, the wind torque is quenched in the inner gap due to this effect of magnetic field lines deflection, while planet-induced and wind torques have opposite signs in the outer gap. This forces a symmetry breaking between the inner and outer gaps. The end result is an outflow that is seemingly quenched from the inner gap because of the bending of magnetic lines.
On the contrary, an enhanced wind is emitted from the outer gap, fed by the positive planet torque. The resulting asymmetry of the wind torque in the gap is somehow at the origin of the step in the accretion rate which in turn leads to a gradual erosion of the outer disk, only visible in our high planet mass ($M_p\geq M_j$) and high magnetization ($\beta_0=10^3$) runs. For weaker magnetisations ($\beta_0=10^4$), the gap's outer drift is small, if not negligible, on the small timescales considered here ($200$ orbits). 


The complex redistribution of angular momentum in the gap related to the planet and wind torques conditions the accretion behavior. In particular, we find that the accretion rate $\dot{M}_{\rm acc}$ is nearly constant and comparable to the disk one in the gap. Furthermore, at high magnetization ($\beta_0=10^3$), the enhanced wind torque induces a top-down non-symmetric nearly-sonic accretion streamer in the planet gap, as proposed by \cite{Frank2014}. 
We note that the accretion in the gap is sporadic, leading to azimuthal asymmetries of the horseshoe region that could generate an unsaturated corotation torque \citep[][]{Ogihara2017,Kimmig2020}. The temporal variability of accretion and magnetic field accumulation is also echoed in the temporal variability of the vertical redistribution of matter in and around the gap. For example, gas is sometimes evacuated directly into the wind from the outer disk, or, only a few orbits later, wraps poloidally around the planet (Appendix~\ref{sec:appendix_a}). Although the accretion rate does not evolve much radially in the disk, it is on average slightly larger in the gap than at the gap edges. Such mismatch in the radial profile of $\dot{M}_{\rm acc}$ leads to strong radial asymmetries of the gap both in width and in depth: there is a mass loss (accumulation) in the outer (inner) gap, which is slightly emptier (denser). More specifically, the outer gap is wider and emptier at higher magnetization due to a stronger wind torque.

Regarding tidal torques exerted on massive planets in magnetized disks, the radial asymmetries of the planet gap gradually lead to a slightly positive gravitational torque exerted by the gas onto the planet. It means that the inward migration of a massive planet in such disk could be slowed down, then stopped, and even reversed. For low-mass planets, the gravitational torque is stochastic because of turbulence, oscillating between positive and negative values. It would seem however that the migration of such planets is slow and directed outwards (Section~\ref{sec:torques}). This is to be confirmed by enabling planet migration in future numerical simulations. 

\subsection{Comparison to works using prescribed wind torques}

We can estimate $\alpha_\nu$ and $\alpha_{dw}$ in our simulations and compare with the Figure~4 of \citet{Elbakyan2022} that shows the dependence of the minimum planet mass needed to
open a gap as a function of $\alpha_\nu$ and $\alpha_{dw}$. In our cases $\beta_0=10^4$, with the exact definitions of $\alpha_\nu$ and $\alpha_{\rm dw}$ in Eq.~(4) and (6) by \citet{Tabone2022} (neglecting $\mathcal{R}_{\rm surf}$ and averaged in time over the last $100$ orbits), we find outside the planet gap that $\alpha_\nu\simeq10^{-3}-10^{-2}$ and $\alpha_{\rm dw}\simeq4\times10^{-2}$. It corresponds to a minimum pessimistic (concerning their model) planet mass needed to open a gap of $\simeq1M_j$, but our run \emph{Ms-$\beta_4$} is able to produce a gap. Likewise, when $\beta_0=10^3$ and outside the planet gap, $\alpha_\nu\simeq3\times10^{-3}-6\times10^{-2}$ and $\alpha_{\rm dw}\simeq6\times10^{-2}-0.9$ (also averaged in time over the last $100$ orbits). When $\beta_0=10^3$, the Maxwell stresses dominate their Reynolds counterparts, both radially and at the disk surface. Looking at Figure~4 of \citet{Elbakyan2022}, it corresponds to a minimum (very) pessimistic planet mass needed to open a gap of $\simeq2M_j$, but here again, the run \emph{Mj-$\beta_3$} is able to produce a gap. One major difference with \citet{Elbakyan2022} is that $\alpha_\nu$ and $\alpha_{dw}$ strongly depend on the magnetic field. Because the distribution of magnetic field relies on self-organized gaps and planet gaps, $\alpha_\nu$ and $\alpha_{dw}$ also depend on the localization in the disk. One aspect that could be improved when prescribing the wind torque is to use a $\alpha_{\rm dw}$ that is not constant in time (accretion bursts) and in radius (especially in the gap, with values $10-100$ times larger). We would also need a way to model the accumulation of magnetic field in the gaps, as it is important to understand the gaps' radial asymmetries. Note that we retrieve a correct order of magnitude for the critical planet-to-primary mass ratio $q_p^C$ to open a gap with the criterion we derived in Eq.~\ref{eq:criterion}.

In \citet{Kimmig2020}, they consider the migration of massive planets in 2D, with a wind torque prescribed via a constant lever arm $\lambda$ and constant mass loss parameter $b$. This latter quantity is related to the vertical mass flux $\dot{\Sigma}_{\rm wind}=\left[\langle\rho v_z\rangle_\phi\right]_{-}^{+}$. However, if we compute $\dot{\Sigma}_{\rm wind}$ from our 3D simulations and use it to define an equivalent mass loss, we find that $b$ is not constant and varies radially. In particular, for (\emph{Mj-$\beta_3$}), $b$ reaches a few $10^{-3}$ in the disk, slightly increasing radially, but an order of magnitude higher in the gap during the episode $\mathcal{E}_2$ (i.e. during the last $100$ orbits), reaching $1.75\times10^{-2}$. Considering Figure~7 of \citet{Kimmig2020} (or the two bottom panels of their Figure~5), it could have a strong influence on the Jupiter-mass planet migration, as in one case we would have almost no migration ($b=10^{-3}$) and in the other case a fast runaway outward migration ($b\simeq10^{-2}$). The lever arm $\lambda$ is more complicated to characterize as there are some inflow at the gap edges, leading to sign reversal of the mass outflow rate. However, in the gap, $\lambda$ is larger than $2.5$, reaching $\simeq7.5$ in some gaps. \citet{Kimmig2020} chose a constant lever arm $\lambda=2.25$, which is coherent with the values we obtain in the disk but underestimated in the planet-induced gaps.

\citet{Lega2022} showed that when the planet-induced torque was larger than the wind torque, matter should pile up at the gap outer edge leading to a fast inward migration and the blocking of the accretion flow. In our simulations, we never observed the blocking of the accretion flow. On the contrary, we find that a large positive planet-induced torque is usually compensated for by a stronger wind torque (Figure~\ref{fig:main_torques_pw}) so as to keep a constant accretion rate through the gap. This discrepancy is probably due to the fact that the wind responds dynamically to the planet-induced torque, but also to the redistribution of magnetic field that changes the efficiency of magnetic wind launching. The overall conclusion is that constant wind torque that are often prescribed in effective 2D or 3D hydrodynamical models are too simplistic to capture the dynamics of planet-disk-wind interaction.

Finally, as observed in \citet{McNally2020}, our runs involving low-mass planets ($M_p=10M_{\oplus}$) confirm that surface accreting layers and midplane seem decoupled regarding accretion, even if we detect highly variable gravitational torques $\Gamma$ exerted on such planets. We would therefore need to activate planetary migration in order to assess the impact of wind-driven accretion on the orbital evolution of low-mass planets.

\begin{acknowledgements}
The authors acknowledge support from the European Research Council (ERC) under the European Union Horizon 2020 research and innovation programme (Grant agreement No. 815559 (MHDiscs)). This work was granted access to the HPC resources of IDRIS under the allocation A0100402231 made by GENCI. The 3D simulations were performed with the version v0.8.0 of {\fontfamily{cmtt}\selectfont IDEFIX}, commit bbffc9390353ab917728734e77543f69041eb003. Some of the computations presented in this paper were performed using the GRICAD infrastructure (https://gricad.univ-grenoble-alpes.fr), which is supported by Grenoble research communities. The data presented in this work were processed and plotted with Python via various libraries, in particular {\fontfamily{cmtt}\selectfont numpy}, {\fontfamily{cmtt}\selectfont matplotlib} and {\fontfamily{cmtt}\selectfont scipy}, but also the following personal projects under development: {\fontfamily{cmtt}\selectfont nonos} to analyze results from {\fontfamily{cmtt}\selectfont IDEFIX} simulations, {\fontfamily{cmtt}\selectfont cblind} for most of the colormaps and {\fontfamily{cmtt}\selectfont lick} for the visualizations in line integral convolution. G.W.F. wishes also to thank Clément Baruteau for fruitful discussions.
\end{acknowledgements}

\bibliographystyle{aa}
\bibliography{biblio}

\begin{thebibliography}{103}
\expandafter\ifx\csname natexlab\endcsname\relax\def\natexlab#1{#1}\fi

\bibitem[{{Andrews} {et~al.}(2018){Andrews}, {Huang}, {P{\'e}rez}, {Isella},
  {Dullemond}, {Kurtovic}, {Guzm{\'a}n}, {Carpenter}, {Wilner}, {Zhang}, {Zhu},
  {Birnstiel}, {Bai}, {Benisty}, {Hughes}, {{\"O}berg}, \&
  {Ricci}}]{Andrews2018}
{Andrews}, S.~M., {Huang}, J., {P{\'e}rez}, L.~M., {et~al.} 2018, \apjl, 869,
  L41

\bibitem[{{Aoyama} \& {Bai}(2023)}]{Aoyama2023}
{Aoyama}, Y. \& {Bai}, X. 2023, arXiv e-prints, arXiv:2302.01514

\bibitem[{{Avenhaus} {et~al.}(2018){Avenhaus}, {Quanz}, {Garufi}, {Perez},
  {Casassus}, {Pinte}, {Bertrang}, {Caceres}, {Benisty}, \&
  {Dominik}}]{Avenhaus2018}
{Avenhaus}, H., {Quanz}, S.~P., {Garufi}, A., {et~al.} 2018, \apj, 863, 44

\bibitem[{{Bae} {et~al.}(2016){Bae}, {Zhu}, \& {Hartmann}}]{Bae2016}
{Bae}, J., {Zhu}, Z., \& {Hartmann}, L. 2016, \apj, 819, 134

\bibitem[{{Bai} \& {Stone}(2013{\natexlab{a}})}]{Bai2013b}
{Bai}, X.-N. \& {Stone}, J.~M. 2013{\natexlab{a}}, \apj, 769, 76

\bibitem[{{Bai} \& {Stone}(2013{\natexlab{b}})}]{Bai&Stone2013}
{Bai}, X.-N. \& {Stone}, J.~M. 2013{\natexlab{b}}, \apj, 769, 76

\bibitem[{{Balbus} \& {Hawley}(1991)}]{Balbus&Hawley1991}
{Balbus}, S.~A. \& {Hawley}, J.~F. 1991, \apj, 376, 214

\bibitem[{{Balbus} \& {Papaloizou}(1999)}]{Balbus&Papaloizou1999}
{Balbus}, S.~A. \& {Papaloizou}, J. C.~B. 1999, \apj, 521, 650

\bibitem[{{Baruteau} {et~al.}(2019){Baruteau}, {Barraza}, {P{\'e}rez},
  {Casassus}, {Dong}, {Lyra}, {Marino}, {Christiaens}, {Zhu}, {Carmona},
  {Debras}, \& {Alarcon}}]{Baruteau2019}
{Baruteau}, C., {Barraza}, M., {P{\'e}rez}, S., {et~al.} 2019, \mnras, 486, 304

\bibitem[{{Baruteau} {et~al.}(2014){Baruteau}, {Crida}, {Paardekooper},
  {Masset}, {Guilet}, {Bitsch}, {Nelson}, {Kley}, \&
  {Papaloizou}}]{Baruteau2014}
{Baruteau}, C., {Crida}, A., {Paardekooper}, S.~J., {et~al.} 2014, in
  Protostars and Planets VI, ed. H.~{Beuther}, R.~S. {Klessen}, C.~P.
  {Dullemond}, \& T.~{Henning}, 667

\bibitem[{{Baruteau} {et~al.}(2011){Baruteau}, {Fromang}, {Nelson}, \&
  {Masset}}]{Baruteau2011}
{Baruteau}, C., {Fromang}, S., {Nelson}, R.~P., \& {Masset}, F. 2011, \aap,
  533, A84

\bibitem[{{Baruteau} \& {Masset}(2013)}]{Baruteau&Masset2013}
{Baruteau}, C. \& {Masset}, F. 2013, in Lecture Notes in Physics, Berlin
  Springer Verlag, ed. J.~{Souchay}, S.~{Mathis}, \& T.~{Tokieda}, Vol. 861,
  201

\bibitem[{{B{\'e}thune} \& {Latter}(2022)}]{Bethune&Latter2022}
{B{\'e}thune}, W. \& {Latter}, H. 2022, arXiv e-prints, arXiv:2206.03917

\bibitem[{{B{\'e}thune} {et~al.}(2017){B{\'e}thune}, {Lesur}, \&
  {Ferreira}}]{Bethune2017}
{B{\'e}thune}, W., {Lesur}, G., \& {Ferreira}, J. 2017, \aap, 600, A75

\bibitem[{{Blandford} \& {Payne}(1982)}]{Blandford&Payne1982}
{Blandford}, R.~D. \& {Payne}, D.~G. 1982, \mnras, 199, 883

\bibitem[{{Carballido} {et~al.}(2017){Carballido}, {Matthews}, \&
  {Hyde}}]{Carballido2017}
{Carballido}, A., {Matthews}, L.~S., \& {Hyde}, T.~W. 2017, \mnras, 472, 3277

\bibitem[{{Combet} \& {Ferreira}(2008)}]{Combet&Ferreira2008}
{Combet}, C. \& {Ferreira}, J. 2008, \aap, 479, 481

\bibitem[{{Crida} {et~al.}(2006){Crida}, {Morbidelli}, \& {Masset}}]{Crida2006}
{Crida}, A., {Morbidelli}, A., \& {Masset}, F. 2006, \icarus, 181, 587

\bibitem[{{Cui} \& {Bai}(2021)}]{Cui&Bai2021}
{Cui}, C. \& {Bai}, X.-N. 2021, \mnras, 507, 1106

\bibitem[{{Cui} \& {Bai}(2022)}]{Cui&Bai2022}
{Cui}, C. \& {Bai}, X.-N. 2022, \mnras, 516, 4660

\bibitem[{{de Valon} {et~al.}(2020){de Valon}, {Dougados}, {Cabrit}, {Louvet},
  {Zapata}, \& {Mardones}}]{deValon2020}
{de Valon}, A., {Dougados}, C., {Cabrit}, S., {et~al.} 2020, \aap, 634, L12

\bibitem[{{Elbakyan} {et~al.}(2022){Elbakyan}, {Wu}, {Nayakshin}, \&
  {Rosotti}}]{Elbakyan2022}
{Elbakyan}, V., {Wu}, Y., {Nayakshin}, S., \& {Rosotti}, G. 2022, arXiv
  e-prints, arXiv:2206.11595

\bibitem[{{Flaherty} {et~al.}(2017){Flaherty}, {Hughes}, {Rose}, {Simon}, {Qi},
  {Andrews}, {K{\'o}sp{\'a}l}, {Wilner}, {Chiang}, {Armitage}, \&
  {Bai}}]{Flaherty2017}
{Flaherty}, K.~M., {Hughes}, A.~M., {Rose}, S.~C., {et~al.} 2017, \apj, 843,
  150

\bibitem[{{Flaherty} {et~al.}(2015){Flaherty}, {Hughes}, {Rosenfeld},
  {Andrews}, {Chiang}, {Simon}, {Kerzner}, \& {Wilner}}]{Flaherty2015}
{Flaherty}, K.~M., {Hughes}, A.~M., {Rosenfeld}, K.~A., {et~al.} 2015, \apj,
  813, 99

\bibitem[{{Frank} {et~al.}(2014){Frank}, {Ray}, {Cabrit}, {Hartigan}, {Arce},
  {Bacciotti}, {Bally}, {Benisty}, {Eisl{\"o}ffel}, {G{\"u}del}, {Lebedev},
  {Nisini}, \& {Raga}}]{Frank2014}
{Frank}, A., {Ray}, T.~P., {Cabrit}, S., {et~al.} 2014, in Protostars and
  Planets VI, ed. H.~{Beuther}, R.~S. {Klessen}, C.~P. {Dullemond}, \&
  T.~{Henning}, 451

\bibitem[{{Fung} \& {Chiang}(2016)}]{Fung&Chiang2016}
{Fung}, J. \& {Chiang}, E. 2016, \apj, 832, 105

\bibitem[{{Fung} {et~al.}(2014){Fung}, {Shi}, \& {Chiang}}]{Fung2014}
{Fung}, J., {Shi}, J.-M., \& {Chiang}, E. 2014, \apj, 782, 88

\bibitem[{{Gardiner} \& {Stone}(2005)}]{Gardiner2005}
{Gardiner}, T.~A. \& {Stone}, J.~M. 2005, Journal of Computational Physics,
  205, 509

\bibitem[{{Gressel} {et~al.}(2013){Gressel}, {Nelson}, {Turner}, \&
  {Ziegler}}]{Gressel2013}
{Gressel}, O., {Nelson}, R.~P., {Turner}, N.~J., \& {Ziegler}, U. 2013, \apj,
  779, 59

\bibitem[{{Guilet} \& {Ogilvie}(2012)}]{Guilet2012}
{Guilet}, J. \& {Ogilvie}, G.~I. 2012, \mnras, 424, 2097

\bibitem[{{Guilet} \& {Ogilvie}(2013)}]{Guilet2013}
{Guilet}, J. \& {Ogilvie}, G.~I. 2013, \mnras, 430, 822

\bibitem[{{Guilet} \& {Ogilvie}(2014)}]{Guilet2014}
{Guilet}, J. \& {Ogilvie}, G.~I. 2014, \mnras, 441, 852

\bibitem[{{Guzm{\'a}n} {et~al.}(2018){Guzm{\'a}n}, {Huang}, {Andrews},
  {Isella}, {P{\'e}rez}, {Carpenter}, {Dullemond}, {Ricci}, {Birnstiel},
  {Zhang}, {Zhu}, {Bai}, {Benisty}, {{\"O}berg}, \& {Wilner}}]{Guzman2018}
{Guzm{\'a}n}, V.~V., {Huang}, J., {Andrews}, S.~M., {et~al.} 2018, \apjl, 869,
  L48

\bibitem[{{Huang} {et~al.}(2018){Huang}, {Andrews}, {Dullemond}, {Isella},
  {P{\'e}rez}, {Guzm{\'a}n}, {{\"O}berg}, {Zhu}, {Zhang}, {Bai}, {Benisty},
  {Birnstiel}, {Carpenter}, {Hughes}, {Ricci}, {Weaver}, \&
  {Wilner}}]{Huang2018}
{Huang}, J., {Andrews}, S.~M., {Dullemond}, C.~P., {et~al.} 2018, \apjl, 869,
  L42

\bibitem[{{Isella} {et~al.}(2018){Isella}, {Huang}, {Andrews}, {Dullemond},
  {Birnstiel}, {Zhang}, {Zhu}, {Guzm{\'a}n}, {P{\'e}rez}, {Bai}, {Benisty},
  {Carpenter}, {Ricci}, \& {Wilner}}]{Isella2018}
{Isella}, A., {Huang}, J., {Andrews}, S.~M., {et~al.} 2018, \apjl, 869, L49

\bibitem[{{Izquierdo} {et~al.}(2022){Izquierdo}, {Facchini}, {Rosotti}, {van
  Dishoeck}, \& {Testi}}]{Izquierdo2022}
{Izquierdo}, A.~F., {Facchini}, S., {Rosotti}, G.~P., {van Dishoeck}, E.~F., \&
  {Testi}, L. 2022, \apj, 928, 2

\bibitem[{{Izquierdo} {et~al.}(2021){Izquierdo}, {Testi}, {Facchini},
  {Rosotti}, \& {van Dishoeck}}]{Izquierdo2021}
{Izquierdo}, A.~F., {Testi}, L., {Facchini}, S., {Rosotti}, G.~P., \& {van
  Dishoeck}, E.~F. 2021, \aap, 650, A179

\bibitem[{{Keith} \& {Wardle}(2015)}]{Keith&Wardle2015}
{Keith}, S.~L. \& {Wardle}, M. 2015, \mnras, 451, 1104

\bibitem[{{Kim} \& {Turner}(2020)}]{Kim&Turner2020}
{Kim}, S.~Y. \& {Turner}, N.~J. 2020, \apj, 889, 159

\bibitem[{{Kimmig} {et~al.}(2020){Kimmig}, {Dullemond}, \& {Kley}}]{Kimmig2020}
{Kimmig}, C.~N., {Dullemond}, C.~P., \& {Kley}, W. 2020, \aap, 633, A4

\bibitem[{{Kratter} \& {Lodato}(2016)}]{Kratter&Lodato2016}
{Kratter}, K. \& {Lodato}, G. 2016, \araa, 54, 271

\bibitem[{{Lega} {et~al.}(2022){Lega}, {Morbidelli}, {Nelson}, {Ramos},
  {Crida}, {B{\'e}thune}, \& {Batygin}}]{Lega2022}
{Lega}, E., {Morbidelli}, A., {Nelson}, R.~P., {et~al.} 2022, \aap, 658, A32

\bibitem[{{Lega} {et~al.}(2021){Lega}, {Nelson}, {Morbidelli}, {Kley},
  {B{\'e}thune}, {Crida}, {Kloster}, {M{\'e}heut}, {Rometsch}, \&
  {Ziampras}}]{Lega2021}
{Lega}, E., {Nelson}, R.~P., {Morbidelli}, A., {et~al.} 2021, \aap, 646, A166

\bibitem[{{Lesur} {et~al.}(2023){Lesur}, {Baghdadi}, \&
  {Wafflard-Fernandez}}]{Lesur2023}
{Lesur}, G., {Baghdadi}, S., \& {Wafflard-Fernandez}, G. 2023, in prep

\bibitem[{{Lesur} {et~al.}(2022){Lesur}, {Ercolano}, {Flock}, {Lin}, {Yang},
  {Barranco}, {Benitez-Llambay}, {Goodman}, {Johansen}, {Klahr}, {Laibe},
  {Lyra}, {Marcus}, {Nelson}, {Squire}, {Simon}, {Turner}, {Umurhan}, \&
  {Youdin}}]{Lesur2022}
{Lesur}, G., {Ercolano}, B., {Flock}, M., {et~al.} 2022, arXiv e-prints,
  arXiv:2203.09821

\bibitem[{{Lesur} {et~al.}(2014){Lesur}, {Kunz}, \& {Fromang}}]{Lesur2014}
{Lesur}, G., {Kunz}, M.~W., \& {Fromang}, S. 2014, \aap, 566, A56

\bibitem[{{Lesur} \& {Papaloizou}(2009)}]{Lesur&Papaloizou2009}
{Lesur}, G. \& {Papaloizou}, J.~C.~B. 2009, \aap, 498, 1

\bibitem[{{Lesur}(2021)}]{Lesur2021}
{Lesur}, G. R.~J. 2021, \aap, 650, A35

\bibitem[{{Leung} \& {Ogilvie}(2019)}]{Leung2019}
{Leung}, P. K.~C. \& {Ogilvie}, G.~I. 2019, \mnras, 487, 5155

\bibitem[{{Lin} \& {Youdin}(2015)}]{LinYoudin2015}
{Lin}, M.-K. \& {Youdin}, A.~N. 2015, \apj, 811, 17

\bibitem[{{Lodato} {et~al.}(2019){Lodato}, {Dipierro}, {Ragusa}, {Long},
  {Herczeg}, {Pascucci}, {Pinilla}, {Manara}, {Tazzari}, {Liu}, {Mulders},
  {Harsono}, {Boehler}, {M{\'e}nard}, {Johnstone}, {Salyk}, {van der Plas},
  {Cabrit}, {Edwards}, {Fischer}, {Hendler}, {Nisini}, {Rigliaco}, {Avenhaus},
  {Banzatti}, \& {Gully-Santiago}}]{Lodato2019}
{Lodato}, G., {Dipierro}, G., {Ragusa}, E., {et~al.} 2019, \mnras, 486, 453

\bibitem[{{Louvet} {et~al.}(2018){Louvet}, {Dougados}, {Cabrit}, {Mardones},
  {M{\'e}nard}, {Tabone}, {Pinte}, \& {Dent}}]{Louvet2018}
{Louvet}, F., {Dougados}, C., {Cabrit}, S., {et~al.} 2018, \aap, 618, A120

\bibitem[{{Lovelace} {et~al.}(1999){Lovelace}, {Li}, {Colgate}, \&
  {Nelson}}]{Lovelace1999}
{Lovelace}, R.~V.~E., {Li}, H., {Colgate}, S.~A., \& {Nelson}, A.~F. 1999,
  \apj, 513, 805

\bibitem[{{Manger} {et~al.}(2021){Manger}, {Pfeil}, \& {Klahr}}]{Manger2021}
{Manger}, N., {Pfeil}, T., \& {Klahr}, H. 2021, \mnras, 508, 5402

\bibitem[{{Martel} \& {Lesur}(2022)}]{Martel&Lesur2022}
{Martel}, {\'E}. \& {Lesur}, G. 2022, arXiv e-prints, arXiv:2205.02126

\bibitem[{{Masset}(2008)}]{Masset2008}
{Masset}, F.~S. 2008, in EAS Publications Series, Vol.~29, EAS Publications
  Series, ed. M.~J. {Goupil} \& J.~P. {Zahn}, 165--244

\bibitem[{{Masset} \& {Papaloizou}(2003)}]{Masset2003}
{Masset}, F.~S. \& {Papaloizou}, J.~C.~B. 2003, \apj, 588, 494

\bibitem[{{McNally} {et~al.}(2020){McNally}, {Nelson}, {Paardekooper},
  {Ben{\'\i}tez-Llambay}, \& {Gressel}}]{McNally2020}
{McNally}, C.~P., {Nelson}, R.~P., {Paardekooper}, S.-J.,
  {Ben{\'\i}tez-Llambay}, P., \& {Gressel}, O. 2020, \mnras, 493, 4382

\bibitem[{Meyer {et~al.}(2014)Meyer, Balsara, \& Aslam}]{Meyer2014}
Meyer, C.~D., Balsara, D.~S., \& Aslam, T.~D. 2014, Journal of Computational
  Physics, 257, 594

\bibitem[{{Morbidelli} {et~al.}(2014){Morbidelli}, {Szul{\'a}gyi}, {Crida},
  {Lega}, {Bitsch}, {Tanigawa}, \& {Kanagawa}}]{Morbidelli2014}
{Morbidelli}, A., {Szul{\'a}gyi}, J., {Crida}, A., {et~al.} 2014, \icarus, 232,
  266

\bibitem[{{Nazari} {et~al.}(2019){Nazari}, {Booth}, {Clarke}, {Rosotti},
  {Tazzari}, {Juhasz}, \& {Meru}}]{Nazari2019}
{Nazari}, P., {Booth}, R.~A., {Clarke}, C.~J., {et~al.} 2019, \mnras, 485, 5914

\bibitem[{{Nelson}(2005)}]{Nelson2005}
{Nelson}, R.~P. 2005, \aap, 443, 1067

\bibitem[{{Nelson} {et~al.}(2013){Nelson}, {Gressel}, \&
  {Umurhan}}]{Nelson2013}
{Nelson}, R.~P., {Gressel}, O., \& {Umurhan}, O.~M. 2013, \mnras, 435, 2610

\bibitem[{{Nelson} \& {Papaloizou}(2004)}]{Nelson&Papaloizou2004}
{Nelson}, R.~P. \& {Papaloizou}, J. C.~B. 2004, \mnras, 350, 849

\bibitem[{{{\"O}berg} {et~al.}(2021){{\"O}berg}, {Guzm{\'a}n}, {Walsh},
  {Aikawa}, {Bergin}, {Law}, {Loomis}, {Alarc{\'o}n}, {Andrews}, {Bae},
  {Bergner}, {Boehler}, {Booth}, {Bosman}, {Calahan}, {Cataldi}, {Cleeves},
  {Czekala}, {Furuya}, {Huang}, {Ilee}, {Kurtovic}, {Le Gal}, {Liu}, {Long},
  {M{\'e}nard}, {Nomura}, {P{\'e}rez}, {Qi}, {Schwarz}, {Sierra}, {Teague},
  {Tsukagoshi}, {Yamato}, {van't Hoff}, {Waggoner}, {Wilner}, \&
  {Zhang}}]{Oberg2021}
{{\"O}berg}, K.~I., {Guzm{\'a}n}, V.~V., {Walsh}, C., {et~al.} 2021, \apjs,
  257, 1

\bibitem[{{Ogihara} {et~al.}(2017){Ogihara}, {Kokubo}, {Suzuki}, {Morbidelli},
  \& {Crida}}]{Ogihara2017}
{Ogihara}, M., {Kokubo}, E., {Suzuki}, T.~K., {Morbidelli}, A., \& {Crida}, A.
  2017, \aap, 608, A74

\bibitem[{{Ogihara} {et~al.}(2015){Ogihara}, {Morbidelli}, \&
  {Guillot}}]{Ogihara2015}
{Ogihara}, M., {Morbidelli}, A., \& {Guillot}, T. 2015, \aap, 584, L1

\bibitem[{{Paardekooper}(2014)}]{Paardekooper2014}
{Paardekooper}, S.~J. 2014, \mnras, 444, 2031

\bibitem[{{Papaloizou} {et~al.}(2004){Papaloizou}, {Nelson}, \&
  {Snellgrove}}]{Papaloizou2004}
{Papaloizou}, J. C.~B., {Nelson}, R.~P., \& {Snellgrove}, M.~D. 2004, \mnras,
  350, 829

\bibitem[{{Pascucci} {et~al.}(2022){Pascucci}, {Cabrit}, {Edwards}, {Gorti},
  {Gressel}, \& {Suzuki}}]{Pascucci2022}
{Pascucci}, I., {Cabrit}, S., {Edwards}, S., {et~al.} 2022, arXiv e-prints,
  arXiv:2203.10068

\bibitem[{{Pepli{\'n}ski} {et~al.}(2008{\natexlab{a}}){Pepli{\'n}ski},
  {Artymowicz}, \& {Mellema}}]{Peplinski2008b}
{Pepli{\'n}ski}, A., {Artymowicz}, P., \& {Mellema}, G. 2008{\natexlab{a}},
  \mnras, 386, 179

\bibitem[{{Pepli{\'n}ski} {et~al.}(2008{\natexlab{b}}){Pepli{\'n}ski},
  {Artymowicz}, \& {Mellema}}]{Peplinski2008c}
{Pepli{\'n}ski}, A., {Artymowicz}, P., \& {Mellema}, G. 2008{\natexlab{b}},
  \mnras, 387, 1063

\bibitem[{{P{\'e}rez} {et~al.}(2019){P{\'e}rez}, {Casassus}, {Baruteau},
  {Dong}, {Hales}, \& {Cieza}}]{Perez2019}
{P{\'e}rez}, S., {Casassus}, S., {Baruteau}, C., {et~al.} 2019, \aj, 158, 15

\bibitem[{{Perez-Becker} \&
  {Chiang}(2011{\natexlab{a}})}]{Perez-Becker&Chiang2011b}
{Perez-Becker}, D. \& {Chiang}, E. 2011{\natexlab{a}}, \apj, 735, 8

\bibitem[{{Perez-Becker} \&
  {Chiang}(2011{\natexlab{b}})}]{Perez-Becker&Chiang2011a}
{Perez-Becker}, D. \& {Chiang}, E. 2011{\natexlab{b}}, \apj, 727, 2

\bibitem[{{Pinte} {et~al.}(2016){Pinte}, {Dent}, {M{\'e}nard}, {Hales}, {Hill},
  {Cortes}, \& {de Gregorio-Monsalvo}}]{Pinte2016}
{Pinte}, C., {Dent}, W.~R.~F., {M{\'e}nard}, F., {et~al.} 2016, \apj, 816, 25

\bibitem[{{Pinte} {et~al.}(2020){Pinte}, {Price}, {M{\'e}nard}, {Duch{\^e}ne},
  {Christiaens}, {Andrews}, {Huang}, {Hill}, {van der Plas}, {Perez}, {Isella},
  {Boehler}, {Dent}, {Mentiplay}, \& {Loomis}}]{Pinte2020}
{Pinte}, C., {Price}, D.~J., {M{\'e}nard}, F., {et~al.} 2020, \apjl, 890, L9

\bibitem[{{Pinte} {et~al.}(2018){Pinte}, {Price}, {M{\'e}nard}, {Duch{\^e}ne},
  {Dent}, {Hill}, {de Gregorio-Monsalvo}, {Hales}, \& {Mentiplay}}]{Pinte2018}
{Pinte}, C., {Price}, D.~J., {M{\'e}nard}, F., {et~al.} 2018, \apjl, 860, L13

\bibitem[{{Pinte} {et~al.}(2019){Pinte}, {van der Plas}, {M{\'e}nard}, {Price},
  {Christiaens}, {Hill}, {Mentiplay}, {Ginski}, {Choquet}, {Boehler},
  {Duch{\^e}ne}, {Perez}, \& {Casassus}}]{Pinte2019}
{Pinte}, C., {van der Plas}, G., {M{\'e}nard}, F., {et~al.} 2019, Nature
  Astronomy, 3, 1109

\bibitem[{{Rabago} \& {Zhu}(2021)}]{Rabago&Zhu2021}
{Rabago}, I. \& {Zhu}, Z. 2021, \mnras, 502, 5325

\bibitem[{{Riols} \& {Lesur}(2019)}]{Riols&Lesur2019}
{Riols}, A. \& {Lesur}, G. 2019, \aap, 625, A108

\bibitem[{{Riols} {et~al.}(2020){Riols}, {Lesur}, \& {Menard}}]{Riols2020}
{Riols}, A., {Lesur}, G., \& {Menard}, F. 2020, \aap, 639, A95

\bibitem[{{Robert} {et~al.}(2018){Robert}, {Crida}, {Lega}, {M{\'e}heut}, \&
  {Morbidelli}}]{Robert2018}
{Robert}, C.~M.~T., {Crida}, A., {Lega}, E., {M{\'e}heut}, H., \& {Morbidelli},
  A. 2018, \aap, 617, A98

\bibitem[{{Shakura} \& {Sunyaev}(1973)}]{Shakura1973}
{Shakura}, N.~I. \& {Sunyaev}, R.~A. 1973, \aap, 24, 337

\bibitem[{{Stoll} \& {Kley}(2014)}]{Stoll&Kley2014}
{Stoll}, M. H.~R. \& {Kley}, W. 2014, \aap, 572, A77

\bibitem[{{Suriano} {et~al.}(2017){Suriano}, {Li}, {Krasnopolsky}, \&
  {Shang}}]{Suriano2017}
{Suriano}, S.~S., {Li}, Z.-Y., {Krasnopolsky}, R., \& {Shang}, H. 2017, \mnras,
  468, 3850

\bibitem[{{Suriano} {et~al.}(2018){Suriano}, {Li}, {Krasnopolsky}, \&
  {Shang}}]{Suriano2018}
{Suriano}, S.~S., {Li}, Z.-Y., {Krasnopolsky}, R., \& {Shang}, H. 2018, \mnras,
  477, 1239

\bibitem[{{Suriano} {et~al.}(2019){Suriano}, {Li}, {Krasnopolsky}, {Suzuki}, \&
  {Shang}}]{Suriano2019}
{Suriano}, S.~S., {Li}, Z.-Y., {Krasnopolsky}, R., {Suzuki}, T.~K., \& {Shang},
  H. 2019, \mnras, 484, 107

\bibitem[{{Svanberg} {et~al.}(2022){Svanberg}, {Cui}, \& {Latter}}]{Svanberg22}
{Svanberg}, E., {Cui}, C., \& {Latter}, H.~N. 2022, \mnras, 514, 4581

\bibitem[{{Szul{\'a}gyi} {et~al.}(2022){Szul{\'a}gyi}, {Binkert}, \&
  {Surville}}]{Szulagyi2022}
{Szul{\'a}gyi}, J., {Binkert}, F., \& {Surville}, C. 2022, \apj, 924, 1

\bibitem[{{Szul{\'a}gyi} {et~al.}(2014){Szul{\'a}gyi}, {Morbidelli}, {Crida},
  \& {Masset}}]{Szulagyi2014}
{Szul{\'a}gyi}, J., {Morbidelli}, A., {Crida}, A., \& {Masset}, F. 2014, \apj,
  782, 65

\bibitem[{{Tabone} {et~al.}(2022){Tabone}, {Rosotti}, {Cridland}, {Armitage},
  \& {Lodato}}]{Tabone2022}
{Tabone}, B., {Rosotti}, G.~P., {Cridland}, A.~J., {Armitage}, P.~J., \&
  {Lodato}, G. 2022, \mnras, 512, 2290

\bibitem[{{Teague} {et~al.}(2019){Teague}, {Bae}, \& {Bergin}}]{Teague2019}
{Teague}, R., {Bae}, J., \& {Bergin}, E.~A. 2019, \nat, 574, 378

\bibitem[{{Thi} {et~al.}(2019){Thi}, {Lesur}, {Woitke}, {Kamp}, {Rab}, \&
  {Carmona}}]{Thi2019}
{Thi}, W.~F., {Lesur}, G., {Woitke}, P., {et~al.} 2019, \aap, 632, A44

\bibitem[{{Uribe} {et~al.}(2011){Uribe}, {Klahr}, {Flock}, \&
  {Henning}}]{Uribe2011}
{Uribe}, A.~L., {Klahr}, H., {Flock}, M., \& {Henning}, T. 2011, \apj, 736, 85

\bibitem[{{Venuti} {et~al.}(2014){Venuti}, {Bouvier}, {Flaccomio}, {Alencar},
  {Irwin}, {Stauffer}, {Cody}, {Teixeira}, {Sousa}, {Micela}, {Cuillandre}, \&
  {Peres}}]{Venuti2014}
{Venuti}, L., {Bouvier}, J., {Flaccomio}, E., {et~al.} 2014, \aap, 570, A82

\bibitem[{{Villenave} {et~al.}(2020){Villenave}, {M{\'e}nard}, {Dent},
  {Duch{\^e}ne}, {Stapelfeldt}, {Benisty}, {Boehler}, {van der Plas}, {Pinte},
  {Telkamp}, {Wolff}, {Flores}, {Lesur}, {Louvet}, {Riols}, {Dougados},
  {Williams}, \& {Padgett}}]{Villenave2020}
{Villenave}, M., {M{\'e}nard}, F., {Dent}, W.~R.~F., {et~al.} 2020, \aap, 642,
  A164

\bibitem[{{Wafflard-Fernandez} \&
  {Baruteau}(2020)}]{Wafflard-Fernandez&Baruteau2020}
{Wafflard-Fernandez}, G. \& {Baruteau}, C. 2020, \mnras, 493, 5892

\bibitem[{{Wardle} \& {Koenigl}(1993)}]{Wardle1993}
{Wardle}, M. \& {Koenigl}, A. 1993, \apj, 410, 218

\bibitem[{{Williams} \& {McPartland}(2016)}]{Williams&McPartland2016}
{Williams}, J.~P. \& {McPartland}, C. 2016, \apj, 830, 32

\bibitem[{{Zhang} {et~al.}(2018){Zhang}, {Zhu}, {Huang}, {Guzm{\'a}n},
  {Andrews}, {Birnstiel}, {Dullemond}, {Carpenter}, {Isella}, {P{\'e}rez},
  {Benisty}, {Wilner}, {Baruteau}, {Bai}, \& {Ricci}}]{Zhang2018}
{Zhang}, S., {Zhu}, Z., {Huang}, J., {et~al.} 2018, \apjl, 869, L47

\bibitem[{{Zhu} {et~al.}(2013){Zhu}, {Stone}, \& {Rafikov}}]{Zhu2013}
{Zhu}, Z., {Stone}, J.~M., \& {Rafikov}, R.~R. 2013, \apj, 768, 143

\bibitem[{{Ziampras} {et~al.}(2022){Ziampras}, {Kley}, \&
  {Nelson}}]{Ziampras2022}
{Ziampras}, A., {Kley}, W., \& {Nelson}, R.~P. 2022, arXiv e-prints,
  arXiv:2212.10639

\end{thebibliography}

\appendix

\section{Temporal variability of the flow}
\label{sec:appendix_a}

\begin{figure*}
    \centering
    \includegraphics[width=0.99\hsize]{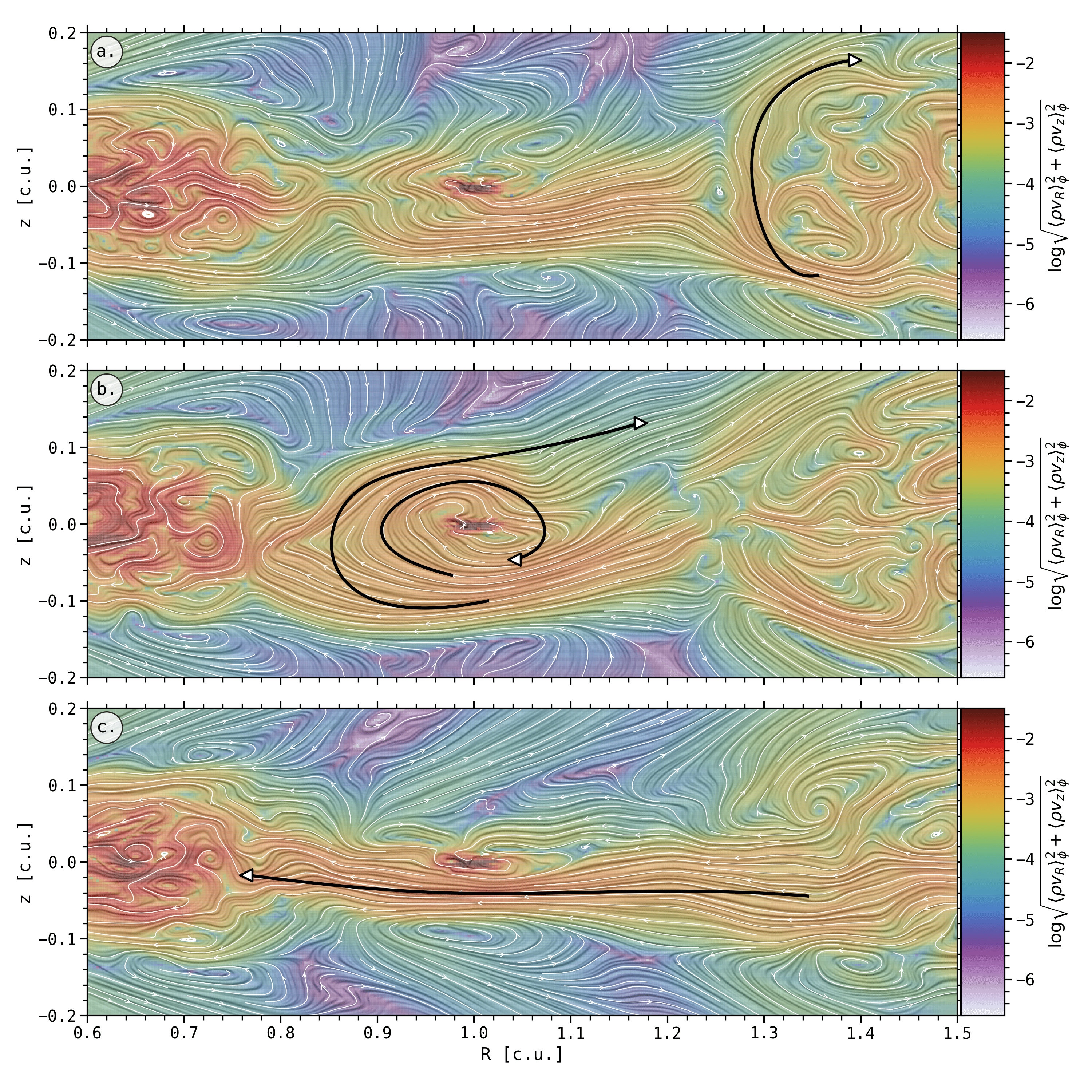} \\
    \caption{Temporal variability of the azimuthally-averaged mass flux after $119.75$ (top panel), $121.25$ (middle panel) and $129.5$ (bottom panel) orbits for the run \emph{$3$Mj-$\beta_3$}, with specific features represented by the black arrows: a. South-North U-turn of material at the transition between the outer disk and the gap. b. South-North U-turn of material at the transition between the gap and the inner disk. The material either rotates in the poloidal plane around the planet, or is evacuated in the upper layer's wind c. Top-down non-symmetrical accretion streamer from the outer disk to the inner disk.}
    \label{fig:appendix_mass_flux}
\end{figure*}

The aim of this appendix is to illustrate the temporal variability of the accretion streamer and of matter transport from the outer disk to the inner disk via the planet gap. In Figure~\ref{fig:appendix_mass_flux}, we show three snapshots between $119.75$ and $129.5$ orbits of the magnitude of the poloidal mass flux $\langle\rho v_p\rangle_{\phi}$ in log scale, averaged azimuthally for the run \emph{$3$Mj-$\beta_3$}. As in Figure~\ref{fig:main_3Mj_meridional_flux}, the texture in LIC and the white arrows indicate the direction of the radial ($\langle\rho v_R\rangle_{\phi}$) and vertical ($\langle\rho v_z\rangle_{\phi}$) components of the mass flux. Black arrows illustrate schematically streamlines of interest for the gas. The mass flux is extremely stochastic, with a fast temporal variability (inferior to the orbital timescale). Panel c. displays how the material sometimes crosses the gap from the outer disk to the inner disk via an accretion streamer at $z<0$ unperturbed by the disk morphology, as described in Section~\ref{sec:HM2_MF}. On the contrary, the flow is sometimes affected by the presence of the gap and is forced to perform sporadic South-North U-turns at the transition with the gap edges. Such material is either evacuated in the wind or redistributed in the outer disk and/or in the gap. In panel a., the U-turns carry material in the outer disk from the lower accretion layer to an upper wind-launching layer, and thus partially prevent the material from entering the planet gap. In panel b., the fast accreting material crosses $R=1$, under the planet, from the outer gap to the inner gap. Then, the South-North U-turns bring a fraction of this material directly to the wind, while the rest seems to accumulate in the partially refilled gap, rotating in a poloidal way around the planet radial location (note that the graphs are azimuthally-averaged maps, not azimuthal cuts). These U-turns are beyond the scope of our study, but could result from the accumulation of magnetic field at the gap edges.

\section{Planet-free disk model, turbulence and stochasticity}
\label{sec:appendix_b}

\begin{figure}[htbp]
    \centering
    \includegraphics[width=0.5\textwidth,keepaspectratio]{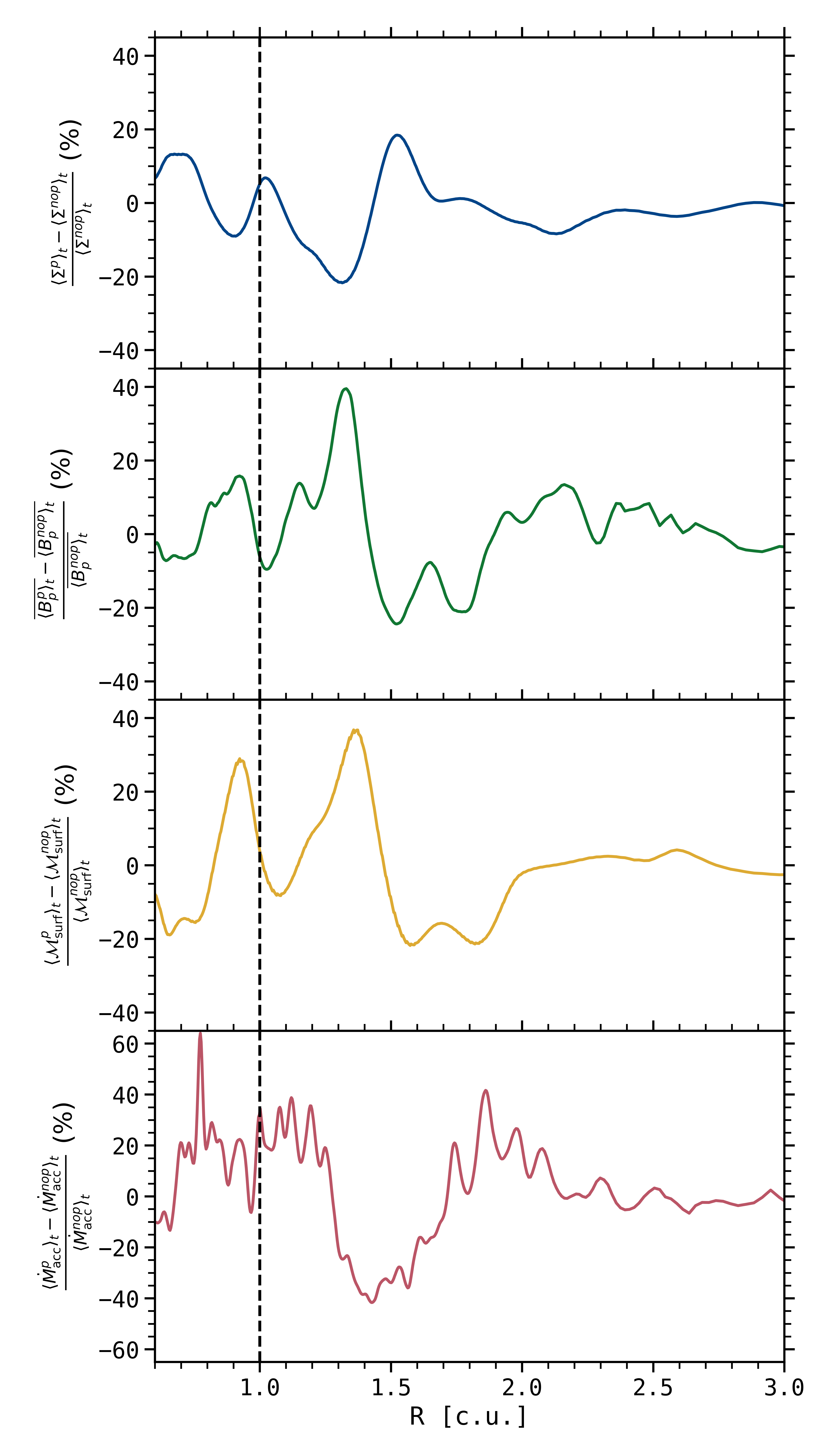}%
    \caption{Radial profiles of different quantities averaged between $50$ and $100$ orbits, compared in percent between a disk model without and with a $10$ Earth mass planet: gas surface density $\Sigma$ in blue (first panel), amplitude of the poloidal magnetic field $B_p$ in green (second panel), wind torque at the disk surface $\mathcal{M}_{\rm surf}$ in yellow (third panel), radial mass flux $\dot{M}_{\rm acc}$ in red (fourth panel). The location of the planet is indicated by the vertical black dashed line. Note that such disk model evolves dynamically, and the structures that are created at a given time will evolve radially and in amplitude, making the comparison difficult.}
    \label{fig:appendix_rhobpmaccmsurf}
\end{figure}

\subsection{Planet-free disk model}
\label{sec:appendix_b1}

In this paragraph, we develop how the protoplanetary disk model behaves in the absence of the planets for an initial $\beta_0=10^3$, and compare with the run \emph{$10$Me-$\beta_3$} in order to determine if the structures observed in that simulation are mainly due to the planet or not. If we look at the structures in the disk, they are similar without and with a low-mass planet, at least concerning the distributions of density and poloidal magnetic field, with in particular the emergence of self-organized structures (see also Section~\ref{sec:GO_overview} and, e.g., the top-right panel of Figure~\ref{fig:main_allsigma}). We can however highlight the fact that the presence of such planet can create rings of underdensity locally on both sides of its radial location (see the first panel in Figure~\ref{fig:appendix_rhobpmaccmsurf}, near $R=0.85$ and $R=1.3$) that accumulate the magnetic field more efficiently (see the two peaks of $\simeq15\%$ and $\simeq40\%$ in the second panel in Figure~\ref{fig:appendix_rhobpmaccmsurf} at those same locations). At the same time, this accumulation of magnetic field compared to a run without planet slightly enhances the wind torque by $\simeq30\%$ (see the third panel in Figure~\ref{fig:appendix_rhobpmaccmsurf}). Note also that the accretion via the radial mass flux is also increased by $\simeq20\%$ in the run \emph{$10$Me-$\beta_3$} around the planet between $R=0.7$ and $R=1.3$, although quite variable in that region (see the fourth panel in Figure~\ref{fig:appendix_rhobpmaccmsurf}). The step in $\dot{M}_{\rm acc}$ between $R=1.2$ and $R=1.4$ seems to indicate that the structures observed are drifting outwards during the episode considered here.

\subsection{Turbulence}
\label{sec:appendix_b2}

\begin{figure}[htbp]
    \centering
    \includegraphics[width=0.5\textwidth,keepaspectratio]{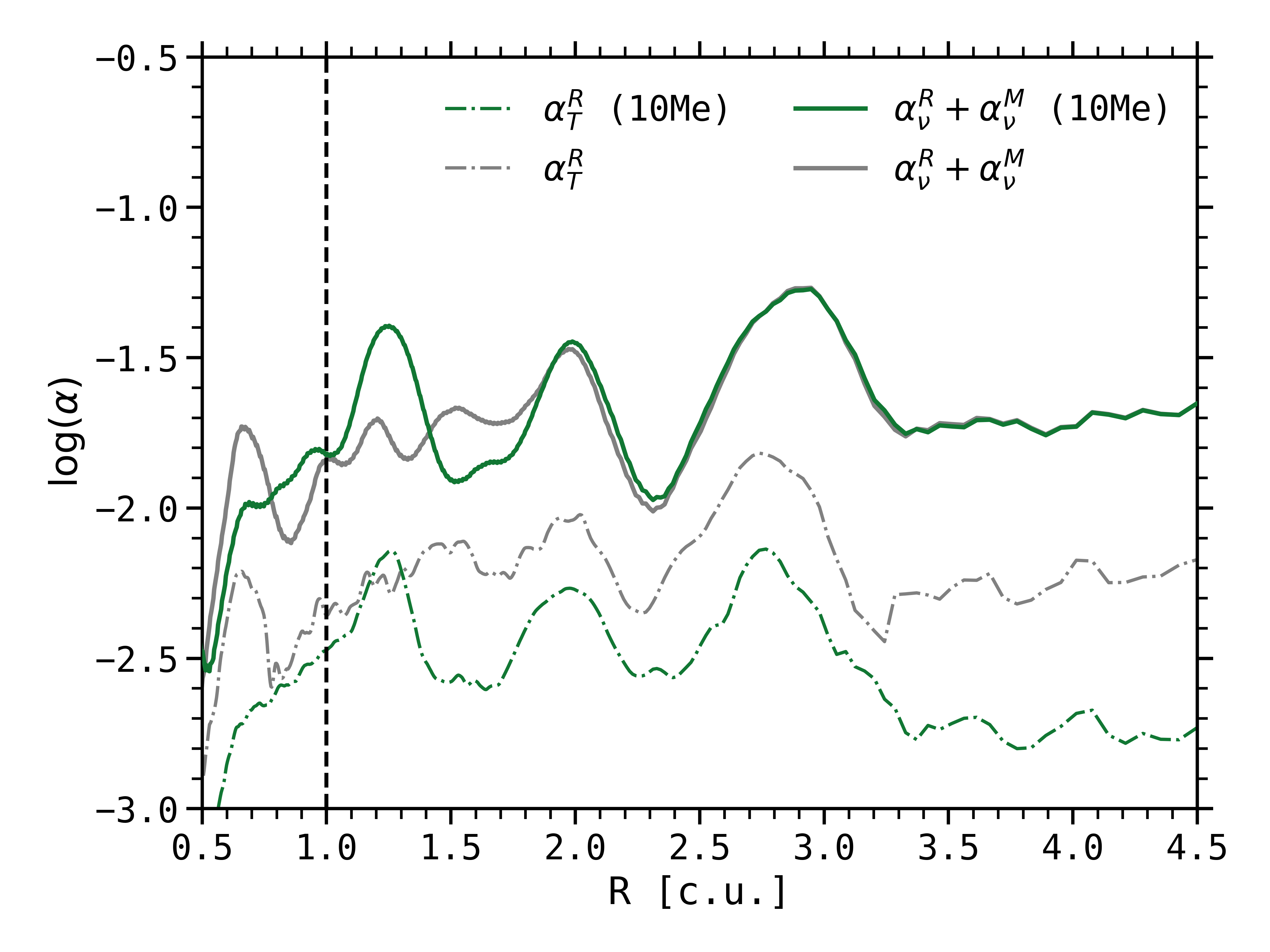}%
    \caption{Radial profiles of the logarithm of the total $\alpha_\nu$ (solid lines) and $\alpha^R_T$ (dash-dotted lines) related to the turbulent Reynolds stress, with (green curves) and without (gray curves) a $10M_{\oplus}$ planet, for an initial $\beta_0$ of $10^3$ and averaged between $50$ and $100$ orbits at $R=1$.}
    \label{fig:appendix_alpha_turb}
\end{figure}

Our non-ideal MHD disk models are susceptible to be MRI and possibly VSI unstable. Applying the VSI cooling timescale criterion of \cite{LinYoudin2015} to our parameters ($\gamma=5/3,\,q=1,\,h=0.05$, with $-q$ being the power-law exponent of the radial temperature profile near the midplane $\tilde{T}(R)$), we need
\begin{equation}
\Omega \mathcal{B} <0.075 
\label{eq:VSI_criterion}
\end{equation}
to get "vigorous" VSI. Hence, our models are likely to have severely quenched (or even suppressed) VSI signatures since we always use $\Omega \mathcal{B}=0.1$.

\begin{figure*}
    \centering
    \includegraphics[width=0.99\hsize]{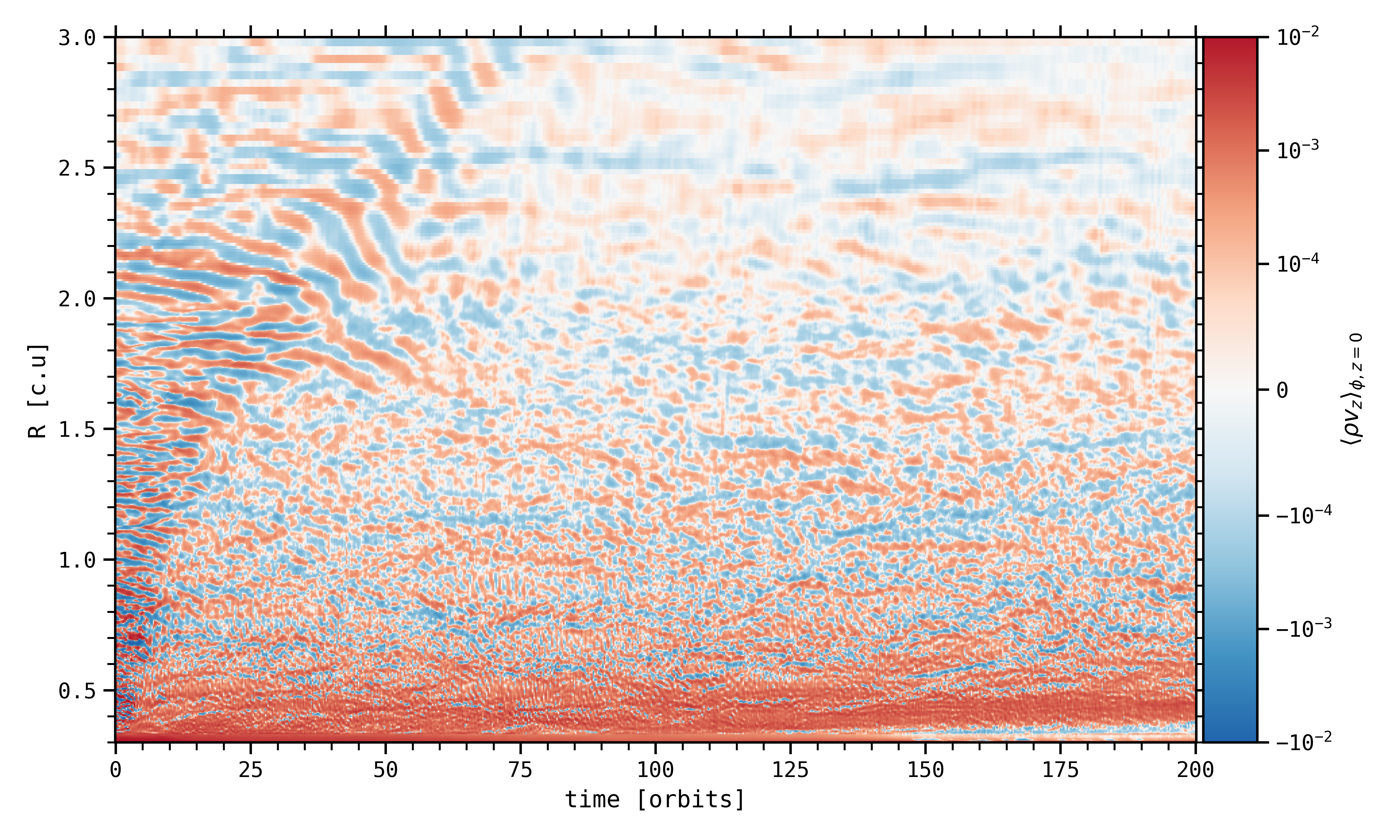}%
    \caption{Space-time diagram in log scale of the vertical component of the mass flux in the midplane and azimuthally averaged $\langle \rho v_z \rangle_{\phi,z=0}$, for the low-mass planet and low magnetization case \emph{$10$Me-$\beta_4$}.}
    \label{fig:appendix_10Me_spacetime_rhovz_vm}
\end{figure*}

In Figure~\ref{fig:appendix_10Me_spacetime_rhovz_vm}, we show the space-time diagram of the vertical component of the mass flux at the midplane in our 3D run with a low-mass planet, averaged in $\phi$ accross the disk. When the VSI is present, one expects to see obvious signatures of large-scale corrugation modes (e.g. fig. 3 in \citealt{Svanberg22}). Our diagram does not exhibit any of these patterns, suggesting that the VSI does not seem to develop in the simulation, which is coherent with the criterion of Eq.~(\ref{eq:VSI_criterion}). We also checked the non-developpment of the VSI at higher magnetization. In particular, we note the absence of large-scale coherent vertical motions most of the time during the simulation. We therefore expect the MRI to be the main driver of turbulence in our models, but not to be the dominant process for angular momentum transport compared to MHD winds since $\tilde{A}_{\rm m}=1$. Other studies observe a low level of turbulence due to MRI for such non-ideal MHD simulations with similar parameters \citep{Cui&Bai2022}.

We show in Figure~\ref{fig:appendix_alpha_turb} how the presence (in green) or absence (in gray) of a low-mass planet impacts the radial profiles of the total $\alpha_\nu$ (solid lines, as defined in Eq.~\ref{eq:alpha_tot}) and $\alpha^R_T$ (dash-dotted lines). This last quantity is linked to the turbulent component of the Reynolds stress, defined as $\alpha^R_T = \alpha_\nu^R - \overline{\langle\rho\rangle_{\phi,t}\langle v_R\rangle_{\phi,t}\langle (v_\phi-v_K)\rangle_{\phi,t}}$, the second term of this expression corresponding to the laminar component of the Reynolds stress. Note that the Reynolds stress is defined from the deviation from Keplerian rotation, but is not supposed to be vanishing when averaged in time \citep[see Eq.~20 in][]{Balbus&Papaloizou1999}. Hence this deviation may be split into a laminar term (i.e. non-zero when averaged in time) and a turbulent term (i.e. zero when averaged in time). We plot the turbulent stress $\alpha^R_T$ in Figure~\ref{fig:appendix_alpha_turb} (dash-dotted lines). We find that there is indeed turbulence in our disk model, both with and without a planet, but at a low level (a few $\alpha^R_T\sim 10^{-3}$) and therefore not sufficient to explain alone angular momentum transport. This confirms the predominance of the $\mathcal{M}_{\rm surf}$ term in the angular momentum budget, as presented in Figure~\ref{fig:main_stress}.

Besides, $\alpha^R_T$ is even not dominant in the total radial transport as stated earlier, (it is about $\simeq5$ times smaller than the total radial stress). In particular, for the run \emph{$10$Me-$\beta_4$}, we checked that $\alpha^R_T\simeq7\times10^{-4}$ for $\beta_0=10^4$, which is similar to the value obtained in \cite{Cui&Bai2022} for their run with instantaneous cooling and $A_{\rm m}=1$ constant within $\pm3.5H$. Far from the planet position ($R>2$), the total radial transport of angular momentum does not seem to be impacted by the planet, with the overlap of the green and gray solid lines. Close to $R=1$, the total stress is slightly larger (see, e.g., near $R=1.25$) in the simulation with planet, which is due to the slight carving of underdensity rings on both sides of the planet compared to a planet-free case. This underdensity is deeper in the outer part ($\simeq20\%$) than in the inner part ($10\%$) compared to the run without planet (see top panel of Figure~\ref{fig:appendix_rhobpmaccmsurf}), which is consistent with a $\alpha_\nu$ larger in the outer part than in the inner part. Note, as indicated in Appendix~\ref{sec:appendix_b1}, that these underdensities are also associated to a slightly more efficient wind torque at these locations for the simulation with planet ($29\%$ and $37\%$ for $\mathcal{M}_{\rm surf}$ respectively in the outer part and the inner part). If now we try to determine the impact of the low-mass planet on turbulence and $\alpha^R_T$, we observe that the level of turbulence is globally lower in the simulation with planet than in the simulation without planet by a factor $2-3$, except close to the planet location (between $R=0.8$ and $R=1.3$) where the level of turbulence is similar in both cases. 

\subsection{Stochasticity}
\label{sec:appendix_b3}

\begin{figure}[htbp]
    \centering
    \includegraphics[width=0.5\textwidth,keepaspectratio]{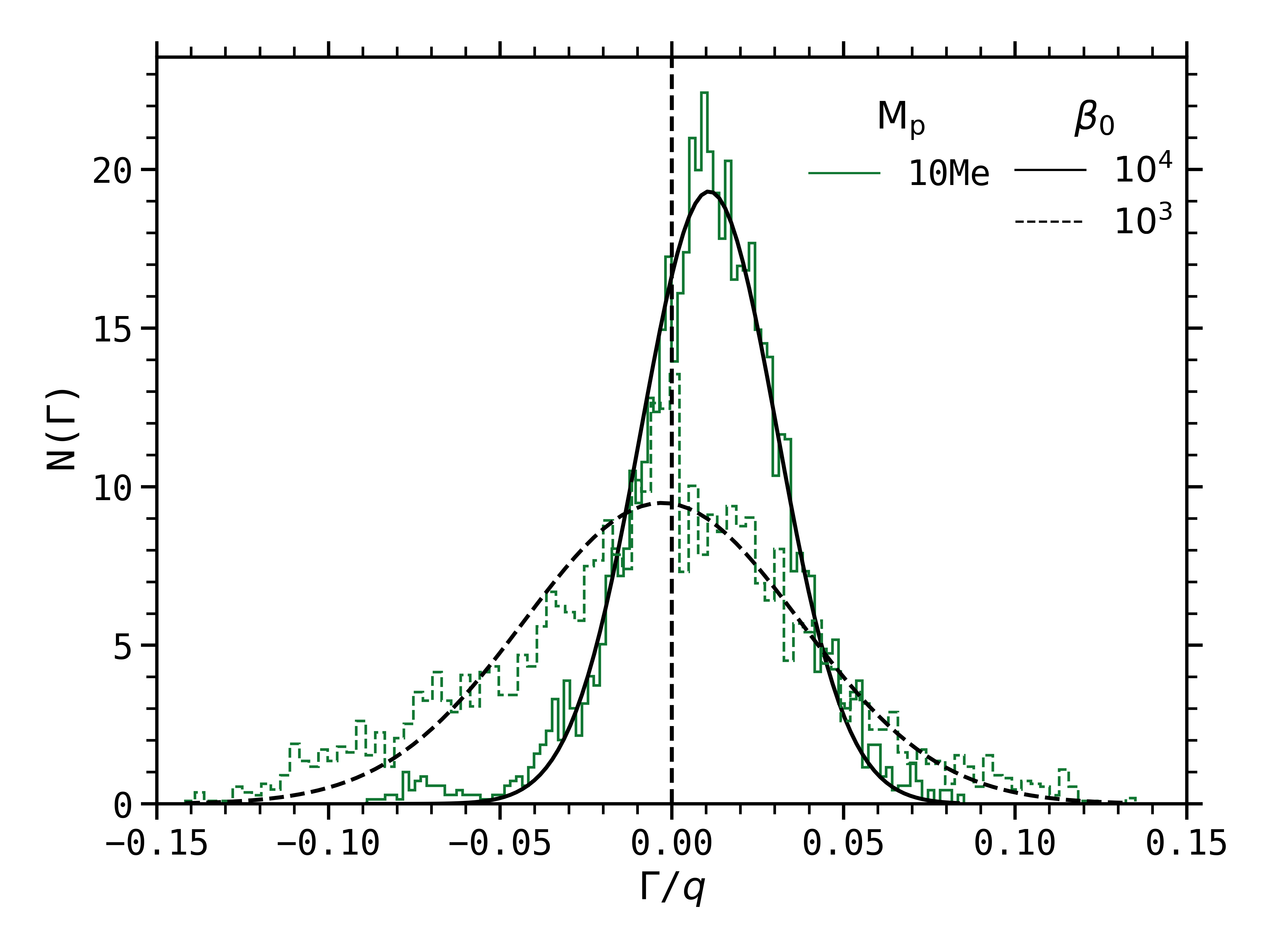}%
    \caption{Frequency distribution of the torques exerted on the least massive planets plotted in green in Figure~\ref{fig:main_torques}, e.g. for the runs \emph{$10$Me-$\beta_3$} and \emph{$10$Me-$\beta_4$}. These two histograms show the stochasticity of the torques experienced by the $10$ Earth mass planets, via a Gaussian distribution with standard deviation of $\sigma_3\simeq4\times10^{-2}$ and $\sigma_4\simeq2\times10^{-2}$ for an initial $\beta_0$ of $10^3$ and $10^4$ respectively.}
    \label{fig:appendix_pdf}
\end{figure}

In this paragraph, we aim at making this link between the level of turbulence in the disk with and without a low-mass planet at $R=1$ and the standard deviation of the probability density function of the planet torque $\Gamma$, and in particular its stochastic nature. To do so, we plot in Figure~\ref{fig:appendix_pdf} the frequency distribution of the planet torques $\Gamma$ for the runs \emph{$10$Me-$\beta_3$} and \emph{$10$Me-$\beta_4$} that were displayed in green in Figure~\ref{fig:main_torques}. We sample these planet torques every twentieth of an orbit during the simulations. We then compute histograms of the values taken by the torques in each bin with a bin width of $\Delta (\Gamma/q)=2.8\times10^{-3}$, with a procedure similar to the one presented in \citep{Nelson2005}. Fitting these distributions with Gaussian profiles, we obtain a standard deviation of $\sigma_3\simeq4\times10^{-2}$ and $\sigma_4\simeq2\times10^{-2}$ for an initial $\beta_0$ of $10^3$ and $10^4$ respectively. On the other hand, if we quantify the Reynolds stresses, and more precisely $\alpha^R_T$ in both runs \emph{$10$Me-$\beta_3$} and \emph{$10$Me-$\beta_4$}, we realize that decreasing the $\beta_0$ by a factor $10$ increases $\alpha^R_T$ at $R=1$ by a factor $\simeq5$, whereas the estimated standard deviation of the probability density function of the planet torque increases only by a factor $\simeq2$. Therefore, the dispersion of the planet torque does not scale directly with $\alpha^R_T$.

As a summary, when quantifying the link between the turbulence and the stochastic torques exerted on low-mass planets, we find that the turbulence has certainly a non-negligible role (the stochasticity seems to increase when the magnetic field strength increases) but not a proportional one, because $\alpha^R_T$ increases faster than the torque dispersion.

\section{Density and magnetization in a gap}
\label{sec:appendix_c}

\begin{figure}[htbp]
    \centering
    \includegraphics[width=0.5\textwidth,keepaspectratio]{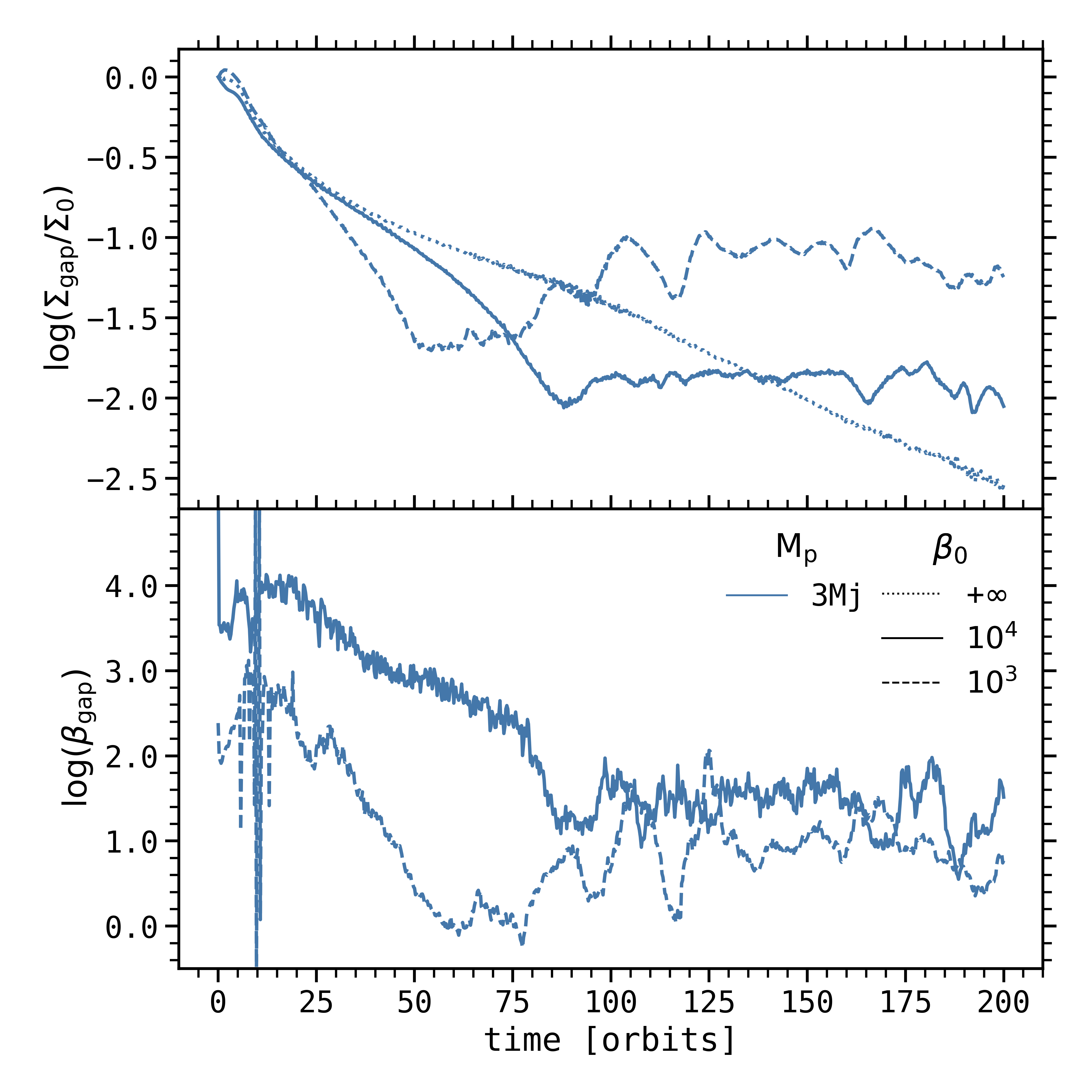}%
    \caption{Temporal evolution of the gas surface density (top panel) and the magnetization (bottom panel), in a gap carved by a $3$ Jupiter mass planet, for three initial magnetizations ($10^3$, $10^4$, and a viscous case).}
    \label{fig:appendix_sbgap}
\end{figure}

We focus in this appendix on the opening of a gap for a $3$ Jupiter mass planet, and especially the evolution of the density and magnetization inside the gap. To do so, we evaluate in the top panel of Figure~\ref{fig:appendix_sbgap} the temporal evolution of the azimuthally-averaged gap density, radially-averaging the annulus spanning from $R_p - 2R_{\rm hill}$ to $R_p + 2R_{\rm hill}$ and excised from $-2R_{\rm hill}/R_p$ to $2R_{\rm hill}/R_p$ in order to remove the contribution of the planet and its circumplanetary disk (CPD). $R_{\rm hill}$ is the planet's Hill radius and is defined in Eq.~(\ref{eq:rhill}). This procedure follows the evaluation of the gap's surface density $\Sigma_{\rm gap}$ presented in \cite{Fung2014} and \cite{Fung&Chiang2016}. We also evaluate in the bottom panel of Figure~\ref{fig:appendix_sbgap} the temporal evolution of the azimuthally-averaged $\beta$ in the gap, considering its minimum in the radial interval $\left[R_p - R_{\rm hill},R_p + R_{\rm hill}\right]$ and also excised from $-2R_{\rm hill}/R_p$ to $2R_{\rm hill}/R_p$. We define here several quantities that can be useful to understand and analyze qualitatively the process of gap opening:
\begin{itemize}
    \item $\Sigma^{\rm eq}_{\rm gap}$ is the quasi-equilibrium value for the density in the gap, i.e. such that the mean value of $\Sigma_{\rm gap}$ does not evolve much with time.
    \item $\delta\Sigma^{\rm eq}_{\rm gap}$ is the variability of the density in the gap once it has reached $\Sigma^{\rm eq}_{\rm gap}$.
    \item $\displaystyle v_{\rm gap}=\left|\frac{d\Sigma_{\rm gap}}{dt}\right|$ is the instantaneous gap opening speed, from the initial density at the planet location to $\Sigma^{\rm eq}_{\rm gap}$.
    \item $t^{\rm eq}_{\rm gap}$ is the time needed by $\Sigma_{\rm gap}$ to reach $\Sigma^{\rm eq}_{\rm gap}$.
\end{itemize}

Using this metrics and focusing on the $3M_{j}$ planet cases, we notice in the top panel of Figure~\ref{fig:appendix_sbgap} that $\Sigma^{\rm eq}_{\rm gap}$ increases with the magnetization. More precisely, $\Sigma_{\rm gap}$ reaches $\simeq 3\times10^{-3}$ of its initial value $\Sigma_0$ (and still decreasing) for \emph{$3$Mj-$\alpha_0$} after $200$ planet orbits, whereas it reaches $\Sigma^{\rm eq}_{\rm gap}\simeq 10^{-2}~\Sigma_0$ for \emph{$3$Mj-$\beta_4$} after $t^{\rm eq}_{\rm gap}\simeq 90$ planet orbits and $\Sigma^{\rm eq}_{\rm gap}\simeq 2\times 10^{-2}~\Sigma_0$ for \emph{$3$Mj-$\beta_3$} after $t^{\rm eq}_{\rm gap}\simeq 50$ planet orbits. This is consistent with our interpretation based on the interplay between the accretion rate $\dot{M}_{\rm acc}$, $\beta$ and $\Sigma$ and presented in the second paragraph of Section~\ref{sec:GO_variability}. $\delta\Sigma^{\rm eq}_{\rm gap}$ also increases with the magnetization, and could be linked to stronger time-variability in the gap when the initial magnetization inceases, as mentioned in the last paragraph of Section~\ref{sec:GO_variability}. We will not focus on these first two quantities here, but rather on $v_{\rm gap}$ and $t^{\rm eq}_{\rm gap}$. On the one hand, the gap opening speed is larger when $\beta_0$ decreases, at least early in the simulations and before $\Sigma_{\rm gap}\simeq\Sigma^{\rm eq}_{\rm gap}$ (see, e.g., between $20$ and $90$ orbits). It would suggest that a torque linked to the accretion is actually helping to form the gap. On the other hand, $t^{\rm eq}_{\rm gap}$ decreases when $\beta_0$ decreases, which means that the gap reaches its minimum density earlier when the magnetization is higher, respectively after $200$, $90$ and $50$ planet orbits for \emph{$3$Mj-$\alpha_0$}, \emph{$3$Mj-$\beta_4$} and \emph{$3$Mj-$\beta_3$} respectively.

In order to reconcile these two processes, we can argue that the magnetic torque linked to the vertical extraction of angular momentum helps the planet-related torque to open a gap, and even more if $\beta_0$ decreases. The deeper the gap is, the more magnetic field accumulates in it, which leads to an increase of the wind torque that helps even more to open the gap. We thus witness a runaway gap opening, particularly visible in the concave segment of the curves just before $\Sigma_{\rm gap}\simeq\Sigma^{\rm eq}_{\rm gap}$ (see the $\beta_0=10^4$ and $10^3$ cases in the top panel of Figure~\ref{fig:appendix_sbgap}). At that moment, when $\beta_{\rm gap}$ reaches a threshold of the order of $1-10$, a quasi steady state is achieved in the gap opening process. This happens earlier when the magnetization is initially lower. It is confirmed by looking at the bottom panel of Figure~\ref{fig:appendix_sbgap}, which shows that $\beta_{\rm gap}$ reaches a value $<10$ near $50$ and $90$ orbits respectively when $\beta_0=10^3$ and $10^4$.

\end{document}